\documentclass[a4paper,12pt]{spieman}  
\usepackage{amsmath,amsfonts,amssymb}
\usepackage{graphicx}
\usepackage{setspace}
\usepackage{tocloft}
\usepackage{ulem}
\usepackage{xcolor}
\usepackage{makecell}
\usepackage{subcaption}
\usepackage{lineno}
\usepackage{hyperref}
\usepackage{comment}
\usepackage{tikz}
\usepackage{gensymb}

\title{Development of ProtoPol: a medium resolution echelle spectro-polarimeter for PRL telescopes, Mt Abu, India - Part II : the data-reduction pipeline, on-sky characterization \& performance verification and first science results}

\author[a,b,*]{Arijit Maiti}
\author[a,*]{Mudit K. Srivastava}
\author[a, $\dagger$]{Vipin Kumar}
\author[a]{Bhaveshkumar Mistry}
\author[a]{Ankita Patel}
\author[a, $\dagger \dagger$]{Vaibhav Dixit}
\author[a,$\ddagger$,$\ddagger \ddagger$]{Ruchi Pandey}
\author[a]{Jay Chitroda}

\affil[a]{Astronomy and Astrophysics Division, Physical Research Laboratory, Ahmedabad, 380009,
India}
\affil[b]{Indian Institute of Technology, Gandhinagar, 382335, India}
\affil[$\dagger$]{\textit{Current affiliation}: I. Physikalisches Institut, Universit\"at zu K\"oln, Z\"ulpicher Stra\ss e 77, 50937, K\"oln, Germany}
\affil[$\dagger \dagger$]{\textit{Current affiliation}: Advanced Engineering Group, Azista Industries, Applewoods Township, Ahmedabad, 380058, Gujarat, India}
\affil[$\ddagger$]{\textit{Current affiliation}: Department of Physics \& Astronomy, Johns Hopkins University, 3400 N. Charles St, Baltimore, MD 21218, USA}
\affil[$\ddagger \ddagger$]{\textit{Current affiliation}: X-ray Astrophysics Laboratory, NASA Goddard Space Flight Center, Code 662, Greenbelt, MD 20771, USA}

\cftpagenumbersoff{figure}
\cftpagenumbersoff{table} 
\begin{document} 

\maketitle

\begin{abstract}
We present the development of ProtoPol - a medium resolution echelle spectro-polarimeter for the PRL 1.2m and 2.5m telescopes at Mt Abu observatory, India. In this second and final part of the paper series, we report on the development of a dedicated data reduction pipeline of ProtoPol along with several characterization,  performance evaluation, and scientific observations to quantify the performance of the instrument. ProtoPol provides a spectral resolution in the range of $\sim$0.4 - 0.75$\AA$ across various orders in the visible wavelength range of 4000-9600$\AA$. On PRL 2.5m telescope, an SNR of 10 is achieved for $m_V\sim13.2$ source in 1 hour of integration time, and its throughput is estimated to be $\sim$6\% including all the contributing factors such as atmospheric transmission, telescope reflectivity, instrument's optics, CCD efficiency etc. ProtoPol achieved a linear polarization accuracy $\delta P \approx 0.1-0.2\%$ in 2 hours of integration time for a source with $m_V\approx8$. The instrumental polarization is determined to be around $0.1\%$.  We also present the first science results with ProtoPol to demonstrate the capabilities of the instrument. A sample of Herbig Ae/Be stars, classical Herbig stars, Symbiotic stars, and AGB/post-AGB stars were observed over the period of one and half years for their spectro-polarimetry measurements covering various physical mechanisms such as intrinsic line polarization in Herbig and classical Be stars, Raman scattered features in Symbiotic stars, as well as continuum polarization in AGB/post-AGB stars to verify the polarization performance of the instrument.
\end{abstract}


{\noindent \footnotesize\textbf{*}Corresponding author(s),  \linkable{arijitmaiti@prl.res.in}, \linkable{mudit@prl.res.in}}

\newcommand\blfootnote[1]{%
  \begingroup
  \renewcommand\thefootnote{}%
  \footnote{#1}%
  \addtocounter{footnote}{-1}%
  \endgroup
}

\begin{spacing}{2}

\section{Introduction}\label{sec1}

Spectro-polarimetry is a powerful technique to probe the morphology of variety of astrophysical situations where limitations of spatial resolutions pose a challenge. The measurement of polarization in various spectral features such as continuum, emission/absorption lines, etc. across the spectrum, their variation across the profile shape, and temporal variability, etc. provide unique information which is generally not possible with other observing techniques \cite{ikeda2003development, arasaki2015very, piskunov2011harpspol, donati2003espadons}. In unresolved stellar systems, spectro-polarimetry is an effective probe of the circumstellar environment of stars on small spatial scales. Any kind of asymmetry in the system, like circumstellar disks, rotationally distorted winds, magnetic fields, and asymmetric radiation fields, etc. can cause a polarization change across a spectral line. Detecting these signatures can help constrain the geometry and density of circumstellar material and help in probing the near-star environments, which would otherwise not be possible \cite{rudy1978polarization, cassinelli1987polarization}. However, spectro-polarimetry, like any polarimetry technique, is a photon-hungry process, and a very high signal-to-noise ratio (SNR) is required to achieve the polarization accuracy to $0.1-0.2\%$ level, which is generally required as typical spectro-polarimeteric signals show only a fractional percent change in polarization across the spectral line. 
\par
Given the significance of spectro-polarimetric measurements in astronomy, a multimode instrument, named M-FOSC-EP (Mt. Abu Faint Object Spectrograph and Camera-Echelle Polarimeter) is currently under development for Physical Research Laboratory (PRL)'s Mt Abu Observatory (having 1.2m and 2.5m optical/NIR telescopes), Mt Abu, India \cite{bastin2022mount, Deshpande1995, kumar2022designs}. M-FOSC-EP is designed to have two channels (a) an intermediate resolution (R$\sim$ 15000) spectro-polarimeter and (b) a low-resolution arm to provide lower resolution (R$\sim$ 500-700) spectroscopy and filter-based imaging. For the development process of M-FOSC-EP, another prototype instrument was conceived to evaluate the development methodology of the spectro-polarimetric arm. The resultant instrument, named ProtoPol, while initially conceived as a prototype was later elevated to the level of a full-fledged back-end instrument for PRL telescopes \cite{srivastava2024development}.  Developed completely with commercially available off-the-shelf components, ProtoPol was designed on the concept of echelle and cross-disperser (CD) gratings having a wavelength range from 4000 to 9600 $\AA$ with a resolution ($\delta \lambda$) in the range of 0.4-0.75 $\AA$. The instrument was successfully developed and subsequently commissioned firstly on PRL 1.2m telescope in December, 2023 and later shifted to PRL 2.5m telescope in February, 2024. Aforementioned rationale behind the instrument, its science objectives, optical and opto-mechanical designs, development methodology, assembly-integration-tests (AIT), laboratory characterization etc. are described in details in Part-I of this paper series (Srivastava et al.; thereafter Paper-I).
\par
This paper, Paper II of the series, focuses on the subsequent activities of the instrument development which includes the development of a dedicated fully-automated spectro-polarimetric/spectroscopic data-reduction pipeline for ProtoPol - which could also be used for similar medium-resolution echelle spectro-polarimeters, on-sky performance evaluation and characterization, and the demonstration of instrument's capabilities by the measurements of polarization across the spectral features in a range of scientifically interesting targets. A variety of suitable sources (such as standard unpolarized and polarized stars, science targets etc.) were observed with ProtoPol and the on-sky performance of ProtoPol has been thoroughly analyzed over two observing seasons, both on PRL 1.2m and 2.5m telescopes. This includes the determination of instrumental polarization, polarization accuracy, polarization ripple,  checks for any systematic polarization error, determination of the instrumental depolarization factor from observations of standard unpolarized stars through a Glan-Taylor prism (GT), etc. Instrumental throughput was determined from the observations of Jupiter, considering it as an extended source, thereby completely filling the slit/aperture and negating the effect of slit-loss. Observations of faint M-Dwarf stars were carried out to determine the spectroscopic magnitude limit of the instrument. Several science observation campaigns were carried out using ProtoPol to verify the spectro-polarimetric performance of the instrument, and are presented as first science results. This includes observations of a bright sample of classical Be stars, a sample of Herbig Ae/Be stars, a sample of Symbiotic stars, and a sample of AGB/post-AGB stars. These sources exhibit considerable polarization in their emission lines and continuum, due to several interesting physical effects, and thus found to be suitable candidates for demonstrating the spectro-polarimetric performance of ProtoPol. The total throughput of the instrument was determined to be $6-7\%$ without accounting for slit-loss. Observations of standard unpolarized stars reveal instrumental polarization of $\sim 0.1\%$. The zero point stability error ($\delta$P) of $0.1-0.2 \%$ per pixel was achieved for V$\sim8$ star with 2 hours of integration time, whereas for spectroscopic observations, SNR of 10 per pixel was achieved for V$\sim13.2$ star on 2.5m telescope.
\par
The paper is organized in the following way: section 2 discusses the complete spectro-polarimetric data reduction pipeline for ProtoPol; On-sky performance and instrument characterization of ProtoPol is presented in section 3, including instrumental polarization, polarimetric error across lines, polarization ripple, polarimetric accuracy, etc.; Section 4 presents the first science results with ProtoPol, which includes samples of Herbig Ae/Be stars, classical Be stars, Symbiotic stars, and AGB/pot-AGB stars; and finally, Section 5 concludes the paper and gives a summary of the achievements derived from ProtoPol project.

\par 
\section{Development of a Data Reduction Pipeline for ProtoPol}\label{sec:ReductionPipeline}

The data reduction pipeline of any scientific instrument plays the most crucial role in converting raw observational data into science-ready measurements data, which can then be analyzed for a variety of purposes. Given the minimal amount of polarization present in astrophysical sources, polarization measurements in general and spectro-polarimeters in particular poses one of the toughest challenges for the development of a robust pipeline for data reduction, which will result into scientifically faithful data for further analysis. The pipeline would remove and correct for various instrumental systematic effects, such as bias subtraction, flat-field response correction, background removal, wavelength calibration, etc, for various echelle orders to generate science-ready spectra for o- and e- rays for all the orders for the determination of Stokes parameters (I, Q, U, V).  There are comparatively fewer spectro-polarimeters around the world \cite{ikeda2003development, arasaki2015very, piskunov2011harpspol, donati2003espadons} each with its own specific design complexities, which also reflect in their data reduction processes \cite{harrington2008spectropolarimetric}. While the underlying philosophy remains the same, such complexities demands subtle instrument-specific changes in the pipeline development, thereby necessitating the development of an independent data reduction pipeline for ProtoPol.
\par
\begin{figure}[htbp]
  \begin{center}
  \includegraphics[width=\textwidth]{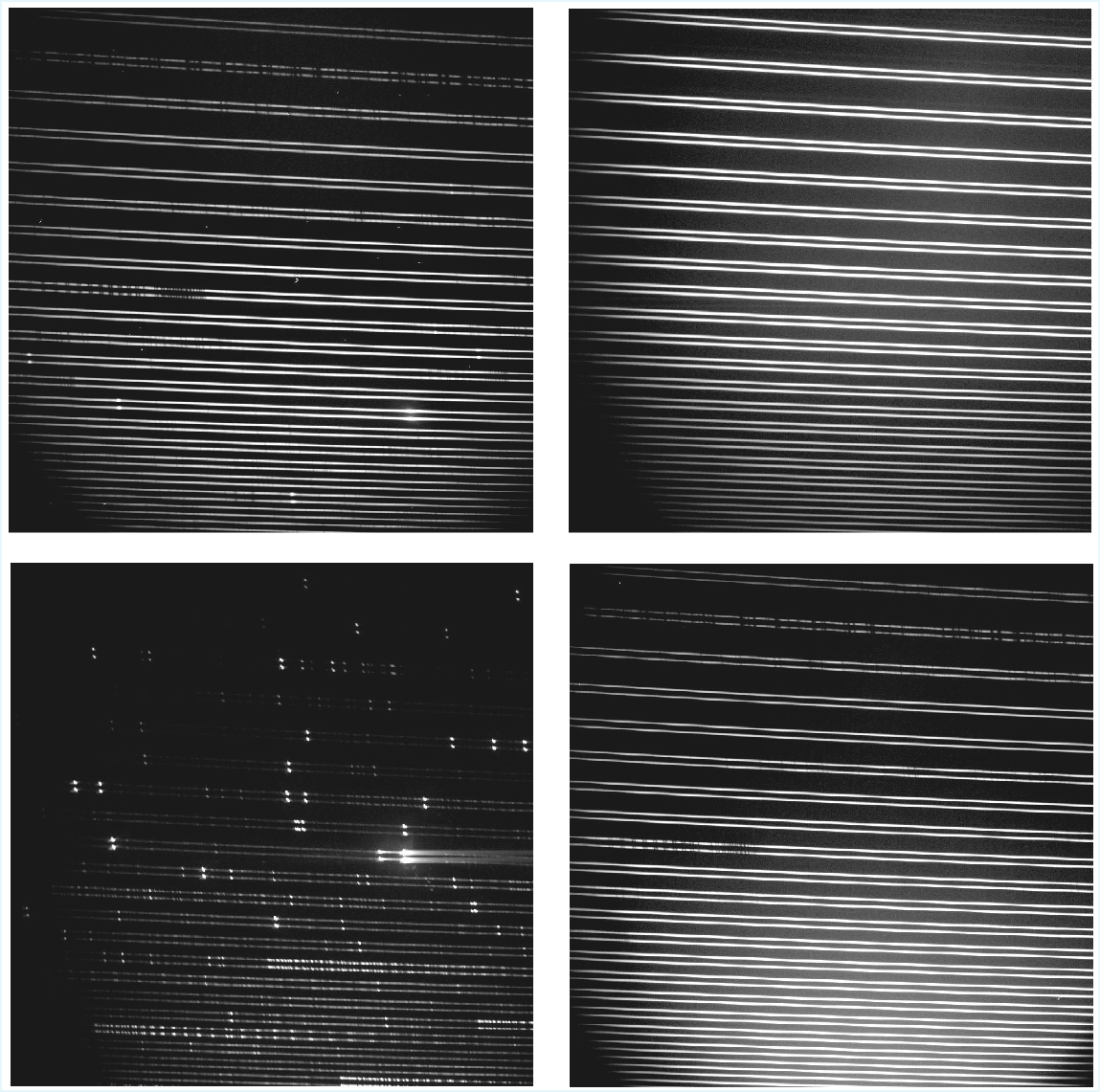}  

  \caption{Examples of raw data frames from ProtoPol. \textit{(Top left)} science frame (RW Hya - a symbiotic star) - various emission and absorption features can be noticed in the frame; \textit{(Top right)} Halogen frame; \textit{(Bottom left)} U-Ar frame; \textit{(Bottom right)} A standard star frame (standard spectro-photometric star - HR 5191)}
  \label{ProtoPol_Data_Frames}
  
  \end{center}
\end{figure}

The data reduction pipeline for ProtoPol serves the purpose of both echelle spectroscopy as well as spectro-polarimetric data reduction. The pipeline is developed to work in an automated way with minimal human intervention. It is developed on \href{https://www.python.org/}{Python 3.11} platform utilizing open source libraries such as \href{https://www.astropy.org/}{ASTROPY}, \href{https://numpy.org/}{NUMPY}, \href{https://scipy.org/}{SCIPY}, etc. The pipeline provides facilities for data reduction and visualization of data reduction at each step, if required.  A polarimetric observation set (for measuring linear polarization, and ignoring circular polarization effects) consists of four data frames collected in the same observing conditions and identical settings of the instrument for 4 positions of the half-wave plate (HWP) at 0, 22.5, 45, and 67.5 degree. Each frame contains the spectra of o- and e- rays in multiple echelle orders. These science data frames are also complemented with additional spectral lamp (Uranium-Argon) calibration frames for wavelength calibration, halogen lamp frames for echelle order trace determination, bias and dark frames etc. Observations of standard spectro-photometric stars are conducted for relative flux calibration on the contemporaneous nights. These are shown in Figure~\ref{ProtoPol_Data_Frames}. The pipeline then generates extracted wavelength-calibrated spectra of the target corrected for various instrumental effects for each of the HWP's positions. Subsequently, the spectra for the two orthogonally polarized o- and e- rays are extracted for each of the orders and then used to determine the Stokes parameters as a function of wavelength. The full pipeline consists of several tasks which is performed in a sequential manner to derive the science-ready spectra and polarization data for the given source. The algorithm of the reduction pipeline can be summarized in the following steps. 

\begin{enumerate}  
\item Subtraction of bias and dark frames from all science, standard star, sky, and calibration lamp frames.
\item Cosmic ray corrections from all frames.
\item Preparing the trace maps of echelle orders using halogen and/or science frames.
\item Preparing the map for the scattered light and its subtraction from the science frames.
\item Extraction of the spectra for each order and each of the o- and e- rays. 
\item Determination of wavelength solution using the data frame of Uranium-Argon (U-Ar) spectral calibration lamps and performing the wavelength calibration of all the science and auxiliary frames.
\end{enumerate}


\begin{figure*}
\begin{center}
    \includegraphics[width=\textwidth]{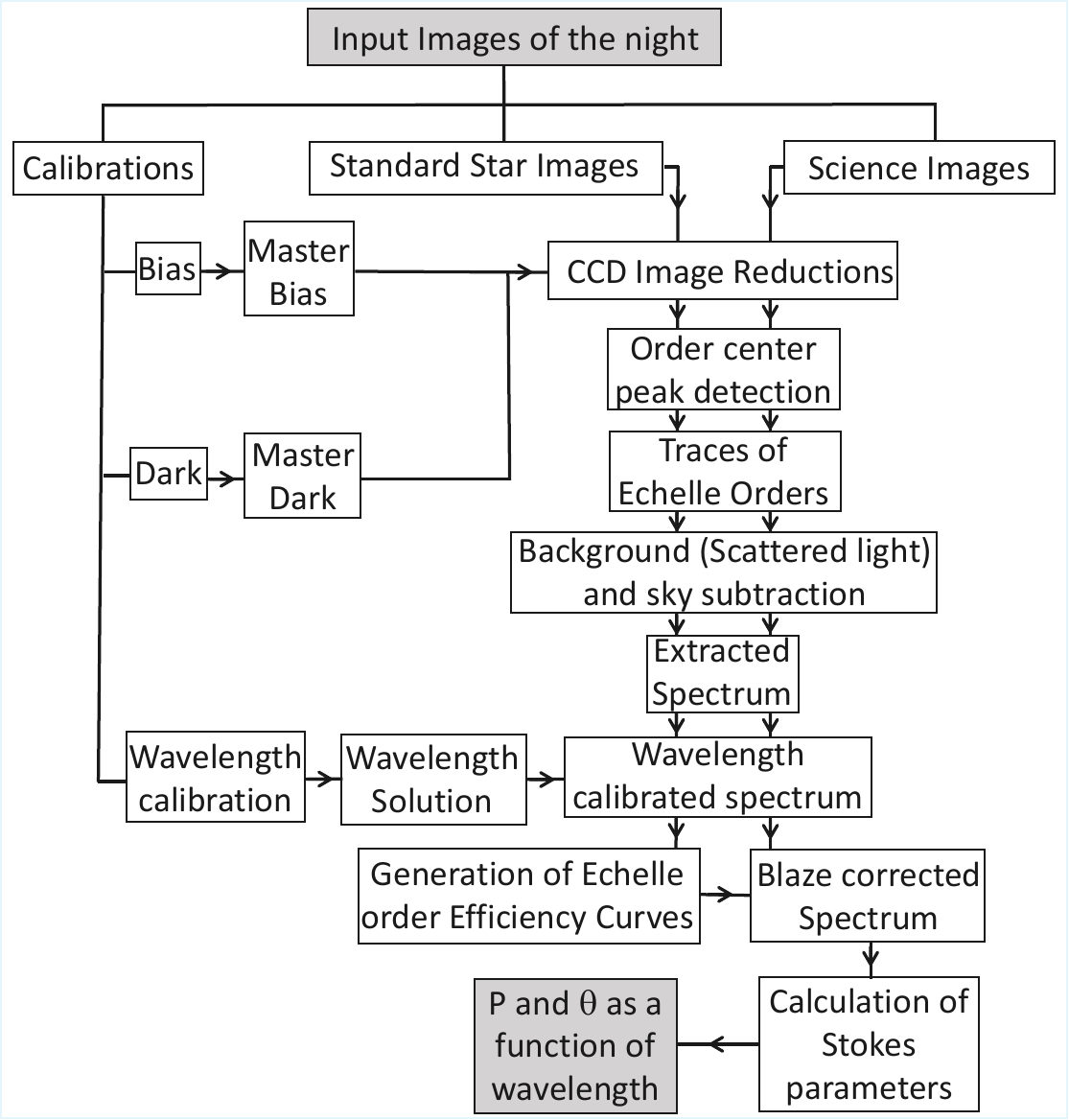}    
    \caption{Flow diagram showing the processes that are in general used by the ProtoPol pipeline for spectro-polarimetric data reduction process. The most important functions at each step are also shown.}\label{ProtoPolAlgo}
\end{center}    
\end{figure*}

The aforementioned steps are applied both on science data frames and on the standard spectro-photometric star's data frames. While the Stokes parameter can directly be determined by comparing the reduced relative intensities of o- and e- rays (see section - \ref{Stokes_Section}), the reduced spectra of standard stars are used to determine the cumulative efficiency of entire observing set-up (i.e. atmospheric transmission, efficiency of telescope and the instrument etc.) by comparing its known archival spectra with the reduced ones. This information is then used to generate the flat field corrected echelle spectra of the science target. In the following sub-sections, each of these steps are explained in greater details. The flow chart of the data reduction pipeline is shown in Figure~\ref{ProtoPolAlgo}.


\subsection{Bias and Dark Subtraction}
 Typically five to ten bias frames are collected during every observing night which are median-combined to construct the master bias frame which is subtracted from all other data frames. The science camera (ANDOR model no. \href{https://andor.oxinst.com/products/ikon-large-ccd-series/ikon-m-934}{iKon M934 1K$\times$1K CCD camera}) is operated at -80 degree C which leads to a typical dark current of 0.017 $\text{electrons}$ pixel$^{-1}$s$^{-1}$ which needs to be considered for longer exposures. Typically five dark frames with the same exposure times as the object frames are also collected during each observing cycle which are median-combined to construct the master dark frame. For observation targets with different exposure times as compared to the acquired dark frames,  the dark frames are scaled appropriately. Bias and dark frame correction is performed by subtracting the master bias frame and the corresponding master dark frame from the science images.

\subsection{Identification of echelle orders}

Multiple echelle orders are to be identified and traced on the CCD frame for their reduction. Usually, a continuum lamp frame, such as halogen, is used to perform this procedure if the spectra of the target stars are not bright enough. The orders are identified firstly near the central columns of the CCD frame due to the higher signal-to-noise-ratio (SNR) at the center of the blaze profiles of the gratings. Several columns (typically ten) about the central column are median combined to construct the master center column, which is then used for order identification. This reference central column is then convolved with a Gaussian kernel to further smoothen out the signal. This allows the rejection of spurious signals which otherwise may lead to false peak identification caused by local pixel-to-pixel signal variation.
\par
\begin{figure}[H]
\begin{center}
    \includegraphics[width=\textwidth]{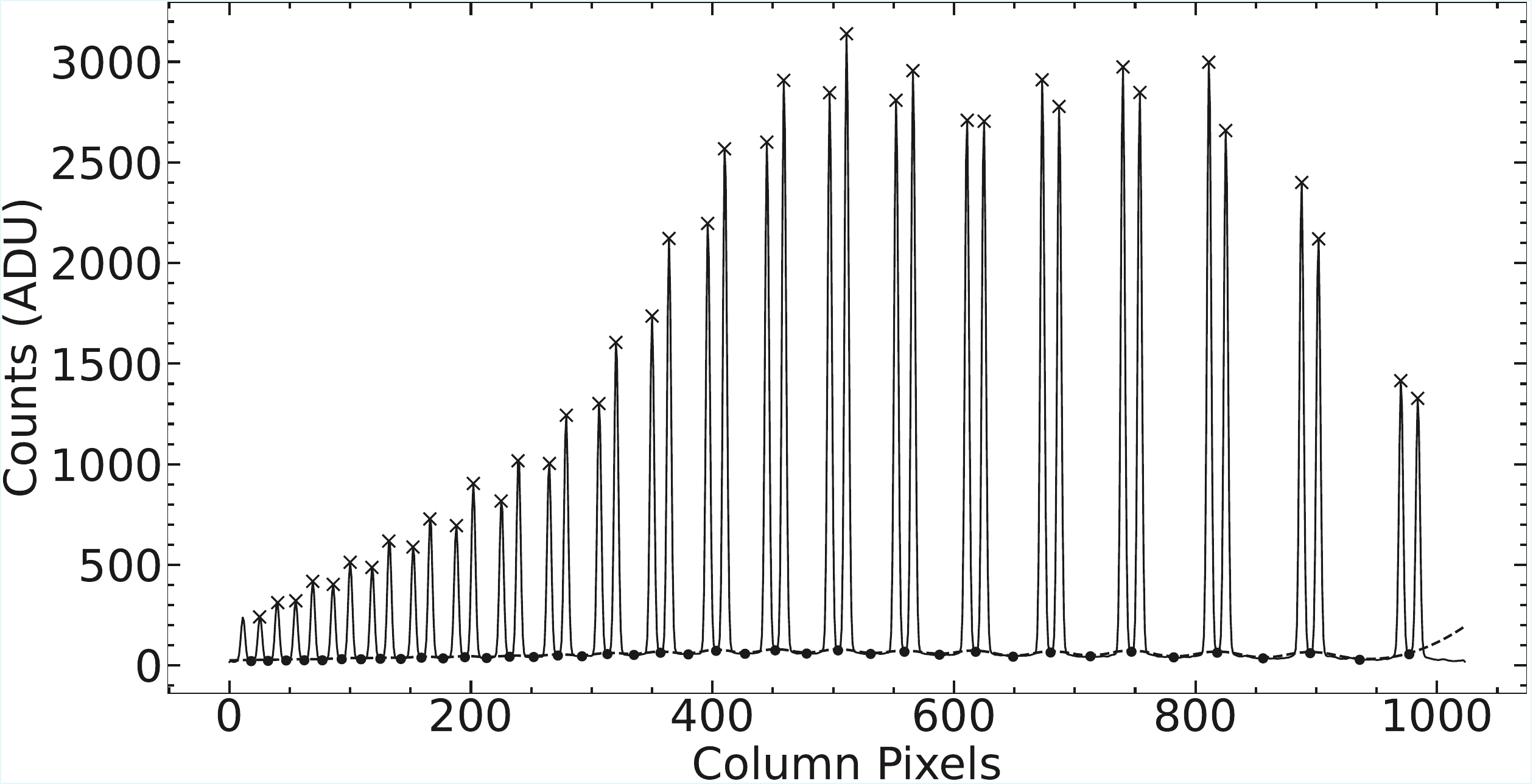}
    \caption{Selected central column spectra for peak detection for a ProtoPol halogen data frame. The black points show the points selected to compute the continuum. The dashed line shows the continuum level after the desired number of detected peaks is achieved, and the black crosses display the detected peak positions.}\label{detected_peaks}
\end{center}    
\end{figure}

Subsequently, all the local peaks of the smoothed reference column are identified. A threshold is set in the peak-detection function such that the function searches for local maxima above this threshold value. This threshold is selected from the underlying background continuum along the reference column. It is computed by selecting mid-points between the initially detected peaks. This background continuum is then set as the initial threshold. The number of peaks detected is then compared to the pre-defined number of peaks for that given cross-disperser grating. If false peaks are detected, the background threshold is increased. This is done iteratively till the correct number of peaks are detected.  As each spectro-polarimetric order contains two traces of two orthogonally polarized rays, we make sure that the number of detected peaks is even. If the total number of detected peaks is odd, the detected peak from such orders are deleted as they cannot be used for spectro-polarimetric data reduction. Furthermore, peaks detected very close to the edge of the detector are also eliminated as part of their order trace would fall outside of the detector. The starting order number is also given such that the order numbers of all the detected peaks are correctly identified. Figure~\ref{detected_peaks} shows the detected peaks of echelle orders after continuum fit. 

\subsection{Tracing of echelle orders}

\begin{figure}[H]
  \begin{center}
    \includegraphics[width=\textwidth]{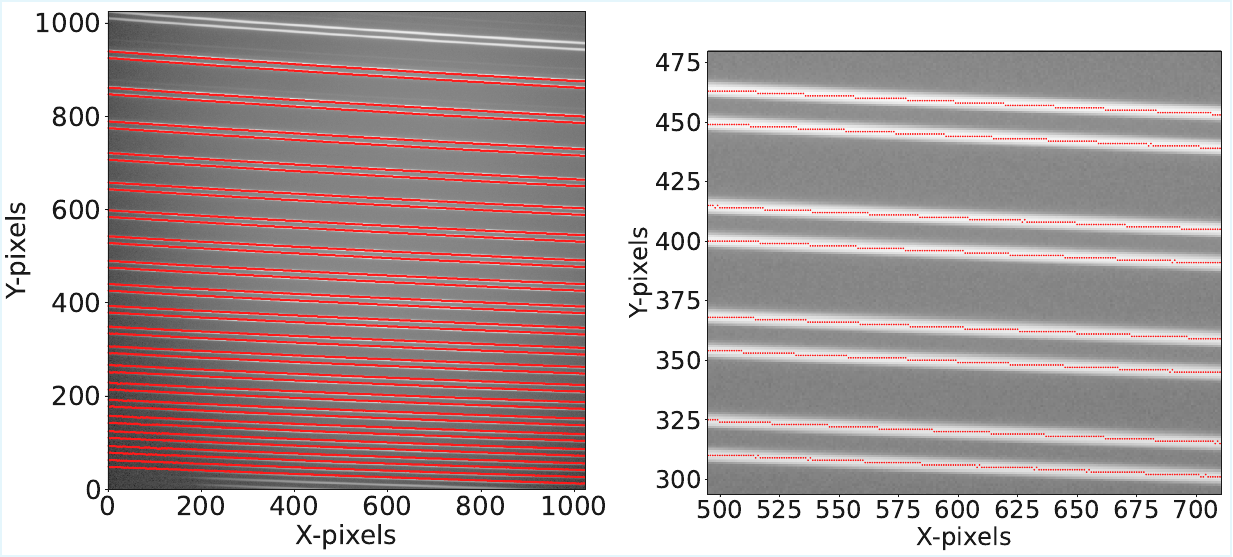}
  \caption{\textit{(Left)} Traced orders of ProtoPol halogen full data frame. \textit{(Right)} Zoomed-in view of the same frame.}\label{order_trace}
  \end{center}
\end{figure}

For most of the echelle spectrographs, the traces of the spectral footprints on the CCD frame show strong curvature \cite{baranne1996elodie, bernstein2015data}. Starting from the central position of each order, as determined in the first step of trace identification, the vertical position (i.e. the row number) of the order in each column is identified by fitting a Gaussian in zones centered on the positions contiguous to the already identified order centers. The width of the zone where the Gaussian is fit must be similar to the approximate cross-dispersion (vertical) extension of the orders. For columns where the SNR is low, like at order edges or at absorption dips where a Gaussian could not be fit are ignored. Finally, a high-order polynomial is fitted to the derived peak positions of the order traces across each CCD column computed for each order, and the coefficients of the fit are saved for extracting the science and calibration spectra. Orders on the extreme edges are ignored if the parameters of the fit lead to negative numbers or numbers greater than detector size as they represent orders only partially falling on the detector. Figure~\ref{order_trace} shows the traced orders of a ProtoPol halogen frame.

\subsection{Scattered Light Subtraction}

\begin{figure}[H]
  \begin{center}
    \includegraphics[width=\textwidth]{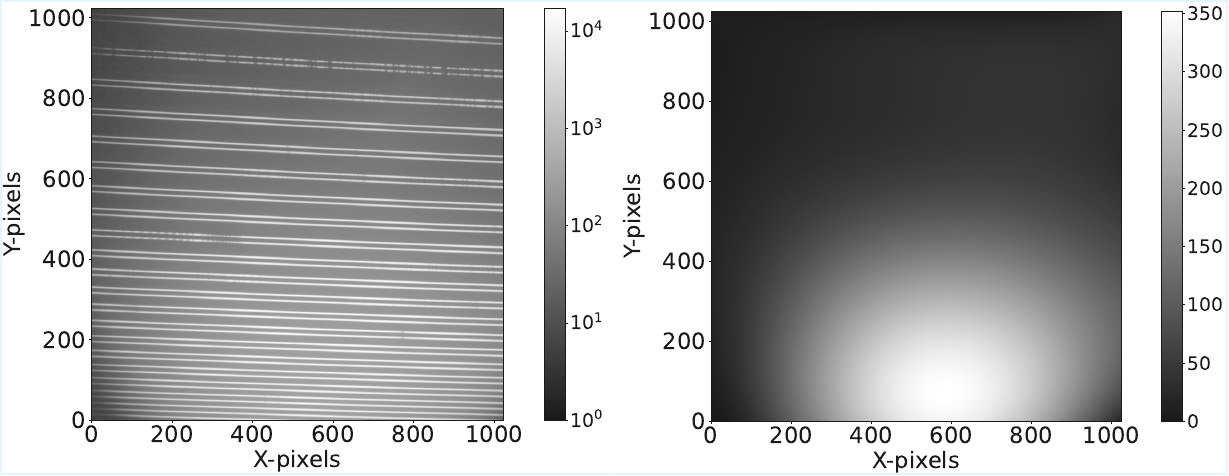}
  \caption{\textit{(Left)} ProtoPol data frame for standard unpolarized star HD 61421 (R band magnitude = -0.05) \textit{(Right)} Extracted scattered light background from the data frame.}\label{scatter}
  \end{center}
\end{figure}

In echelle spectrographs, in addition to the dispersed light registered in each order, the scattered light would also fall on the CCD due to various sources such as gratings, the roughness of the optical surfaces, stray light inside the instrument, etc \cite{piskunov2021optimal, churchill1995treatment}. This contamination produces a smooth background that has to be removed before extracting the spectra in order to conserve the sanity of the spectral features such as the true depth of the absorption lines etc. The effect is more prevalent in brighter targets compared to fainter ones. The extent and magnitude of this scattered background is estimated from the data points in-between the o- and e- ray traces and between adjacent orders (typically outside of $\pm4\sigma$ of the detected peak position of a spectral trace). Thereafter, a 2D polynomial fit is made to interpolate the scattered background values to the locations of the spectral traces, in order to generate the background map for the full CCD frame. To smoothen out the effects of spurious signals, a two-dimensional median filter is applied to the constructed scattered background frame, which is then subtracted from each original science image. The scattered light background map for a bright star HD 61421 (R band magnitude $\sim -0.05$) is shown in Figure~\ref{scatter}. The peak value of the scattered background is estimated to be around 1-2$\%$ of the peak signal value. 


\subsection{Removal of cosmic rays footprints}
Unexpected signals from cosmic rays were removed using  \href{https://lacosmic.readthedocs.io/en/stable/}{lacosmic} Python package based on Laplacian edge detection methods \cite{van2001cosmic}. The parameters of the function are to be appropriately selected to avoid misidentifying an emission line as a cosmic ray signal. 

\subsection{Order Extraction}
\begin{figure}[H]
\begin{center}
    \includegraphics[width=\textwidth]{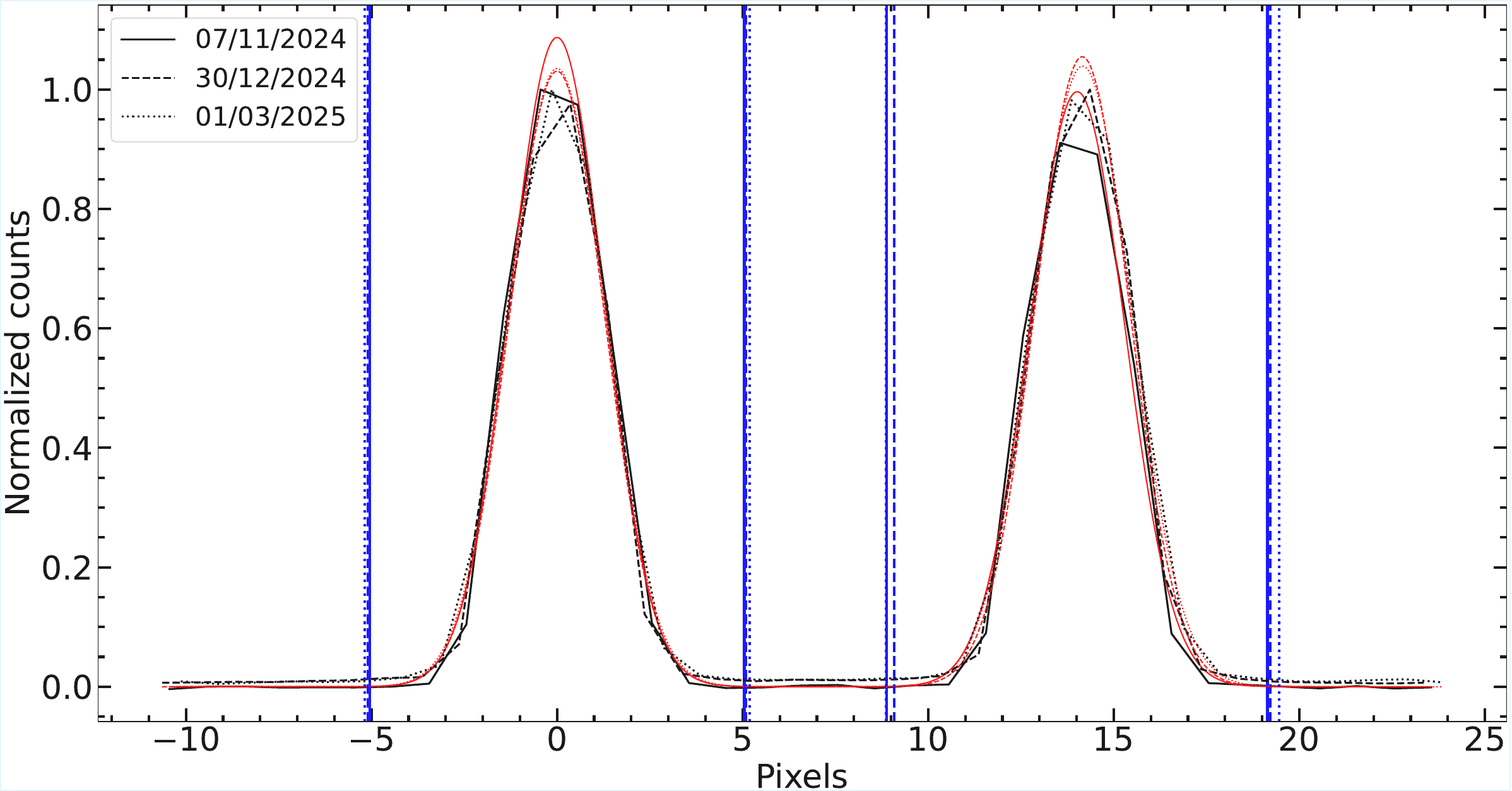}
    \caption{Gaussian profiles (red color) fitted to central column o- and e- ray intensity profile (black color) of an echelle order. The $\pm4 \sigma$ limits for each fitted Gaussian is marked by blue lines. The line styles of the intensity, the Gaussian fit, and the $\pm4 \sigma$ limits are maintained constant for a given observation date, as shown in the legend. The x-axis origin is defined at the peak position of the first peak from the left of the plots. As can be clearly seen, there is no overlap between the extraction regions of o- and e- rays for a given echelle order. Furthermore, no changes are seen in the cross-dispersion Gaussian profiles of o- and e- rays over a period spanning almost 4 months, therefore the $\pm4 \sigma$ summation range has safely been used for spectral reduction.}\label{echelle_gaussian}
\end{center}    
\end{figure}

The extraction of a spectrum refers to the process of adding up all the signal around the trace in the direction perpendicular to the dispersion, thus going from a 2D image to a 1D spectrum. In the case of an echelle spectrograph, the extraction process produces a 1D spectrum for each order. Before adding up the signal, all the systematic effects must be removed, which means that master bias and dark frames must be subtracted from the science image, bad columns must be corrected, cosmic rays removed, and the scattered light background must be subtracted as described above. In the reduction pipeline algorithm, ``rectangular extraction'' was employed to extract the flux for each order. The algorithm involves taking a simple sum of the flux of all the pixels contained in a vertical window of width defined by the user. We have defined this window from $-4\sigma$ to $+4\sigma$ about the detected peak position for the given order and column. As ensured by the optical design of the instrument \cite{srivastava2024development}, the separation of the traces of o- and e- rays are 15 pixels ($\sim$ 5.2 arc-seconds) in the cross dispersion direction and the minimum separation between o- and e- rays of adjacent orders is $>$ 15 pixels (even for highest Blue CD orders incident on the CCD where the inter-order separation is smallest). Therefore, for a typical seeing condition of 1.0 to 1.5 arc-seconds, an aperture width of $\pm4\sigma$ would ensure maximum extraction of the signal with no contamination from adjacent traces. $\sigma$ is determined for each of the science frames by fitting a Gaussian along the cross-dispersion profile of the spectra for any order about the central column. The fitted Gaussian profiles to the o- and e- ray spectra of an echelle order to estimate the $\sigma$ value for that frame and hence extract the intensity spectra is given in Figure~\ref{echelle_gaussian}. As the inter-order separation between o- and e- ray traces remain constant, determination of the value of $\sigma$ for any one order is valid for the entire frame (Blue/Red CD). As spectro-polarimetric reduction techniques deal with intensity ratios of the two mutually orthogonally polarized spectra of each order, employing simple rectangular extraction techniques is sufficient, as has been demonstrated in the later sections. 
Sky frames are also collected separately on the same night as the night of the science observations in four positions of HWP, same as the science frames. Usually, the exposure times of the sky frames are kept equal to the science exposures, but if not, the sky data frames are scaled by the ratio of their exposure times. After subtraction of master bias, master dark, and scattered light frames, if the pixel value of any of the science or sky frames becomes negative, those pixel values are set to zero. The sky spectra are also extracted using the same order traces as the science frames. However, before the extraction of the sky frames, for each pixel contributing to the sky flux extraction, the pixel value is replaced by the median of a $3\times3$ pixel square centered at that pixel. This is done to correct for the effect of any hot pixels or cosmic rays that might be present in the sky frames. Finally, the extracted sky spectra are subtracted from the science spectra for each order. Figure~\ref{wave-calib} shows the extracted intensities for o- and e-rays for an order containing $H\alpha$ absorption feature of a standard unpolarized star HD 47105. Spectra reduced from all four positions of the HWP (i.e. total eight spectra) are shown. Given the unpolarized nature of the star, the apparent difference in intensities is due to the difference in the instrument's efficiencies for o- and e- rays. This is discussed in later sections.

\subsection{Aligning the Spectra}

Any slight misalignment of the axis of Wollaston prism with the spatial axis would result into different positions of o- and e- rays spots in the dispersion direction. This would then cause an offset in their wavelength solution. Minor random flexure effects may also cause such offsets. Such offsets can be corrected prior to wavelength calibration to align both o- and e- rays spectra in their pixel space itself. A simple cross-correlation and integer-pixel shift is then used. This is necessary as a misalignment between the two orthogonally polarized spectra may lead to false polarization signatures. The same cross-correlation technique is also employed to align the o- and e- ray spectra as extracted from other HWP exposures to the same pixel value. After aligning all eight spectra to the same pixel value, wavelength calibration is performed on all spectra, and Stokes parameters are determined.  


\subsection{Wavelength Calibration}

\begin{figure}[H]
\begin{center}
    \includegraphics[width=\textwidth]{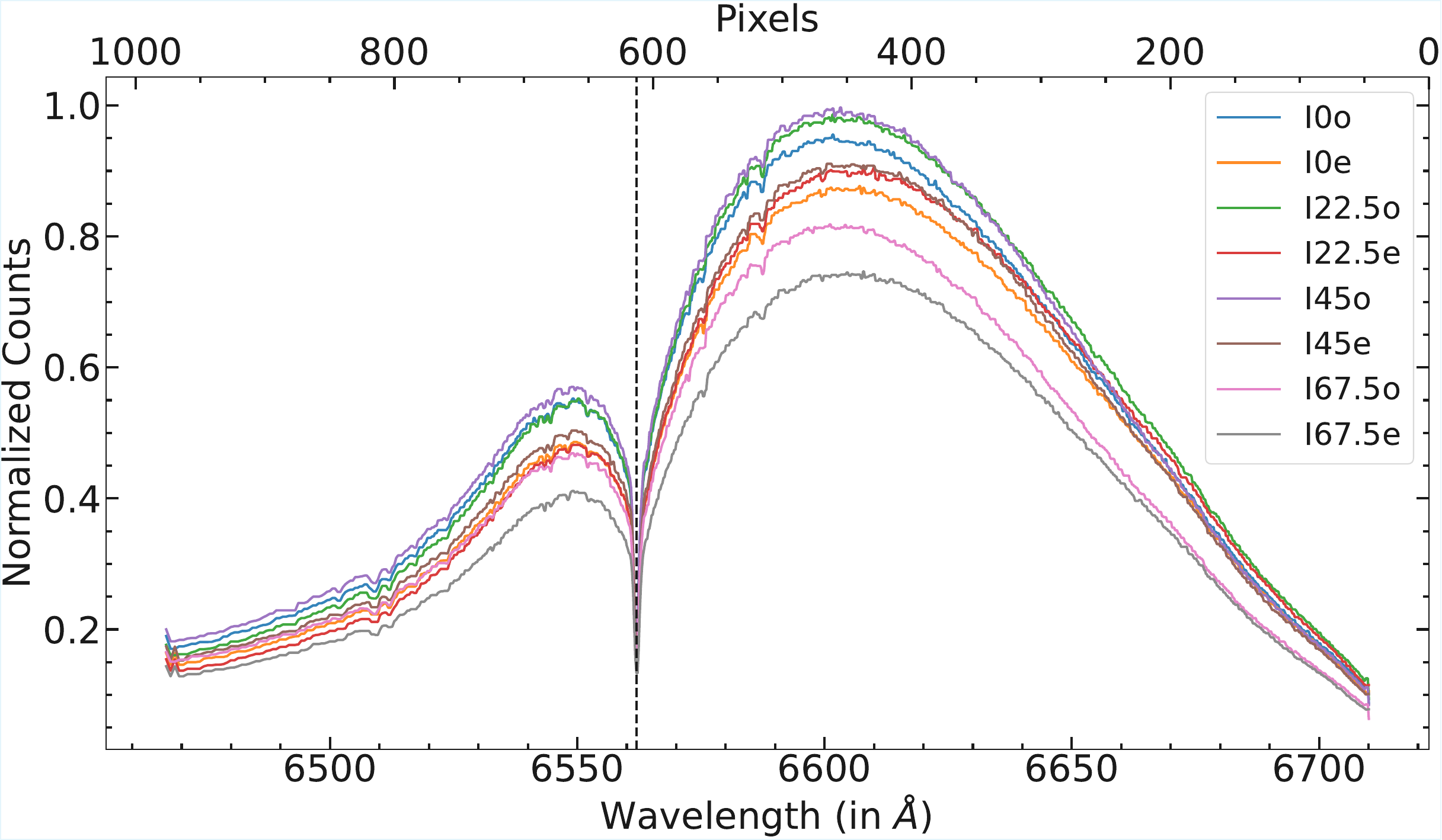}
    
    \caption{Extracted wavelength calibrated intensities of o- and e-rays for four HWP positions (0.0, 22.5, 45.0, and 67.5 degrees) for standard unpolarized star HD 47105 for echelle order containing H$\alpha$ line. The corresponding pixel scale is shown at the top of the plot. The H$\alpha$ absorption dip at $6562.8\: \AA$ is shown by the black dashed line. }\label{wave-calib} 
\end{center}    
\end{figure}

The calibration unit of ProtoPol (see Paper-I) is equipped with a Uranium Argon (U-Ar) arc lamp for wavelength calibration purposes in addition to a halogen lamp. U-Ar wavelength calibration frames were collected during science exposures immediately after each of the science frames were collected. The same order traces as the science frame were used and ``rectangular extraction" algorithm was employed to extract the spectra of the U-Ar frames for each order, similar to that of the science spectra. The reference wavelength map/template for each of the orders were created during the tests and calibration of the instrument in the laboratory. The reduced emission line spectra of the U-Ar lamp are then cross-correlated with the template to assign the correct wavelength values as obtained from their corresponding line lists \cite{sarmiento2018comparing} and \cite{lovis2007new}, respectively. Mean position of each of the emission lines is determined to sub-pixel accuracy by fitting a Gaussian to the line profile, and subsequently, the wavelength solution is created for each of the orders using a third-degree polynomial. The wavelength calibrated H$\alpha$ order for standard unpolarized star HD 47105 is shown in the right panel of Figure~\ref{wave-calib}.


\subsection{Determination of Stokes parameters}\label{Stokes_Section}

The Stokes parameters are calculated for each order from the intensity ratio of the o- and e- rays. Following the standard Stokes parameter formalism for dual-beam spectro-polarimetry \cite{arasaki2015very}, Stokes q and u parameters are obtained as follows:

$$q = \frac{Q}{I(\lambda)} = \frac{A_q - 1}{A_q  + 1}$$ where 
$$A_q = \sqrt{\frac{I_o(\lambda,0)}{I_e(\lambda,0)} \times \frac{I_e(\lambda,45)}{I_o(\lambda,45)}}$$ \\
and 
$$u = \frac{U}{I(\lambda)} = \frac{A_u - 1}{A_u  + 1}$$ where
$$A_u = \sqrt{\frac{I_o(\lambda,22.5)}{I_e(\lambda,22.5)} \times \frac{I_e(\lambda,67.5)}{I_o(\lambda,67.5)}}$$

Subsequently, the degree of polarization (p) and the angle of polarization ($\theta$) are calculated as:
$$p = \sqrt{q^2 + u^2}$$
and 
$$\theta = \frac{1}{2}\tan^{-1} \left( \frac{u}{q} \right) \times \frac{180}{\pi}  degree$$


\subsection{Dynamic Binning}

The polarization variation across the polarized line profile shapes are typically of the order of the fraction of a percent \cite{oudmaijer1999halpha, vink2005probing, ikeda2004polarized}. Therefore, to have statistically significant deduction, it is essential to have polarization error of the order of  $\sim$0.1-0.2$\%$ which is often the limit of instrumental polarization \cite{arasaki2015very, ikeda2003development}. The deduced error in degree (and angle) of polarization ($\sigma_P$) is directly related to the signal-to-noise-ratio (SNR) of the spectral resolution element \cite{patat2006error}, as:
$$\sigma_P = \frac{1}{\sqrt{N/2}(S/N)}$$ 
where N is the number of HWP positions in which the star has been observed for Stokes parameter determination and (S/N) is the required SNR. For ProtoPol, with N = 4, the typical polarization errors of $\sim$0.1-0.2$\%$ demand the SNR to be in the range of $\sim$300-700 for the spectral resolution elements.

\par 

However, while determining the polarization values across a typical emission line, the line/continuum ratios may vary from edge of the line profile to the center of the emission peak. Therefore, even though sufficient SNR is achieved near the peak of the line, at the line wings or absorption troughs (in the case of P-Cygni profiles or double-peaked line structures, etc.), the SNR is significantly lower \cite{oudmaijer1999halpha, vink2002probing}. Thus, a dynamic binning procedure is developed to ensure the SNR per binned pixel is roughly constant over the line profile at the expense of a decrease in spectral resolution. Dynamic binning is achieved by increasing a particular bin size dynamically and adding more flux elements to that bin till a given SNR is achieved, thereby altering the spectral bin-size across the line profile. While computing the continuum polarization too, the same technique is employed to determine the polarization with the required accuracy.


\subsection{Joining Echelle orders and combining different sets of data}

The reduced spectra of adjacent orders are joined to create a single 1-D spectrum. As successive echelle orders tend to have slightly different dispersion values, an extended wavelength scale is established with finer wavelength bins. Intensities from various orders are then suitably interpolated to fill the finer wavelength grid spanning the full spectral range. In regions where orders overlap, the segment with the higher signal-to-noise ratio (SNR) is retained, while the corresponding lower-SNR segment is discarded.


\subsection{Relative flux calibration}\label{subsec:RelativeFluxCalib}

The relative flux calibration of the derived spectra is done by observing standard spectro-photometric stars on the same or contemporaneous observing nights as science targets at a similar air-mass. Their reduced spectra would then be compared with the known one to generate the cumulative efficiency curves for each of the echelle orders. The reduced spectra of the science target would then be corrected for the efficiency values across the wavelengths to obtain the relative flux-calibrated spectra of the target for each of the echelle orders.  While this approach works well in typical low-resolution spectroscopy (e.g. MFOSC-P instrument on PRL 1.2m telescope \cite{srivastava2021design},  in the case of ProtoPol, it resulted in incorrect blaze corrections at the order's edges, where SNR tends to be on the lower side. Such inaccuracies in modeling the cumulative blaze profile - including all other factors, such as responses of cross-disperser (CD) gratings, optical chain of the instrument along with telescope optics, atmospheric response, etc. - are known to the community and have been seen in several other echelle spectrographs as well \cite{vskoda2008investigation, suzuki2003relative}. Here, in the case of ProtoPol, a major source of such inaccuracy seems to be apparent wavelength-dependent vignetting of the beam within the camera system. This vignetting not only reduces the system throughput, but can also introduce false spectral polarization signatures (both zero-point polarization continuum and rotation/circular polarization components) by altering the intensity ratios of the orthogonal beams. Some of these effects may be noticed in the low flux regions of the derived polarization plots (see section~\ref{sec-science}, Figure~\ref{Symbiotic1}). As noted in Paper-I, ProtoPol uses off-the-shelf \href{https://www.canon-europe.com/lenses/ef-200mm-f-2l-is-usm-lens/}{Canon make EF 200mm f/2L IS USM} camera lens system to image the dispersed spectra onto the CCD. While such lens systems have been used in other astronomical instruments \cite{harding2016chimera}, they have also noted that  EF 200 lens system expects an entrance pupil on its internal stop, which could cause geometrical vignetting. In ProtoPol, such behavior can cause a wavelength-dependent vignetting. Further, considering that a pin-hole is used at the focal plane of the telescope in the instrument, any difference in observing conditions of the science target and standard star (such as slight difference in air-mass, atmospheric transparency, etc.) may also give rise to wavelength dependent efficiency factor e.g. differential atmospheric refraction \cite{Filippenko1982Atmospheric, gubler1998differential}; though we expect such effects to be minimal. Nevertheless, such factors could lead to wavelength-dependent vignetting, thereby the effective efficiency over the 2-D face of the CCD detector would vary, as we discuss below, leading to wrong blaze corrections.
\par
\begin{figure}[H]
  \begin{center}
    \includegraphics[width=\textwidth]{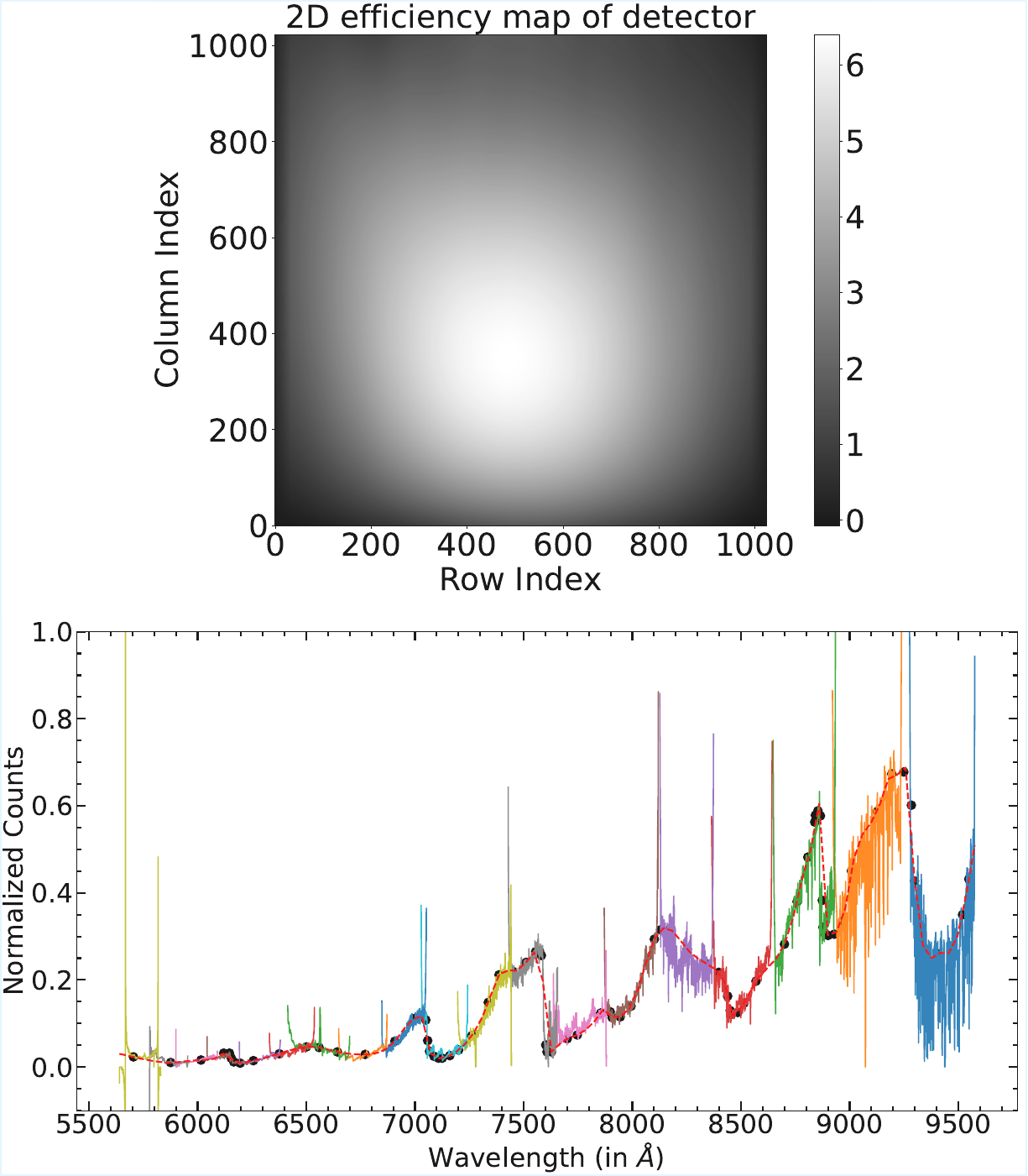}

  \caption{\textit{(Top)} 2-D efficiency map on the face of the CCD, incorporating the effects of CD efficiency, echelle blaze, and optics vignetting. \textit{(bottom)} The derived continuum of symbiotic star Y Gem obtained by applying the above 2-D efficiency map. User-defined points can be seen in each order to trace the true continuum. Spurious features may be noticed at the order edges, which are removed by suitably truncating the order edges as shown in Figure~\ref{RelativeFluxCalib}. See text for details.}
  \label{2DEffMap}

  \end{center}
  
\end{figure}

The cumulative efficiency of the instrument at any point of the CCD is morphed by a combination of three efficiencies given in the equation below:

$$\eta_{Cumulative} = \eta_{CD_{optics}} \times \eta_{Echelle_{Blaze}} \times \eta_{Vignetting}$$ 

where $\eta_{CD_{optics}}$ is the efficiency of the entire observing set-up, including atmospheric transmission, telescope optics, instrument optics, including cross-disperser grating, and detector, etc., but excluding the blaze efficiency of the echelle grating. $\eta_{Echelle_{Blaze}}$ of the blaze efficiency of the echelle grating, and $\eta_{Vignetting}$ is the space-wavelength dependent vignetting factor expressed in terms of a separate efficiency. Therefore, as a first-order correction, a 2-dimensional cumulative efficiency $\eta_{Cumulative}$ map on the CCD surface was generated in the following manner. First, a third-degree polynomial was fit to the reduced data of the standard star for each of the columns of the CCD pixels (i.e. along the direction of cross-dispersion). This will provide a low-resolution spectrum for each column. Later, these low-resolution column-wise spectra were used, in conjunction with a known archival spectrum of the standard star, to generate an efficiency curve for each of the columns. Thus, a cumulative 2-D efficiency map of the instrument was derived over the face of the CCD (Figure~\ref{2DEffMap}). This efficiency map is then used to correct the commutative response of the optics, cross-disperser, and blaze function from the reduced data of the science target.
\par
Though the above correction provides a good tentative estimate of the true continuum of the science target while stitching the reduced spectra of all the orders in wavelength space, minor deviations can still be noticed at the edges of the echelle orders. Therefore, another polynomial fit is made by taking the median of the center 100 pixels of each of the orders to trace the true continuum of the star in an iterative manner. The first iteration is automated, wherein a fitted continuum is presented to the user. In successive iterations, the user may choose to further refine the determined continuum of the stitched spectra by manually selecting the points on the reduced spectra for polynomial fitting, preferably avoiding points near the order edges. Such User intervention is usually required when a very broad emission/absorption feature (e.g., for AGB stars or novae, etc.) falls near the order center, rendering the selected continuum point for that order wrong. The process is shown in the bottom plot of Figure~\ref{2DEffMap}. 
\par
\begin{figure}[H]
\begin{center}
    \includegraphics[width=\textwidth]{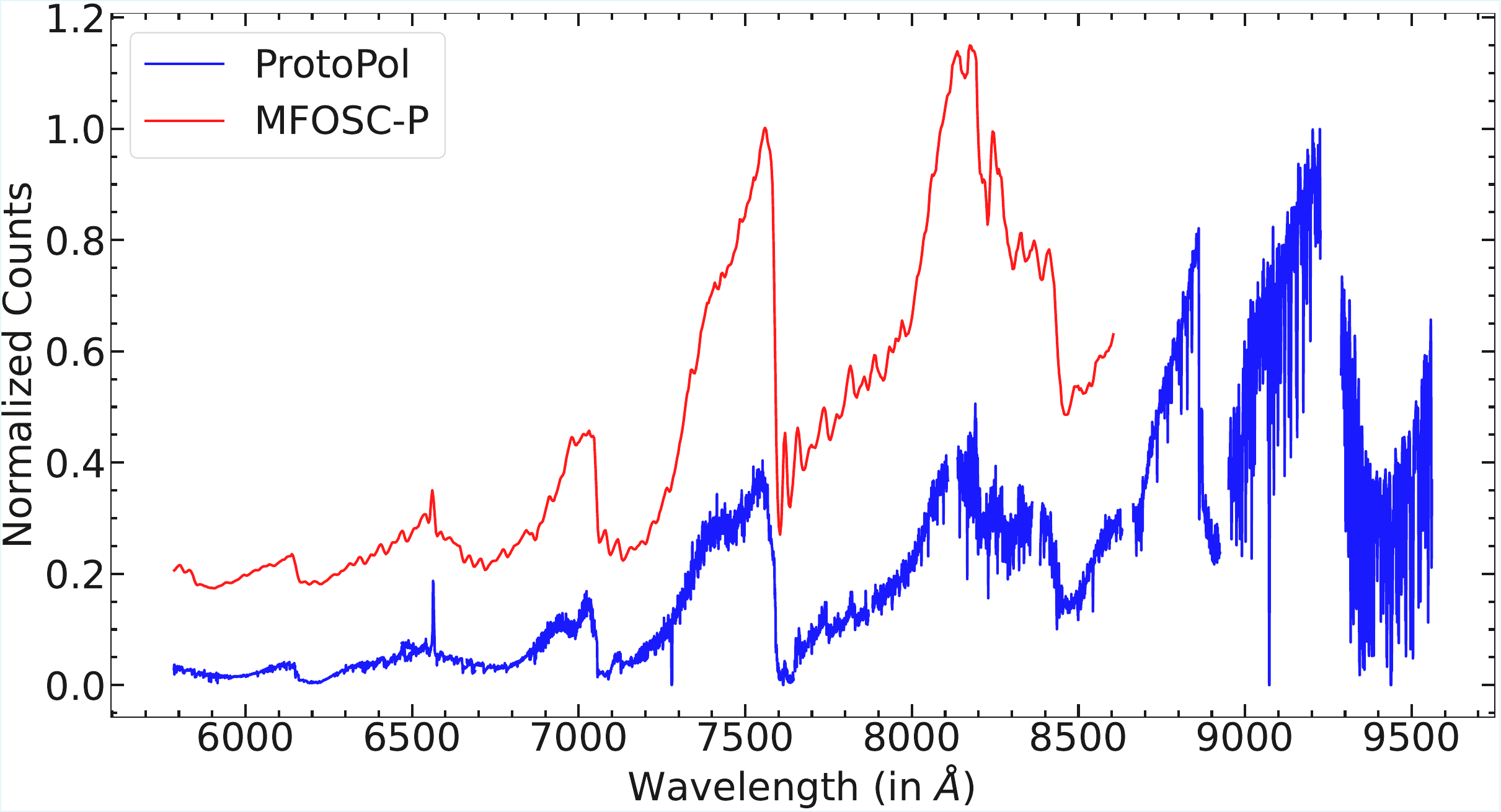}    
    \caption{Relative flux calibrated spectrum of symbiotic star Y Gem \textit{(blue)} obtained by scaling the fitted pseudo-continuum for each order to the true continuum. The low-resolution spectrum (R$\sim$500) of Y Gem \textit{(red)} as obtained from MFOSC-P instrument \cite{srivastava2021design} on PRL 1.2m telescope has also been shown for comparison. The MFOSC-P spectrum has been given a 0.15 offset (in arbitrary units) along y-axis for better clarity.} 
    \label{RelativeFluxCalib}
\end{center}    
\end{figure}

Once the shape of the true continuum of the star is established, the reduced intensities of the echelle orders are to be scaled appropriately to follow this shape. A polynomial is, therefore, fit to each order's spectra, and the scale factor (ratio of the true continuum to the polynomial) is derived for each pixel position. For any overlap between orders, the overlapping part of the order with the lower SNR is truncated. While the process is automated, an iterative user interface is also provided for visual inspection of the polynomial fit. Such features are useful while normalizing the echelle orders consisting of a very broad emission/absorption feature or stray signals at the order edges, wherein manual selection of the polynomial fit-points are needed. The relative flux calibrated spectrum for a symbiotic star Y Gem is shown in Figure~\ref{RelativeFluxCalib}. A low-resolution spectrum of the same star obtained with MFOSC-P instrument \cite{srivastava2021design} on PRL 1.2m telescope is also shown for comparison. It can be seen that the above data reduction methodology is very successful in deriving the true shape of the complex continuum. 


\section{On-sky Performance of ProtoPol}
\label{sec-performance}

The laboratory characterization and performance of ProtoPol has been discussed in Paper-I. Here we present the results obtained with ProtoPol on PRL 1.2m and 2.5m telescopes. The observations presented here spans a period from December 2023 to May 2025. ProtoPol was first commissioned on PRL 1.2m telescope in December 2023 and later moved to 2.5m telescope in February 2024. A variety of sources (to be discussed later in the section), including multiple standard unpolarized (UPs) / standard polarized (SPs) stars, spectro-photometric standard stars, etc were observed to characterize the instrumental polarization and polarimetric accuracy of the instrument. Observations of Jupiter were conducted to determine instrumental throughput without slit loss. Many science targets like Symbiotic stars, Herbig stars, and Be stars were observed to check for line polarization profiles while observations of AGB/post-AGB stars were conducted to determine the continuum polarization of the targets. The spectra of U-Ar lamp for spectral calibration and halogen lamp for order tracing were routinely recorded immediately before or after the science exposure. Other auxiliary data such as bias, dark, sky frames, etc were also regularly recorded in the process. In subsequent sub-sections we describe various instrumental characterization aspects of ProtoPol.

\subsection{Spectral Resolution}
\label{subsec-SpecRes}

\begin{figure}[htbp]
  \begin{center}
    \includegraphics[width=\textwidth]{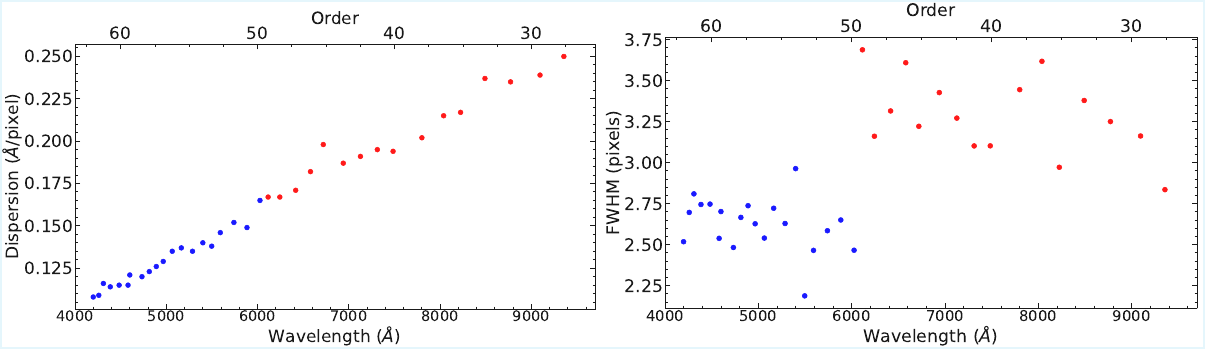}

  \caption{Measured median dispersion (in $\text{\AA}$ per pixel) and FWHMs (in pixels) of U-Ar lines for each order of ProtoPol. The blue and red color-coded circles represent data from the Blue and Red CDs, respectively.} 
    \label{resolution_ProtoPol_1}
    \end{center}
\end{figure}

\begin{figure}
\begin{center}
    \includegraphics[width=\textwidth]{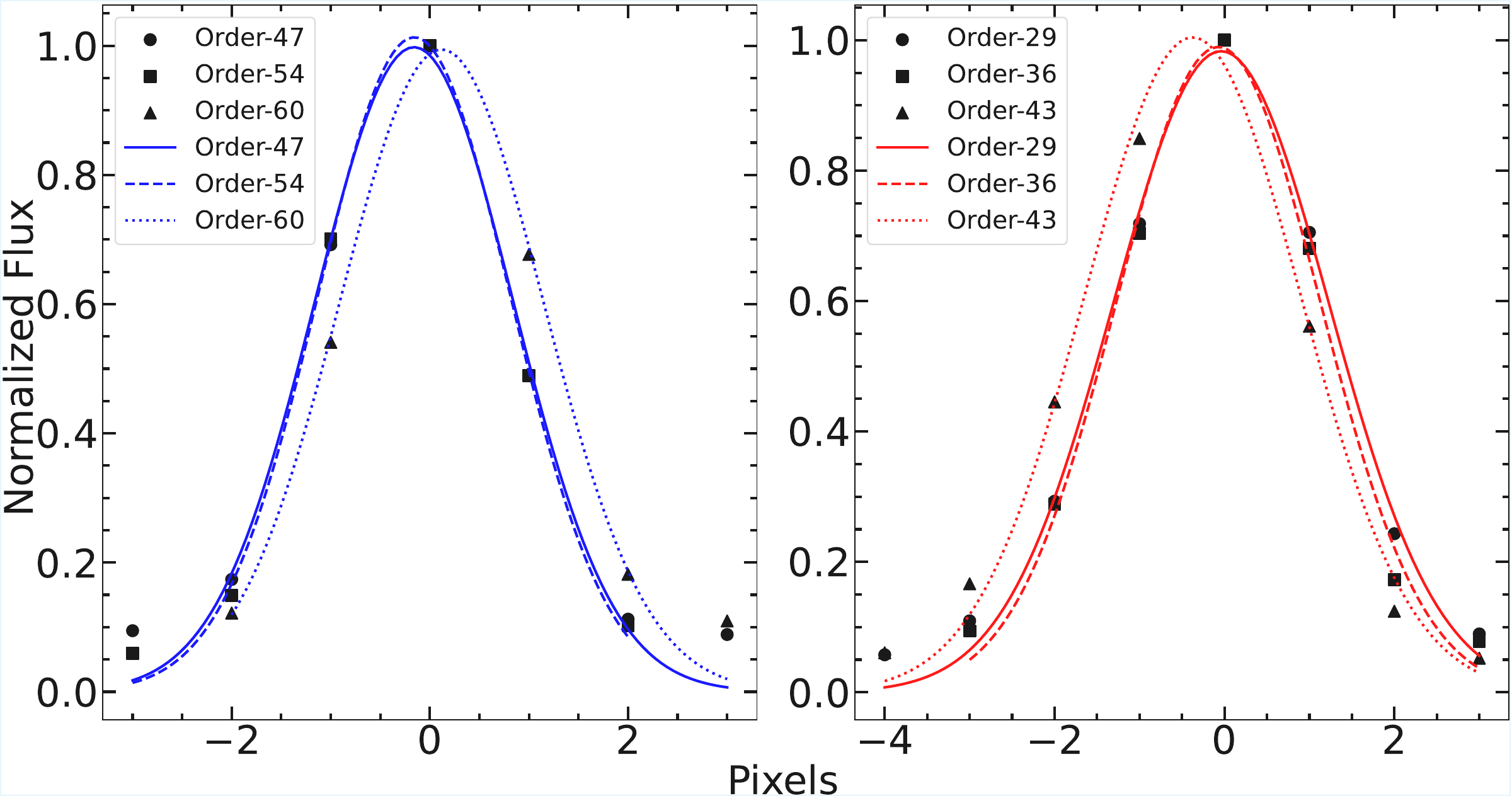}    
    \caption{\textit{(Left)} Figure shows the U-Ar emission line profiles for orders 47, 54, and 60 of Blue CD. The black markers represent the extracted data from U-Ar spectra, while the blue lines represent their corresponding Gaussian fit. \textit{(Right)} The same is shown for the Red CD for orders 29, 36, and 43. These orders represent the edges and centers of the echelle data-frames for Blue and Red CDs. No significant change in line FWHMs is observed across the orders for a given CD.} 
    \label{resolution_ProtoPol_2}
\end{center}    
\end{figure}

ProtoPol is designed to provide slit limited resolution ($R = \delta /\delta \lambda$). It uses a pinhole of diameter 150$\mu$m kept at 45$\degree$, which projects onto $\sim$3.6 pixels in the dispersion direction of echelle orders considering all the design effects such as slit-tilt due to echelle's out-of-plane angle (see paper-I). Figure~\ref{resolution_ProtoPol_1} shows the measured median dispersion (in $\text{\AA}$) and full width at half maximum (FWHMs) (in pixels) of U-Ar lines for each echelle order of ProtoPol for both Blue and Red CD. Figure~\ref{resolution_ProtoPol_2} shows the FWHM (in pixels) of the emission lines in the orders 47, 54, and 60 for Blue CD and orders 29, 36, and 43 of red CD to explicitly check for any variation of FWHMs across the selected orders spanning over the entire CCD. The dispersion of various echelle orders in both the cross-disperser gratings is determined to be in the range of 0.12-0.25$\AA$ per pixel from blue to red (Figure~\ref{resolution_ProtoPol_1}) wavelengths. The FWHM of emission line profiles of U-Ar spectral lamp are used to determine the spectral resolution across various orders. Emission lines are taken from near the order's edges and order's centers to determine any variation in FHWMs across the order. The measured FHWMs are found to be in the range of 2.6-2.8 pixels and 3.0-3.6 pixels for Blue and Red CD, respectively, throughout the entire spectral region (Figure~\ref{resolution_ProtoPol_1}). This translates to $\delta\lambda$ in the range of $\sim 0.3-0.4\:\AA$ and $\sim 0.6-0.75\:\AA$ for Blue and Red CD, respectively. The variation in the U-Ar line profiles for different orders spanning the entire CCD with their corresponding Gaussian fits to determine their respective FWHMs is explicitly shown in Figure~\ref{resolution_ProtoPol_2}. No significant change in line FWHMs is observed across the orders for a given CD.


\subsection{Determination of Instrumental Polarization}
\label{subsec-InstPol}

To establish the instrumental polarization of ProtoPol is of paramount importance to ensure the polarization measurement of science targets. Several standard polarized and unpolarized stars were observed repeatedly over the period of around 15 months (from December 2023 - March 2025) in varying conditions, such as air-mass,  observing time etc to ensure randomness in the deduced results for such analysis. All such observations were carried out during the dark nights. For smaller degrees of polarization ($<1\%$), random errors (for example, photon noise) and zero-point instabilities of the Q-U plane affect the polarimetric accuracy \cite{ikeda2003development, arasaki2015very, chakraborty2003optical, harrington2008spectropolarimetric_a}. On the other hand, for object having a larger degree of polarization,  degree of polarization-dependent system errors may become dominant despite calibrations and corrections \cite{arasaki2015very}. This is because systematic errors scale with the intrinsic polarization signal; so their contribution grows larger than random/statistical errors when observing highly polarized sources, causing the polarization-dependent system errors to dominate the error budget, even after correcting for depolarization. To address these issues, the polarimetric accuracy of the instrument is determined in both cases. Observations of standard unpolarized stars are used to address the former issue, while the latter is addressed by observations of standard polarized stars. The observational log of standard unpolarized and polarized stars is summarized in Table~\ref{Polarized_Standard_Stars}.
\par
\begin{figure}[htbp]
  \begin{center}
    \includegraphics[width=\textwidth]{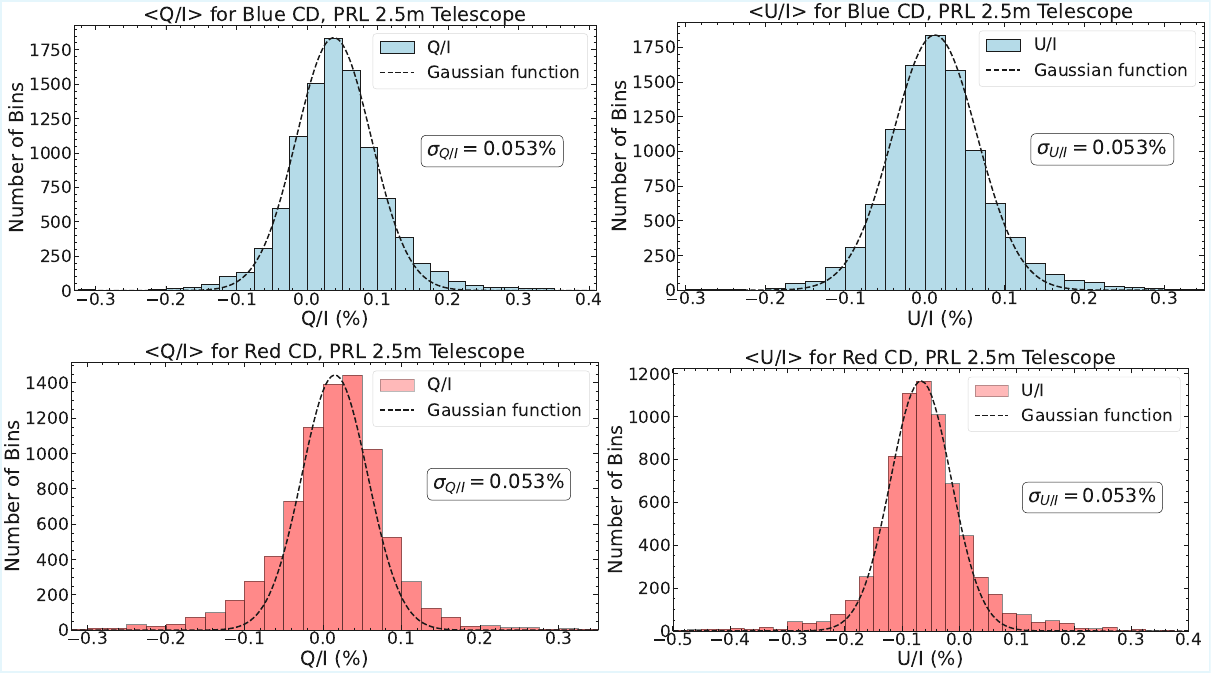}
  \caption{Combined histograms of Q/I \textit{(left)} and U/I \textit{(right)} of several standard unpolarized stars for observations from PRL 2.5m telescope for both Blue \textit{(top row plots)} and Red \textit{(bottom row plots)} cross-dispersers. The plots have been generated by taking the average Stokes parameter values of the center 500 pixels for each order for both the cross-disperser gratings. The dashed line shows the Gaussian fit to the data. The estimated standard deviation is $0.053\%$ for $\langle Q/I \rangle$ and $\langle U/I \rangle$ determined for Blue CD and Red CD.}
  \label{instrumental_polarization_histogram}

  \end{center}
\end{figure}
\begin{figure}[htbp]
  \begin{center}
    \includegraphics[width=\textwidth]{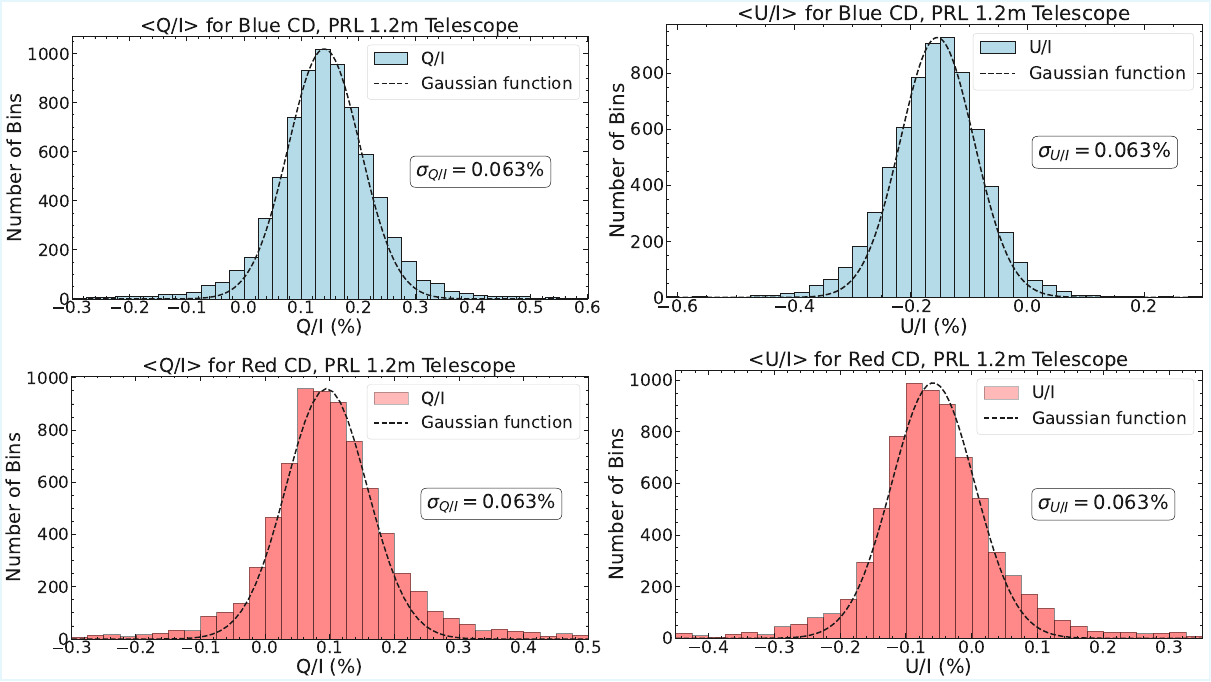}

  \caption{Same as \ref{instrumental_polarization_histogram} but for observations taken from PRL 1.2m telescope.}
  \label{instrumental_polarization_histogram1}

  \end{center}
\end{figure}

Instrumental polarization was determined with the on-sky observation of several standard unpolarized stars during several dark nights over the observing season. Only good sections of the echelle orders (typically the central 500 pixels) with high SNR were considered for the evaluation. A 3-pixel rolling median was applied to the data/spectra to smoothen out the intensities. Histograms are constructed for Q/I and U/I Stokes parameters from the combined data of the standard unpolarized star observations for all the orders for both Blue and Red cross-dispersers for observations taken from both PRL 1.2m and 2.5m telescopes, separately.  A Gaussian was fitted to the histograms to determine the mean (indicating the value of the instrumental polarization) and standard deviation (indicating the error associated with measuring the degree of polarization) of the Stokes parameters $\langle Q/I \rangle$ and $\langle U/I \rangle$. The process was repeated by binning the spectral data by 3, 5, 7, 9, etc. pixels till the photon noise-limited case was achieved. In most cases, the photon noise limit was achieved at 5-pixel binning, at which point there was no further change in the standard deviation of the fitted Gaussian. Figure~\ref{instrumental_polarization_histogram} shows these final histograms for both cross-dispersers for observations taken from PRL 2.5m telescope. Similar histogram for ProtoPol on PRL 1.2m telescope is presented in Figure~~\ref{instrumental_polarization_histogram1}. The mean values for $q \:(Q/I)$ and $u \:(U/I)$ are determined to be in the range of $0.02\:\text{to}\:0.14\%$ and $0.01\:\text{to}\:-0.16\%$  respectively, while the corresponding standard deviation $(\sigma)$ are determined around $0.05-0.06\%$. This translates into a typical instrumental polarization error of $0.1\%$. The determined mean and sigma values of Stokes parameters for both cross-dispersers and on both the telescopes are summarized in Table \ref{instrumental_polarization_table}.


\begin{table*} 
\caption{Summary of observed unpolarized (UPs) and standard polarized (SPs) stars} \label{Polarized_Standard_Stars}
\centering
\setlength{\tabcolsep}{1.5pt}
\renewcommand{\arraystretch}{1.2}
	\begin{tabular}{cc cc cc cc cc cc cc cc}
           \hline
		  \hline

\textbf{Object} & \textbf{Object} & \textbf{Spectral} & \textbf{$m_V$} & \textbf{$P_V$}  & \textbf{Date} & \textbf{Telescope} & \textbf{Integration}\\
\textbf{Type} & \textbf{Name} & \textbf{Type} &  &  &  &  & \textbf{Time}\\
\hline
UPs & HD 47105 & A0 IV &  1.93 & $0.0022\%$ & 2024 Jan 07 & 1.2m & 180s $\times$ 4 $\times$ 3 sets\\
&  &  &  &  & 2024 Apr 28 & 2.5m & 300s $\times$ 4 $\times$ 1 set\\
&  &  &  &  & 2024 Nov 06 & 2.5m & 60s $\times$ 4 $\times$ 2 sets\\
&  &  &  &  & 2025 Jan 24 & 2.5m & 240s $\times$ 4 $\times$ 2 sets\\
&  &  &  &  & 2025 Mar 02 & 2.5m & 60s $\times$ 4 $\times$ 2 sets\\
&  &  &  &  &  &  & \\
UPs & HD 61421 & F5 IV &  0.37 & $0.005\%$ & 2023 Dec 31 & 1.2m & 300s $\times$ 4 $\times$ 2 sets\\
&  &  &  &  & 2024 Apr 29 & 2.5m & 15s $\times$ 4 $\times$ 1 set\\
&  &  &  &  & 2024 Nov 07 & 2.5m & 30s $\times$ 4 $\times$ 1 set\\
&  &  &  &  & 2024 Dec 01 & 2.5m & 20s $\times$ 4 $\times$ 1 set\\
&  &  &  &  & 2025 Jan 24 & 2.5m & 20s $\times$ 4 $\times$ 2 sets\\
&  &  &  &  & 2025 Mar 02 & 2.5m & 15s $\times$ 4 $\times$ 2 sets\\
&  &  &  &  &  &  & \\
UPs & HD 95418 & A1 V &  2.37 & $0.0018\%$ & 2024 Jan 12 & 1.2m & 180s $\times$ 4 $\times$ 3 sets\\
&  &  &  &  & 2024 Apr 26 & 2.5m & 180s $\times$ 4 $\times$ 1 set\\
&  &  &  &  & 2025 Jan 23 & 2.5m & 180s $\times$ 4 $\times$ 2 sets\\
&  &  &  &  &  &  & \\
UPs & HD 39587 & G0 V &  4.40 & $0.013\%$ & 2024 Nov 07 & 2.5m & 600s $\times$ 4 $\times$ 1 set\\
&  &  &  &  & 2025 Jan 23 & 2.5m & 600s $\times$ 4 $\times$ 2 sets\\
&  &  &  &  &  &  & \\
SPs & HD 7927 & F0 Ia &  4.99 & $3.41\%$ & 2024 Jan 12 & 1.2m & 600s $\times$ 4 $\times$ 3 sets\\
&  &  &  &  & 2024 Jan 16 & 1.2m & 600s $\times$ 4 $\times$ 2 sets\\
&  &  &  &  & 2024 Nov 08 & 2.5m & 300s $\times$ 4 $\times$ 1 set\\
&  &  &  &  &  &  & \\
SPs & HD 43384 & B3 Ia &  6.27 & $3.01\%$ & 2024 Feb 09 & 1.2m & 900s $\times$ 4 $\times$ 3 sets\\
&  &  &  &  & 2024 Apr 29 & 2.5m & 300s $\times$ 4 $\times$ 1 set\\
&  &  &  &  &  &  & \\
SPs & HD 21291 & B9 Ia &  4.12 & $3.53\%$ & 2024 Jan 06 & 1.2m & 600s $\times$ 4 $\times$ 2 sets\\
&  &  &  &  & 2024 Nov 07 & 2.5m & 300s $\times$ 4 $\times$ 1 set\\
&  &  &  &  &  &  & \\
SPs & HD 198478 & B3 Ia &  4.83 & $2.72\%$ & 2024 Apr 30 & 2.5m & 600s $\times$ 4 $\times$ 1 set\\
&  &  &  &  &  &  & \\
SPs & HD 147084 & A5 II &  4.54 & $4.46\%$ & 2024 Apr 30 & 2.5m & 300s $\times$ 4 $\times$ 1 set\\
&  &  &  &  &  &  & \\
SPs & HD 25443 & B0.5 III &  6.74 & $5.17\%$ & 2024 Nov 08 & 2.5m & 600s $\times$ 4 $\times$ 1 set\\
		\hline
		\hline
	\end{tabular}

\end{table*}


\begin{table*} 
\caption{Instrumental polarization values for Red and Blue CD for PRL 1.2m and 2.5m telescopes as derived from the data of unpolarized standard stars} \label{StdUnpol_1pt2_2pt5_Combined}
\centering
\setlength{\tabcolsep}{10pt}
\renewcommand{\arraystretch}{1.2}
	\begin{tabular}{cc cc cc cc cc cc }
           \hline
		  \hline

\textbf{Telescope} & \textbf{Grating} & \textbf{Mean q} & \textbf{Sigma q} & \textbf{Mean u} & \textbf{Sigma u} \\
\hline
1.2m & Blue CD & $0.14\%$ &  $0.06\%$ & $-0.16\%$ & $0.06\%$ \\
1.2m & Red CD & $0.09\%$ &  $0.06\%$ & $-0.06\%$ & $0.06\%$ \\
2.5m & Blue CD & $0.04\%$ &  $0.05\%$ & $0.01\%$ & $0.05\%$ \\
2.5m & Red CD & $0.02\%$ &  $0.05\%$ & $-0.07\%$ & $0.05\%$ \\
        
		\hline
		\hline
	\end{tabular}
        \label{instrumental_polarization_table}
\end{table*}


\subsection{Checks on any residual polarization across spectral features and polarization ripple}
\label{subsec-PolErr}

\begin{figure}[H]
\begin{center}
    \includegraphics[width=\textwidth]{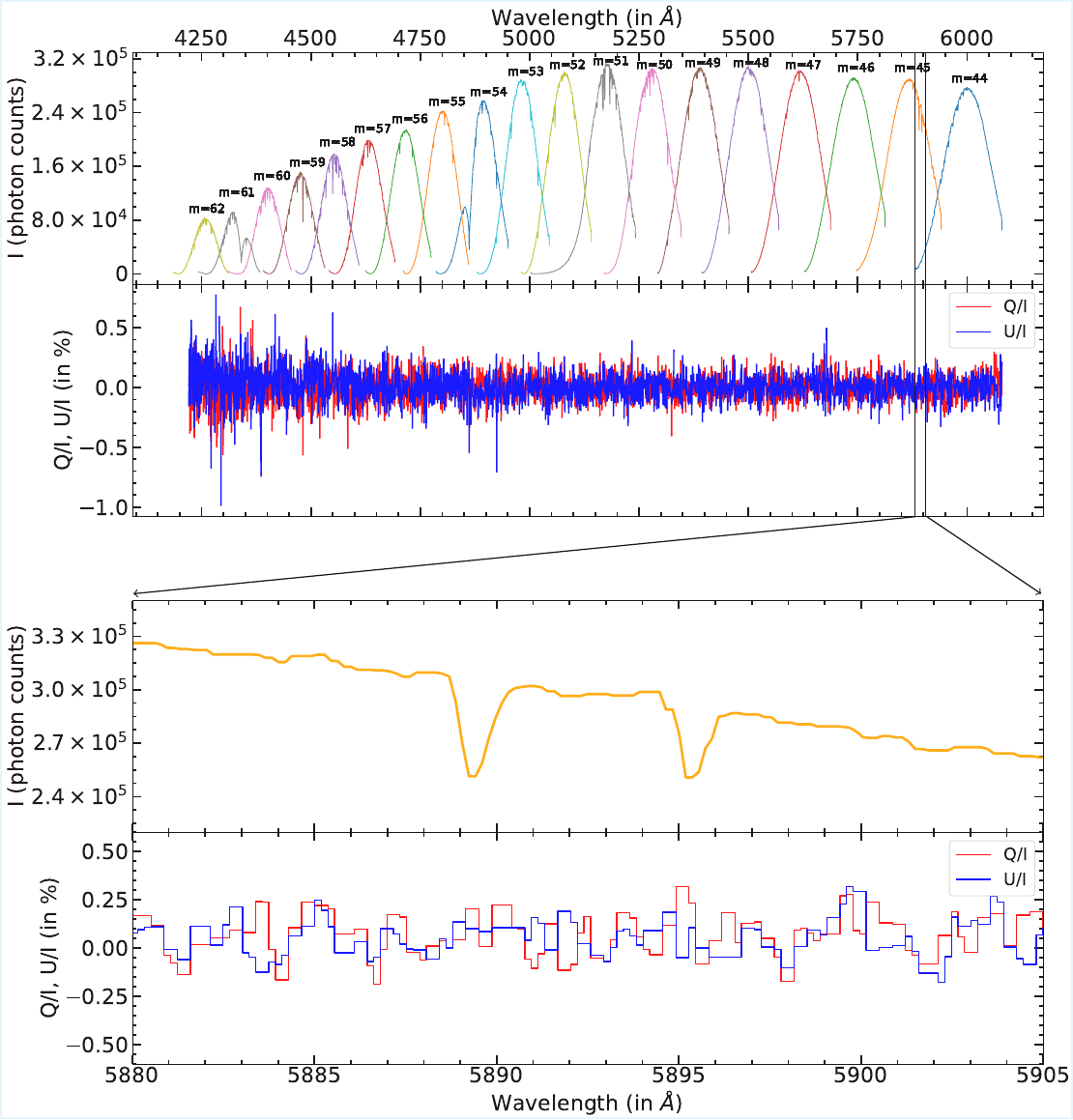}

\caption{Observed spectra of HD47105 (UP). The upper panel shows the observed counts (top) and Stokes Q/I and U/I spectra for orders 44 to 62 for Blue CD (bottom). The lower panel shows a close-up of the spectra shown in the upper panel around the sodium doublet absorption lines ($5890 \: \AA$ and $5896 \: \AA$). As can be seen, any anomalous patterns due to the reduction process are not observed in the polarization spectra across the absorption lines.}
\label{pol_line_variation}

\end{center}
\end{figure}

The upper panel of Figure~\ref{pol_line_variation} shows the intensity spectrum of unpolarized standard star HD 47105 across all orders for Blue CD. The order intensities have not been corrected for blaze of the echelle and the efficiency of the cross-disperser. The standard deviation of the Stokes parameters was calculated to be $0.14\%$, considering the central 500 pixels for each order. The lower panel of Figure~\ref{pol_line_variation} shows magnified spectra across the sodium doublet lines ($5890 \: \AA$ and $5896 \: \AA$). No anomalous polarization signals were observed, which could be produced in the data reduction process due to errors in wavelength calibration of o- and e-ray spectra or misalignment of the different spectra as obtained from different HWP positions.
\par
A magnified region of the Stokes spectra in the wavelength range $5960-6040 \: \AA$ was investigated to determine the presence or absence of polarization ripple. There does not seem to be a polarization ripple. However, a small-amplitude pseudo-periodic polarization variation is noticed with a period of $2-3 \: \AA$ with amplitudes larger than the $3\sigma$ error as determined for the photon shot noise limited case (Table~\ref{instrumental_polarization_table}). Similar polarization ripples have also been reported in other spectro-polarimeter devices like LIPS, VESPolA,  ISIS dual-beam Cassegrain spectrograph, etc \cite{ikeda2003development, arasaki2015very, harries2002spectropolarimetry}. Typically, secondary beams, generated by multiple reflections, coherent to the primary beam but with wavelength-dependent phase difference, give rise to the polarization oscillations \cite{semel2003spectropolarimetry, clarke2005effects}. For superachromatic retarders, like the one used in ProtoPol, rotational misalignment between the various layers, errors in physical thickness of the crystals, or variations in the material birefringence could be some of the causes of the spectral polarization fringes \cite{harrington2017polarization, harrington2018polarization}. Additionally, the refractive index mismatch between boundary layers, effects of anti-reflection coating on the optics, or non-normal incidence of the beam on the polarization optics surface, could all be possible causes of polarization fringes/ripples \cite{harrington2020polarization}. Furthermore, the amplitude and phase of the ripples do not remain stable throughout the night due to ambient temperature changes, causing changes in the refractive indices of the layers. For a laminated 5-layer PMMA retarder from AstroPribor, there are several factors that could affect the fringes/ripple behavior, such as cover window thickness, coefficient of thermal expansion (CTE) etc. A detailed fringe calculations for a multi-layer polymer retarder can be found in section 2 of \cite{harrington2020polarization}. Figure 24 of the same paper shows refractive index models for the polymer retarder components. However, in our case, these ripples are to be treated as random errors similar to VESPolA instrument \cite{arasaki2015very}. 

\par
A Glan-Taylor (GT) prism produces $100\%$ polarization. Hence, we wanted to determine the accuracy of polarization measurement by taking spectro-polarimetric observations of standard unpolarized stars through a GT prism. Figure~\ref{Star_through_GT} shows the instrumental depolarization and position angle of the optical axis of HWP for observations with red CD of the standard unpolarized star HD 61421 through GT prism. The obtained polarization of 101-102$\%$ should be considered as a scale error, different from the zero-point error determined earlier. This could be resulting from ignoring the HWP's circular retardance or the circular polarization generated by the telescope. Though such errors are typically considered to be smaller, nevertheless, in a recent study to calibrate the polarization of the Keck telescope and the LRISp instrument (supposed to have zero circular polarization response), the daytime sky revealed significant linear-to-circular terms \cite{harrington2015correcting}.
\par
The issue of polarization cross-talk between Stokes parameters and their possible causes has been discussed in detail in several references \cite{casini2012analysis, liu2022study, breckinridge2015polarization, chipman2015polarization}, such as seeing-induced polarization, reflections in the telescope’s primary and secondary mirrors, etc., and we would refer readers to check these.  In the present case of ProtoPol, we have conducted extensive characterization of the instrument’s polarization using several standard unpolarized and polarized stars, as documented in paper-2. The one-sigma variation around the zero-point, estimated from the random statistical measurements, was considered as the polarization accuracy. It was determined to be around 0.1-0.2 percent at the photon noise limit. Therefore, we expect any cross-talk between Stokes parameters to be well within this error range. On a separate note, there are comparatively fewer astrophysical sources that show a strong circular polarization as comparable to the linear one \cite{serkowski1974many}, and we would indeed advise a further careful calibration of the instrument should one attempt to observe such a source with a strong circular polarization component.  But for the majority of astrophysical sources with an insignificant circular component, we would expect the aforementioned error to suffice.


\begin{figure}[H]
\begin{center}
    \includegraphics[width=\textwidth]{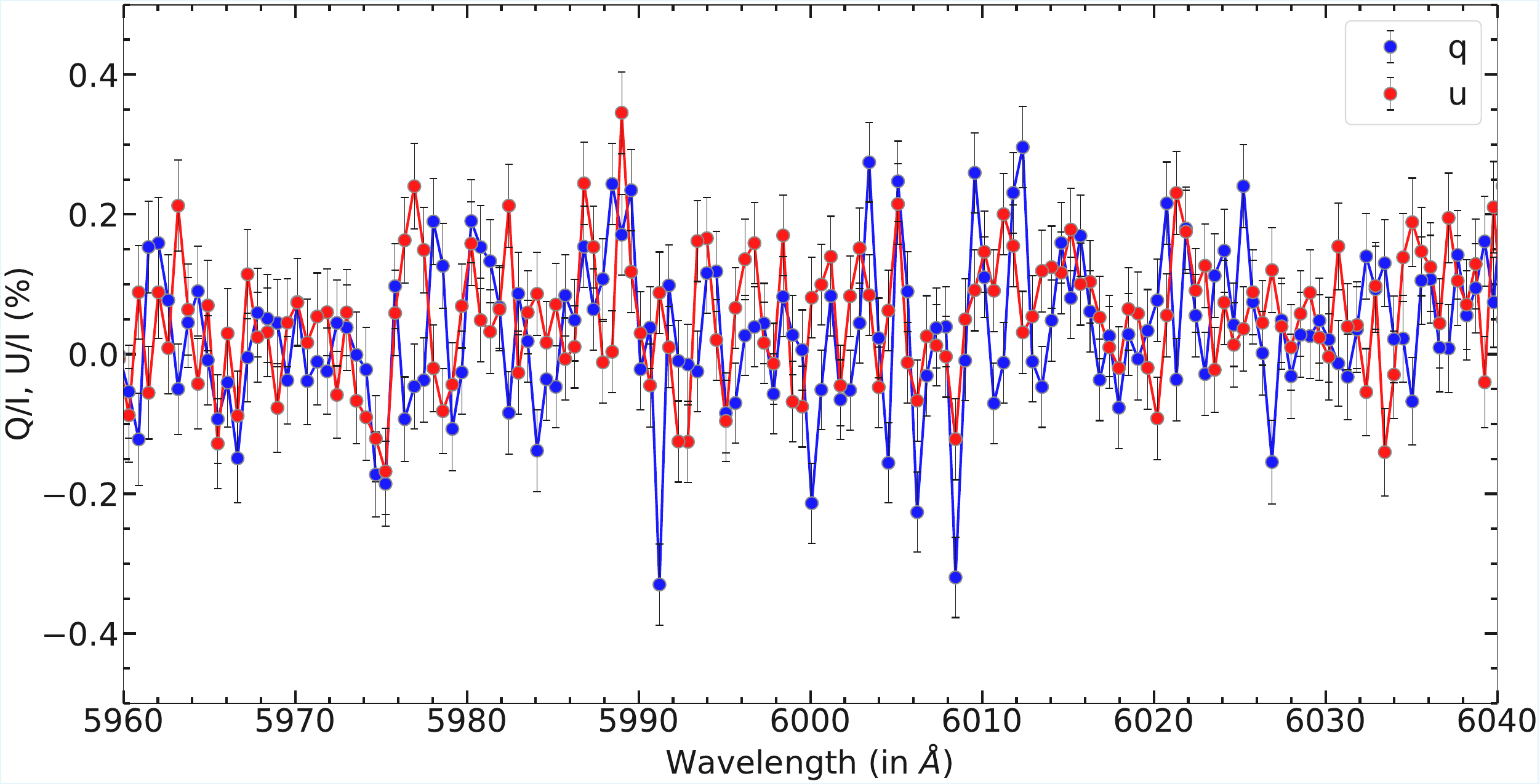}
    
    \caption{A magnified view of the Stokes spectrum of standard unpolarized star HD 47105 in the wavelength range $5960-6040 \: \AA$ to show the magnitude of the pseudo-periodic polarization variation, as discussed in the text. The data has been binned to wavelength intervals equivalent to the spectral resolution of the instrument ($\lambda/\delta \lambda$)} 
    \label{pol_ripple}
\end{center}    
\end{figure}

\begin{figure}[]
\begin{center}
    \includegraphics[width=\textwidth]{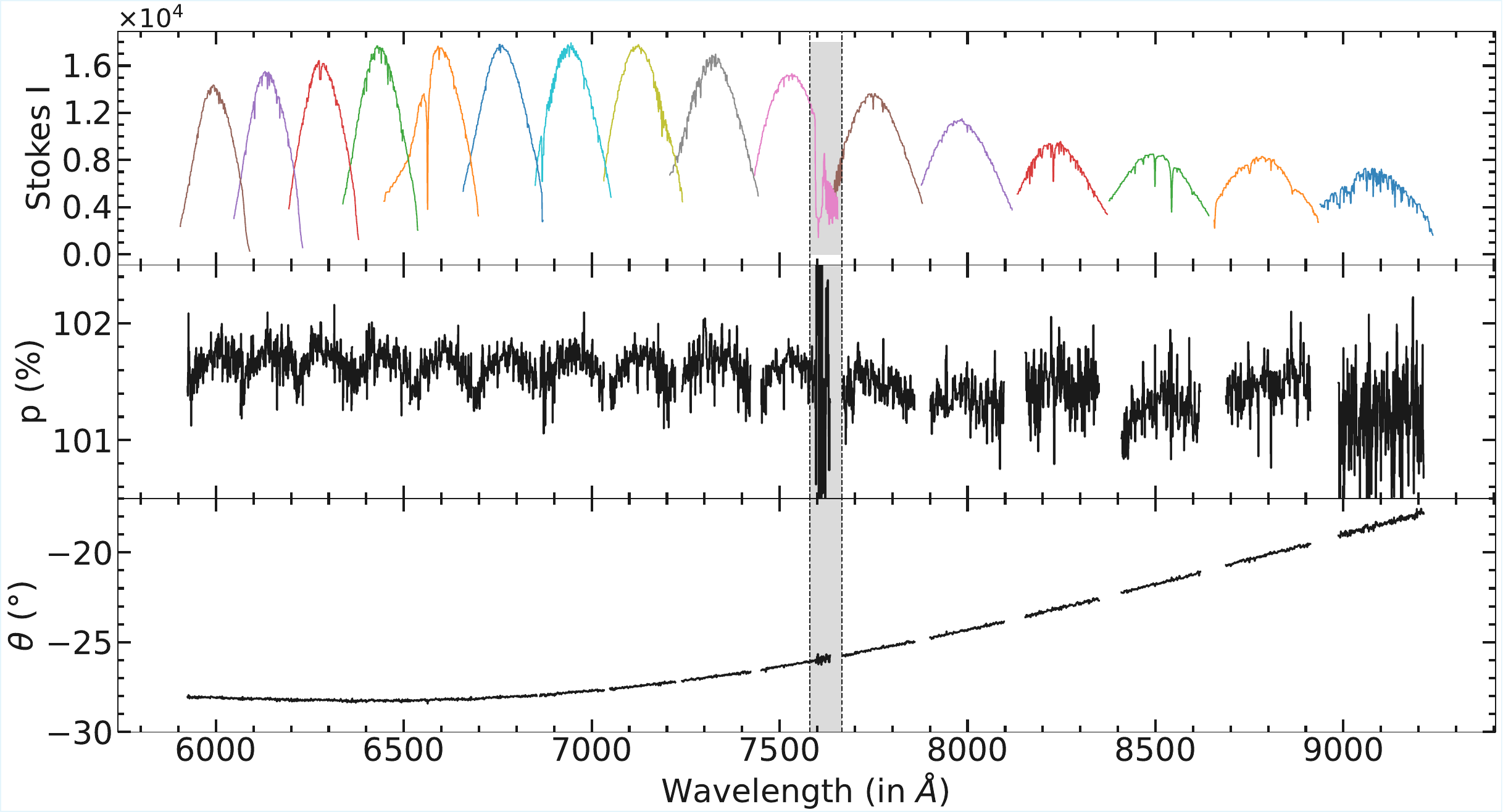}
    
    \caption{ProtoPol instrumental constants as derived from observations of standard unpolarized star HD 61421 through a GT prism. The top plot shows the raw intensity spectrum for Red CD, not corrected for instrumental blaze or CD efficiency. The middle plot shows the depolarization factor in percentage. The bottom plot shows the wavelength dependence of the position angle of the optical axis of the half-wave plate. The grey strip shows the telluric absorption region about $7602 \: \AA$ where polarization calculations should be ignored.} \label{Star_through_GT}
\end{center}    
\end{figure}


\subsubsection{The Serkowski curve Analysis}

\begin{figure}[H]
\begin{center}
    \includegraphics[width=\textwidth]{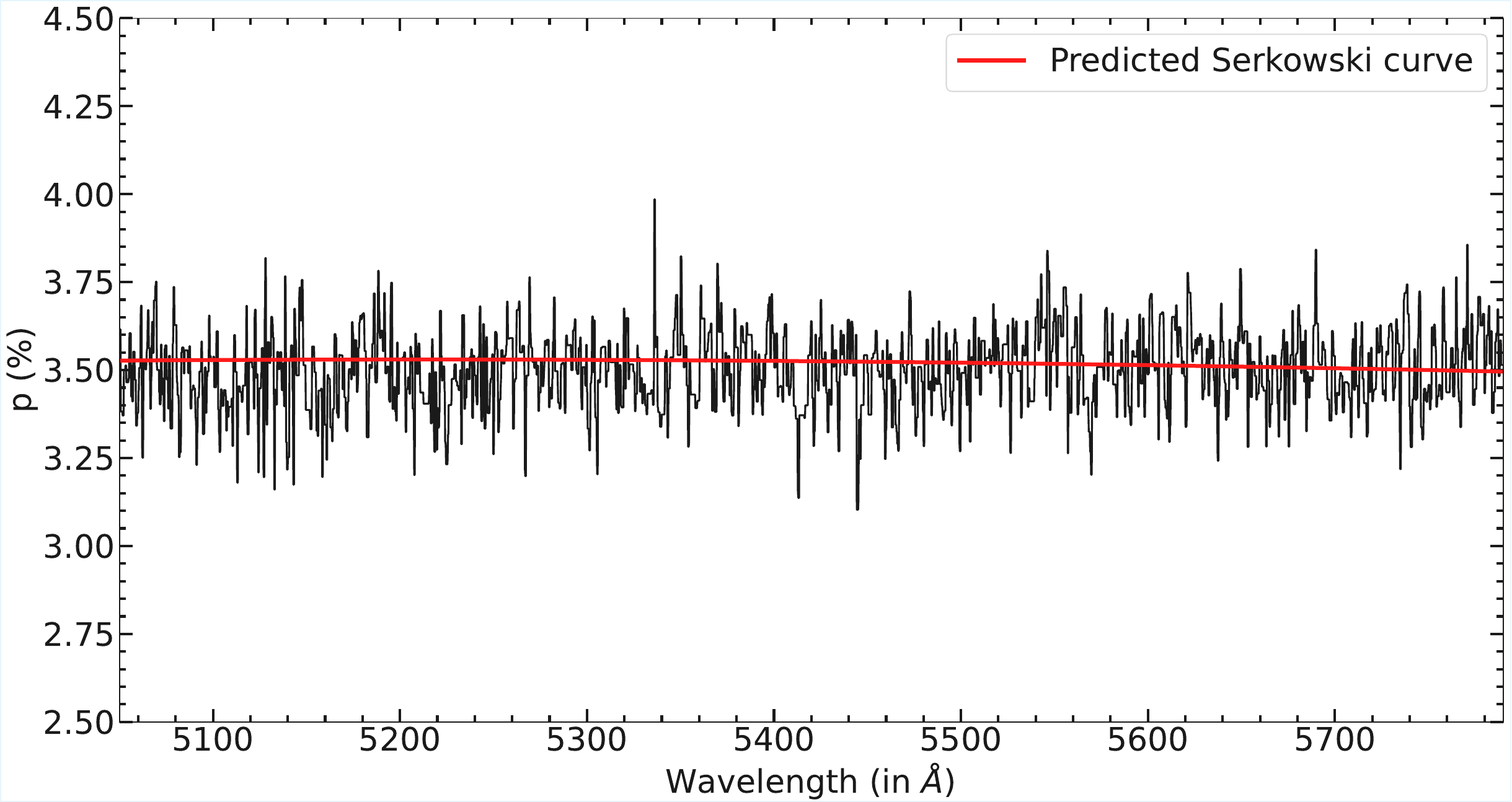}
    
    \caption{The polarization spectra of standard polarized star HD 21291. The predicted Serkowski curve, as shown by the red line, is reproduced using $\lambda_{max}$ and $p_{max}$ values as reported by \cite{hsu1982standard}} \label{Serkowski}
\end{center}    
\end{figure}

The dependence of polarimetric error on polarization degree was evaluated by conducting observations of several standard polarized stars as listed in Table~\ref{Polarized_Standard_Stars}. The polarized spectrum of one of the standard polarized stars HD 21291, together with its predicted Serkowski curve, is shown in Figure~\ref{Serkowski}. The Serkowski curve is predicted from the empirical formula  \cite{whittet1992systematic} \\

$$p_{ser}(\lambda) = p_{max} exp \left[-Kln^2\frac{\lambda_{max}}{\lambda}\right]$$ \\

where $K = 1.66(\lambda_{max}/1000nm) + 0.01$. Here, $p_{max}$ represents the peak polarization at wavelength $\lambda_{max}$. All $p_{max}$ and $\lambda_{max}$ values were obtained from existing photometric polarimeters \cite{hsu1982standard}. As it can be noticed from Fig~\ref{Serkowski}, the agreement between the polarization spectrum and the predicted Serkowski curve is good. 

\begin{figure}[]
\begin{center}
    \includegraphics[width=\textwidth]{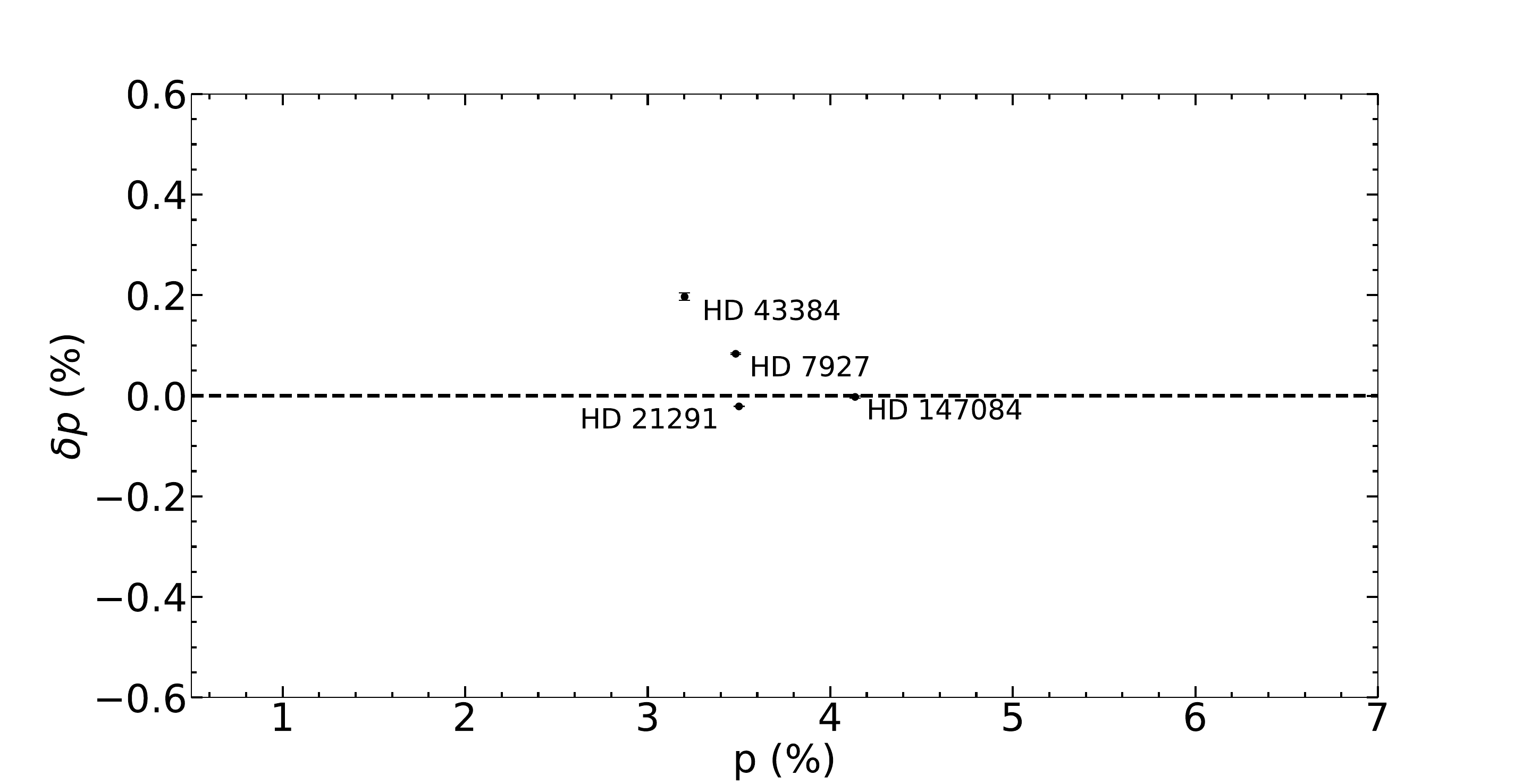}
    
    \caption{Polarimetric error $\langle \delta p \rangle$ representing the depolarization of five standard polarized stars observed by ProtoPol. $\langle \delta p \rangle$ is the average of $\delta p = p - p_{ser}$ as evaluated in the wavelength range 5000 - 5800 $\AA$} \label{Serkowski_Diff}
\end{center}    
\end{figure}

For quantitative analysis of the polarimetric error, the average difference of the polarization spectrum from the predicted Serkowski curve $\langle \delta p \rangle$ over a wavelength range of 5000 - 5800 $\AA$ was evaluated as shown in Figure~\ref{Serkowski_Diff}. The standard deviation of $\langle \delta p \rangle$ was determined to be $0.18\%$. No clear dependence of $\langle \delta p \rangle$ with degree of polarization was seen for p $< 5\%$. The $\langle \delta p \rangle$ scattering seems to be larger than the zero-point error as estimated from standard unpolarized star observations. The error could be due to uncertainties in determining background, such as sky-subtraction, inter-order scattering of the echelle spectra, etc.
\par
In summary, from the standard unpolarized and standard polarized star observations, we conclude that the polarimetric accuracy of ProtoPol is $\delta p \sim 0.1\%$ and there is practically no polarization dependence for objects with degree of polarization $< 5\%$ .


\subsection{Total throughput}
\label{subsec-ThptLimMag}

\begin{figure}[H]
\begin{center}
    \includegraphics[width=\textwidth]{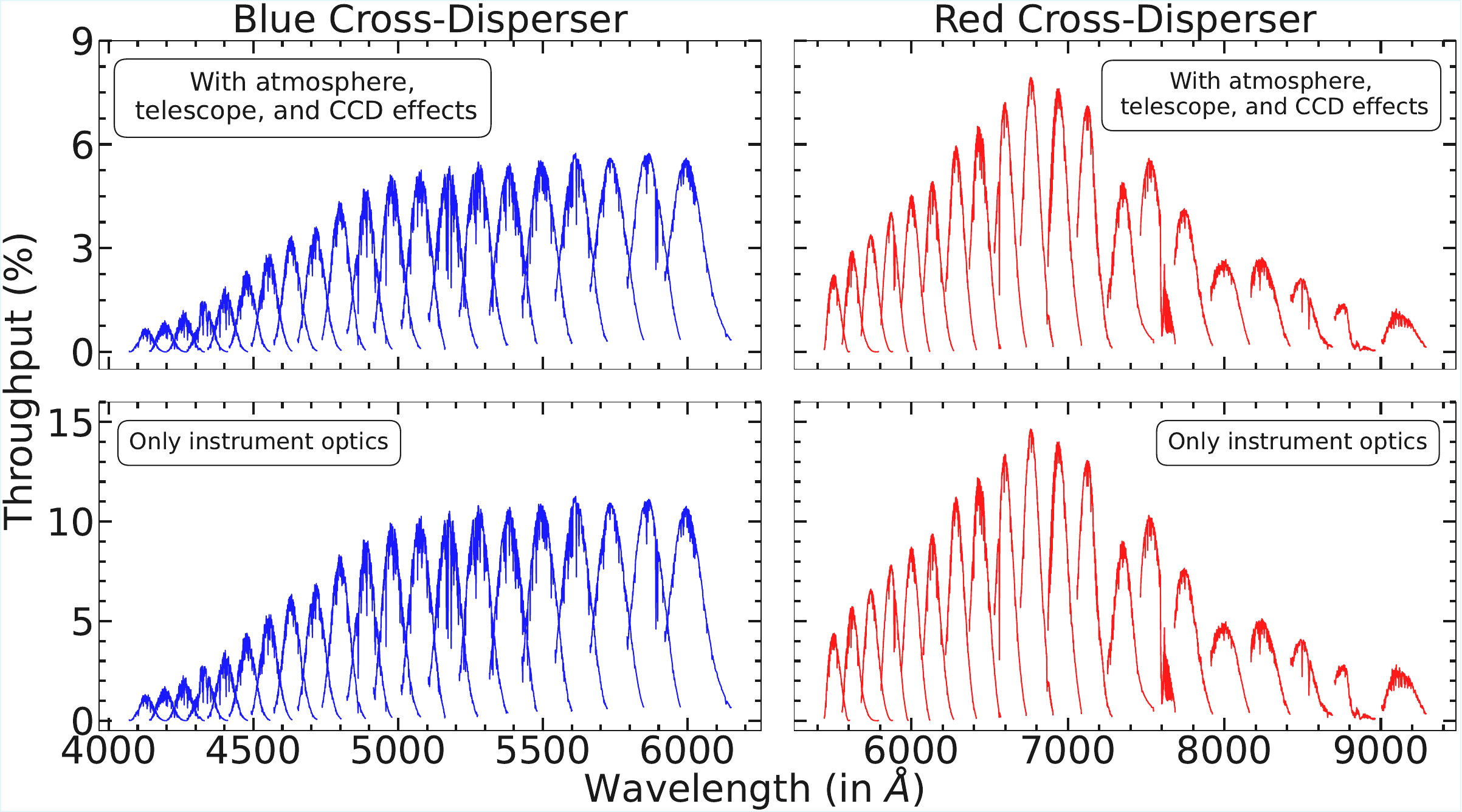}
    
    \caption{\textit{(Top)} The complete throughput of ProtoPol for both Blue \textit{(left)} and Red \textit{(right)} cross-dispersers, including the effects of CCD quantum efficiency, telescope reflectivity, and atmospheric transmission. \textit{(Bottom)} The throughput of the optical chain of ProtoPol only, by eliminating the effects of CCD, telescope, and atmospheric transmission. See text for details. The throughput is estimated by observing an extended source (Jupiter); thereby the above is without any slit-losses. Maximum throughput (for o-ray or e-ray trace) of $\sim 15\%$ as shown in the figure.} \label{throughput}
\end{center}    
\end{figure}

The optical throughput of ProtoPol was obtained from the observations of Jupiter an extended source, so that the throughput calculation does not take into account the slit-loss effect. These throughput characterization observations were conducted on PRL 1.2m telescope on 2024-01-15.675 UT. The V-band surface brightness of Jupiter was calculated to be 5.34 mag arc-second$^{-2}$ or $2.74 \times 10^{-11}$ ergs s$^{-1}$cm$^{-2}$ $\AA ^{-1}$ arc-second$^{-2}$. On PRL 1.2m, $f/13$ telescope, the plate scale is 76$\mu$m per arc-second, so the projection of the 45-degree tilted $150\:\mu m$ diameter pinhole on the sky-plane is $\sim$1.4 arc-seconds. Jupiter's spectrum was represented as a blackbody of temperature equal to that of the Sun ($\sim$5600 K) and normalized to the V-band surface brightness. This was compared with the recorded spectra to estimate the cumulative throughput of the system, including sky transmittance, telescope efficiency, and ProtoPol's efficiency. The resultant throughput is shown in Figure~\ref{throughput}. The maximum throughput is estimated to be around $8\%$ for order m = 39.

\par 
The standalone efficiency of ProtoPol can be estimated by removing the quantum efficiency of the CCD (as obtained from the datasheet of the detector), the estimated reflectivity of the telescope primary and secondary mirrors (assumed to be $85\%$ for each mirror), and the atmospheric transmittance (assumed to be $80 \%$ \cite{arasaki2015very}). This is shown in Figure~\ref{throughput} for both Blue and Red cross-dispersers. The maximum throughput is estimated to be around $15\%$ for order m = 39.
\par 


\begin{figure}[H]
\begin{center}
    \includegraphics[width=\textwidth]{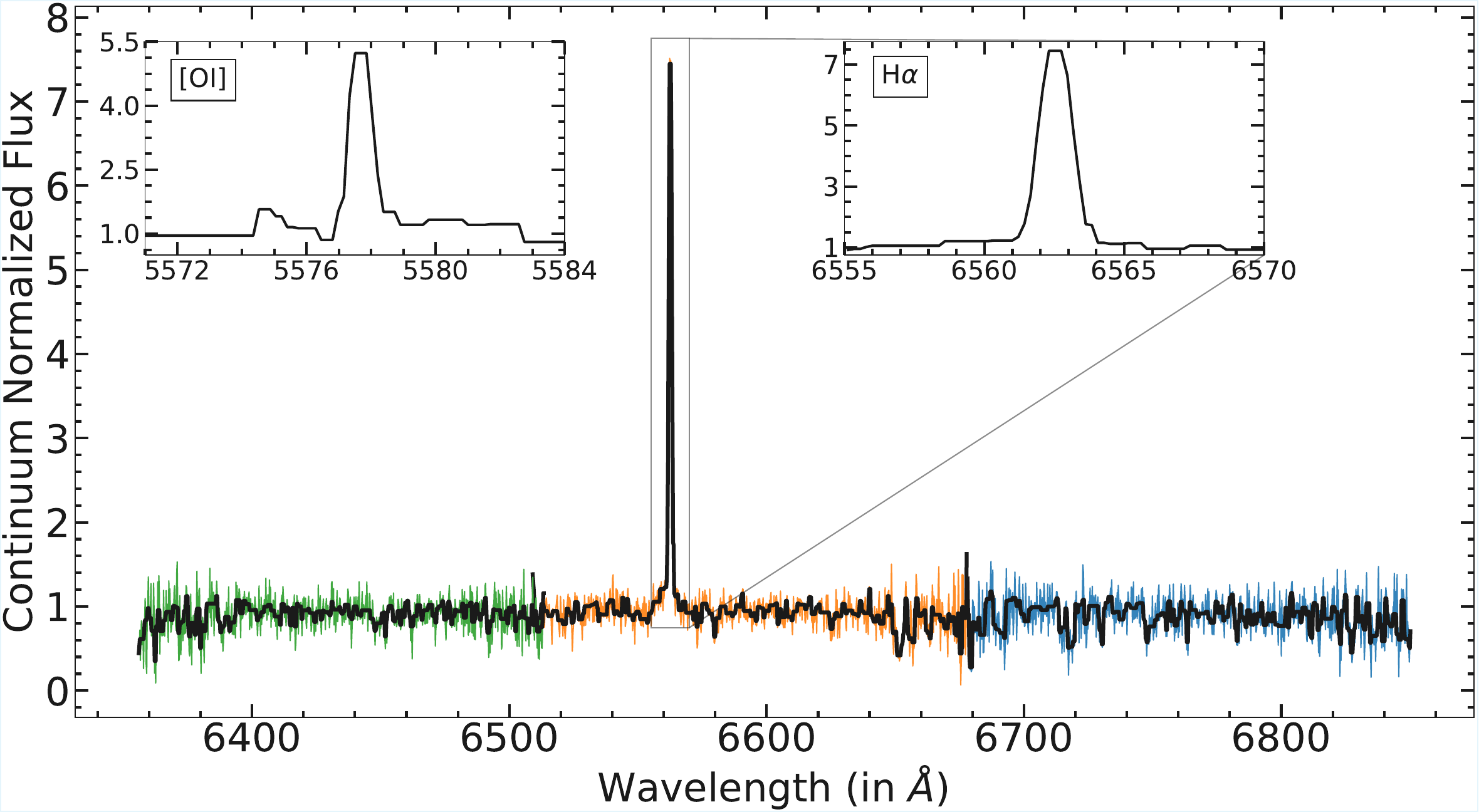}
    
    \caption{Continuum normalized spectra of M-Dwarf star Wolf 47 (V=13.21, R=12.4) for Red CD orders - 39, 40, and 41 for 30 minutes of integration time taken from ProtoPol mounted on PRL 2.5m telescope. The thick black line shows the smoothed spectrum after applying a 5-pixel rolling median. The right inset shows the zoomed-in view of the H$\alpha$ emission line profile after smoothening. The left inset shows the smoothed spectra of the same star across the [OI] emission line at 5577$\AA$ obtained using Blue CD in 1 hour of integration time.} \label{wolf}
\end{center}    
\end{figure}

The continuum normalized spectra of M-dwarf star Wolf 47 (V=13.21, R=12.4) for 3 Red CD orders - 39, 40, and 41 is shown in Figure~\ref{wolf}. SNR of $\sim$ 20-25 per pixel is achieved at the order centers. The H$\alpha$ emission line at 6562.8 $\text{\AA}$ can be seen in the orange spectra. The overlapping wavelengths of each order have been truncated due to a lack of signal at the order edges. The corresponding SNR in Blue CD is $\sim$10 for orders close to V-band center. The overplotted, smoothed spectrum (by applying a 5-pixel rolling median) is shown by the thick black line. The left inset shows the smoothed spectra of the same star across the [OI] emission line at 5577$\AA$ obtained using Blue CD in 1 hour of integration time.


\subsection{Signal to Noise Ratio (SNR)}
\label{subsec-SNR}
\begin{figure}[H]
\begin{center}
    \includegraphics[width=\textwidth]{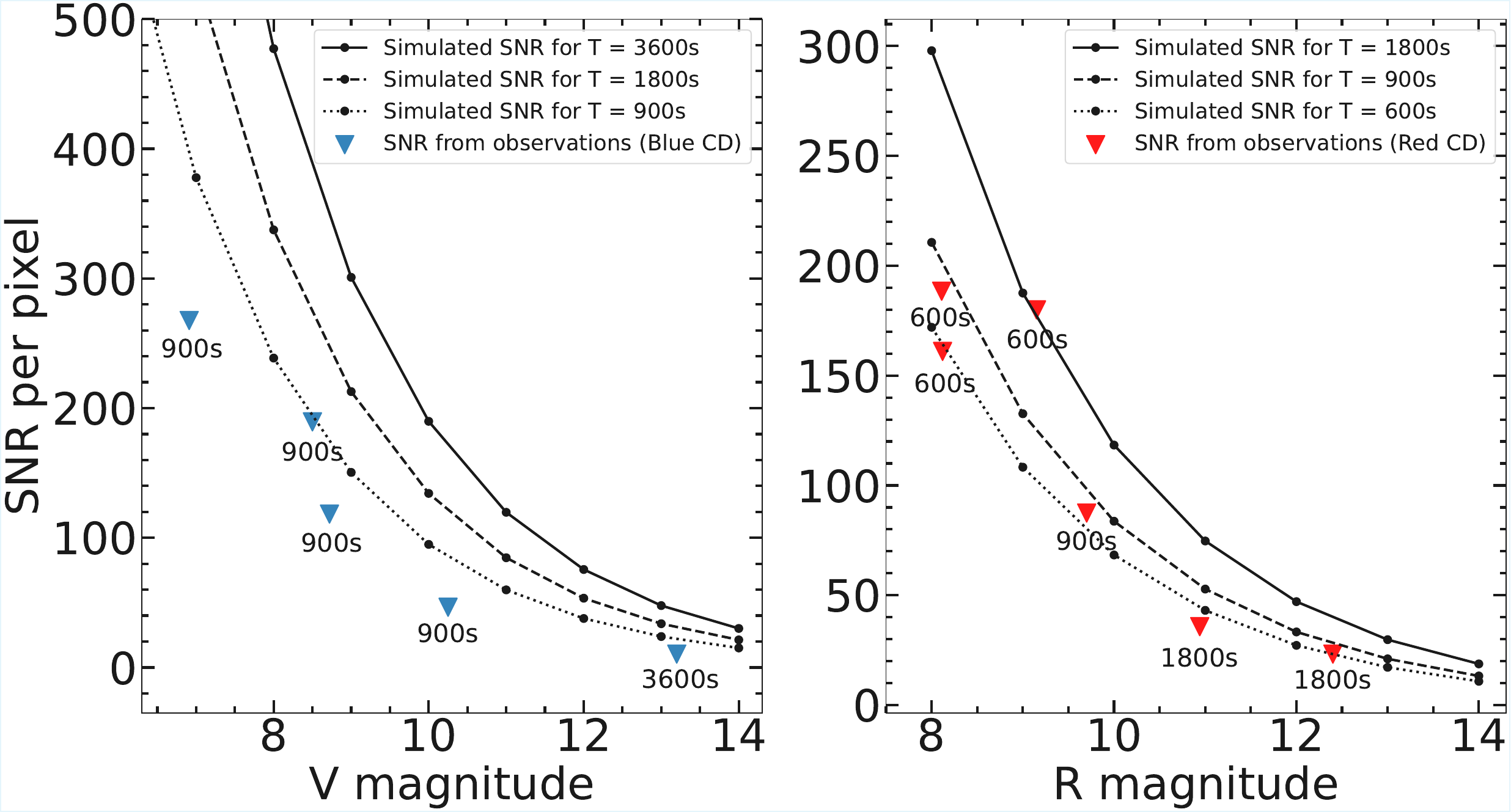}
    
    \caption{The plots show the expected SNRs from ProtoPol for various V and R band magnitudes, for Blue (left) and Red (right) CDs. The expected SNRs (solid circles) are derived from theoretical simulations for exposure times of 900 s (dotted line), 1800 s (dashed line), and 3600 s (solid line). Observationally achieved SNRs, obtained from ProtoPol data at different magnitudes and integration times, are shown as solid triangles (blue for the Blue CD and red for the Red CD). The respective integration times for each observational data point are also indicated. The observed SNRs are found to be in close agreement with the model predictions. Any discrepancy between model and observed data is due to the difference in the combined throughput of the instrument under different observation conditions.}\label{SNR}
\end{center}    
\end{figure}

The Signal-to-noise ratio (SNR) was computed for ProtoPol from several observations of stars having different magnitudes with different integration times. To compute the SNRs, the orders with wavelength centers about 5500 $\text{\AA}$ and 7000 $\text{\AA}$ were considered, for Blue and Red CDs, respectively, and the median signal of the central 100 pixels of the order was considered. A simple theoretical model was constructed to compare the SNRs achieved from observations with theoretical results. The theoretical model computes the expected number of photons incident on the CCD for a star of any magnitude by incorporating the details of telescope area, integration time, resolution of the instrument, and instrument throughput. This instrument throughput incorporates the effects of atmospheric throughput, telescope mirror reflectivity, slit loss, instrument optics throughput, grating efficiencies, and CCD quantum efficiency as discussed in section~\ref{subsec-ThptLimMag}. Since the throughput estimated above is without slit-loss, we have incorporated this effect in the following way. A typical median seeing is taken as a 2-D Gaussian profile of FHWM 1 arc-second. This would then translate into the slit coupling efficiency of $\sim 0.6$, therefore, we have considered the cumulative throughput to be 3.5\%. The observed and theoretically expected SNRs are shown in Figure~\ref{SNR}. In general, a polarization accuracy $\delta p \sim 0.1-0.2\%$ (SNR $\sim$350-700) per pixel can be achieved for a Vmag $\sim$8 source with two hours of integration time.
\par
The observed SNRs match well with the model predictions, in particular for integration times of 900 seconds or below for relatively brighter objects ($m\sim8-9$). For fainter objects, we observe a deviation of expected SNR towards lesser side,  which could be explained by a little mismatch in the alignment/identification of the source's center with the pinhole or variations in the seeing/atmospheric transmission during the long exposure time.
\par
The overall technical performance of ProtoPol is summarized in Table~\ref{protopol_performance_table}.


\begin{table*} 
\caption{Overall achieved performance of ProtoPol} \label{protopol_performance_table}
\centering
\setlength{\tabcolsep}{5pt}
\renewcommand{\arraystretch}{1.2}
	\begin{tabular}{cc cc }
           \hline
		  \hline

\textbf{Criterion} & \textbf{Value} \\
\hline
\hline

Spectral resolution & $\sim$0.40-0.75$\AA$ \\
Spectral coverage & 4000 - 6200 $\AA$ in Blue cross-disperser\\
                  & 5800 - 9600  $\AA$ in Red cross-disperser\\
Instrumental polarization & $\sim 0.1\%$ \\
Instrument throughput (maximum) & $\sim 15.0\%$ (Only instrument's optics) \\
Instrument throughput (maximum) & $\sim 8.0\%$ (Including all factors) \\
Limiting magnitude (spectro-polarimetry)  &  $m_V = 8$ \\
 &  ($\delta p \sim 0.1-0.2\%$ per pixel in 2 hours)  \\
                                    
Limiting magnitude (spectroscopy) &  $m_V = 13.2$\\
                              &  ($SNR \sim 10$ per pixel in 1 hour) \\

		\hline
		\hline
	\end{tabular}
\end{table*}



\section{Science verification of ProtoPol : The first science results}
\label{sec-science}

A variety of astrophysical objects were chosen for the science verification of the instrument. These science observational campaigns were conducted in parallel to the instrument characterization observations ever since the first light of the instrument in December 2023, so that these sample would prove useful for performance characterization in various astrophysical science cases and in the subsequent data reduction pipeline development. Many of the sample stars were chosen whose prior spectro-polarimetric observational data were available in the literature, so that they could be used as a reference in various science cases. Three different kind of samples were chosen for science verification (1.) Herbig Ae/Be stars / classical Be stars, (2.) Symbiotic Stars, and (3.) Red giant stars. While the first two sample provide the scientific test cases of emission line polarization having varying physical origins (to be discussed later), the sample of cool red giants provides the test case of dust-induced continuum polarization measurements. The aim of this exercise is to establish that ProtoPol could be successfully used to probe polarization induced by varying physical processes. Further, Herbig stars, classical Be stars, or Symbiotic stars are all dynamical systems, and their spectral/spectro-polarimetric profiles are expected to change over the years, so any change or similarity of polarization profile with reference to the literature data would be a valuable comparison. Below, we shall be describing each of these cases. While we have observed a good number of candidates in each of the sample, we shall demonstrate the polarization-specific physical effects only in few of them while the data of the rest will be presented in our subsequent works.
\par
The data presented in this section have not been corrected for instrumental polarization, as it would only have a polarization bias effect on the data. For the same reason, the data have also not been corrected for interstellar polarization (ISP), as the ISP varies with wavelength over a very broad wavelength range. As such, across an emission/absorption line, it can be considered constant.


\subsection{Herbig Ae/Be Stars}

Herbig Ae/Be stars are optically visible pre-main-sequence stars that are the more massive counterparts of T Tauri stars with typical masses ranging from 2-10 $M_\odot$. This group of stars was first classified by Herbig in 1960. Due to their intermediate masses, they play an essential role in bridging the gap between the formation mechanism of low-mass stars and high-mass stars. While the low-mass stars form by cloud collapse and magnetically controlled accretion \cite{bouvier2007magnetospheric}, similar formation mechanisms are found not to be consistent with high-mass star formation scenarios \cite{alecian2013high}. Although clear-cut evidence of the presence of circumstellar gas and dust in such systems has been reported by \cite{waelkens1997comet}, there is no general consensus about the geometry of such systems. Spectro-polarimetry across emission lines of spectra is, thus, a very powerful tool to probe the inner circumstellar discs around the stars on scales of a few stellar radii. 
\par
A sample of Herbig Ae/Be stars were selected from the Herbig stars catalog of \cite{de1994new}. The main criterion for selection was the relative brightness of the stars (V $\lesssim$ 11) and their relative position on the sky during the period of observation. The observation log of the sample is given in Table~\ref{HerbigTable}, along with their V magnitudes as noted from \href{https://simbad.u-strasbg.fr/simbad/sim-fbasic}{SIMBAD} database, the exposure times for the observations, and the presence/absence of polarization line effect. All these targets had been observed for their spectro-polarimetry \cite{oudmaijer1999halpha, vink2002probing, harrington2008spectropolarimetric} earlier and hence provide a useful means to compare the obtained results from ProtoPol with the literature data over a period of time. There are mainly four types of polarization effects seen across the spectral emission features (mostly in the emission lines of hydrogen Balmer series, H$\alpha$, H$\beta$ etc.) of Herbig stars. Below we discuss these effects and present the results obtained from ProtoPol for a few sample stars.


\begin{table*} 
\caption{Sample of Herbig Ae/Be Stars. } \label{HerbigTable}
\centering
\setlength{\tabcolsep}{5pt}
\renewcommand{\arraystretch}{1.2}
	\begin{tabular}{cc cc cc cc cc cc }
           \hline
		  \hline

\textbf{Name} & \textbf{Vmag} & \textbf{Date} & \textbf{Exposure (s)} & \textbf{Line Effect?} \\
\hline
\hline
MWC 442 & 7.7 & 01-12-2024 & 900s x 4 x 3 sets & No \\
AB Aur & 7.1 & 28-11-2024 & 900s x 4 x 4 sets & Yes\\
MWC 480 & 7.7 & 30-11-2024 & 900s x 4 x 5 sets & Yes \\
MWC 758 & 8.3 & 02-12-2024 & 900s x 4 x 4 sets & Yes \\
MWC 120 & 7.9 & 29-11-2024 & 900s x 4 x 4 sets & Yes \\
FS CMa & 8.1 & 04-01-2025 & 900s x 4 x 3 sets & Yes \\
MWC 158 & 6.6 & 29-11-2024 & 900s x 4 x 1 set & Yes \\
        &     &  04-01-2025 & 900s x 4 x 2 sets & Yes \\
GU CMa & 6.6 & 02-12-2024 & 600s x 4 x 2 sets & No \\
27 CMa & 4.7 & 02-12-2024 & 900s x 4 x 2 sets & Yes \\
HD 58647 & 6.8 &  28-12-2024 & 1800s x 4 x 1 set & Yes \\
MWC 137 & 11.2 & 28-12-2024 & 900s x 4 x 2 sets & Yes \\
MWC 147 & 8.8 & 28-12-2024 & 1800s x 4 x 2 sets & Yes \\
        &     & 06-01-2025 & 900s x 4 x 3 sets  & No \\
IL Cep & 9.3 & 30-12-2024 & 900s x 4 x 2 sets & Yes \\
        
		\hline
		\hline
	\end{tabular}
\end{table*}

\subsubsection{No Line Polarization}

\begin{figure}[H]
  \begin{center}
    \includegraphics[width=\textwidth]{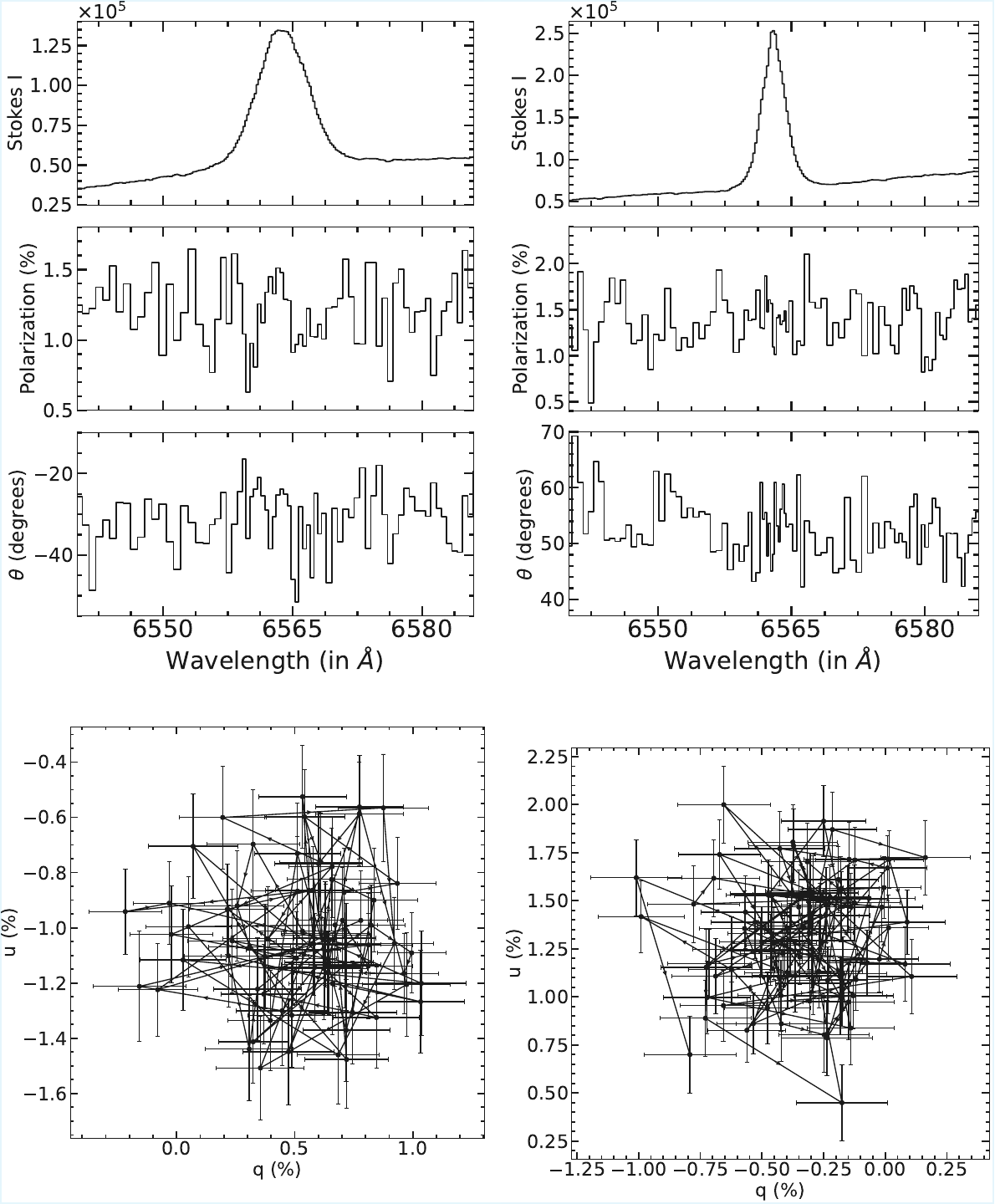}

  \caption{The Polarization spectra for Herbig stars GU CMa \textit{(left)} and MWC 442 \textit{(right)}  across H$\alpha$ emission and their corresponding (Q, U) diagrams \textit{(bottom panel)}. The Stokes I spectrum, derived degree of polarization ($p$), and the angle of polarization are shown in the top, middle, and bottom panels, respectively. The data for both stars are binned to a polarization accuracy of $0.2\%$ per binning element. No line-specific polarization is detected, which is consistent with earlier reported results for these sources.}
  \label{Herbig1}

\end{center}
\end{figure}

In this case, no change in polarization is observed across the H$\alpha$ emission line as compared to the continuum. This would represent a circular projection of the circumstellar environment in the plane of the sky. If the geometry of the system is spherically symmetric or if the disc is viewed pole-on, no change of polarization is expected to be observed \cite{ababakr2016linear}. In the Q-U plane, this effect would be represented by a knot of points at a single dot due to continuum polarization. The derived polarization measurements across H$\alpha$ emission with ProtoPol's observations of two such stars GU CMa and MWC 442 are presented in Figure~\ref{Herbig1} and are discussed below:
\par 
\textbf{GU CMa:} It is a B2 type Herbig star showing no clear line effect across H$\alpha$.  However, the star does possess significant continuum polarization of $1.15\%$ and a line/continuum ratio of 3 (\cite{oudmaijer1999halpha}. Given that GU CMa shows minimal infrared continuum excess (\cite{hillenbrand1992herbig}, it could be interpreted as a classical Be star viewed at a low inclination and located behind a dense interstellar medium. This interpretation is supported by the presence of single-peaked H$\alpha$ emission. GU CMa was a non-detection in the archival ESPaDOnS data of 2006 February 9 and the HiViS observations by \cite{harrington2008spectropolarimetric}. Observations of GU CMa from ProtoPol reveal a non-detection of line effect across H$\alpha$, like previous results and similar continuum polarization values.
\par 
\textbf{MWC 442:} It is B5 type Herbig star, showing no clear evidence of line effect but similar continuum polarization values and single-peaked H$\alpha$  structure to GU CMa, suggesting a similar low-inclination orientation in the sky plane behind dense interstellar medium. \cite{harrington2008spectropolarimetric} reported no line polarization detection. Observations from ProtoPol also reveal the star showing no line effect and 


continuum polarization values of $\sim 1.2\%$.


\subsubsection{Depolarization} 
\begin{figure}[H]
  \begin{center}
    \includegraphics[width=\textwidth]{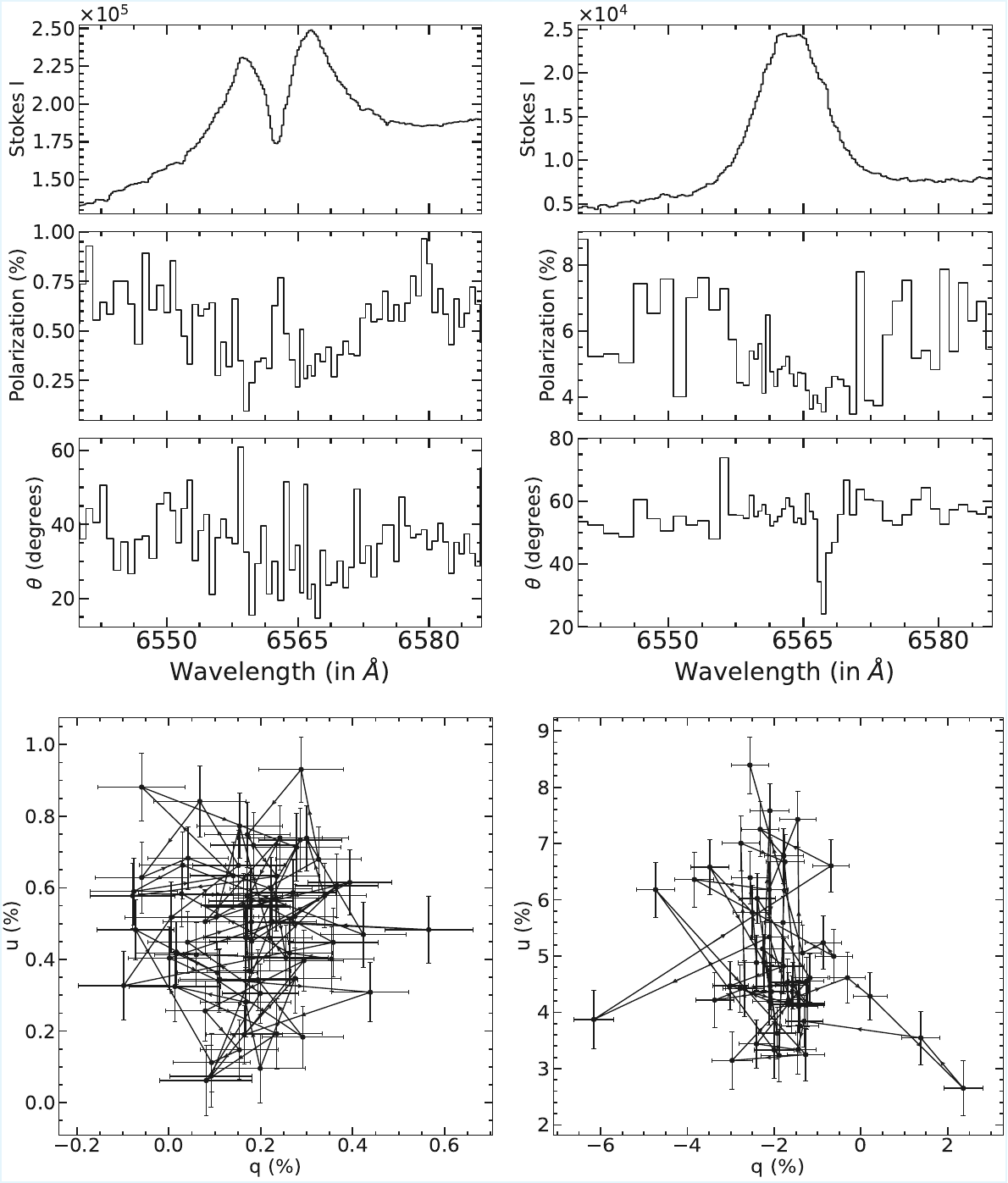}

  \caption{Same as Fig. \ref{Herbig1} but for Herbig stars 27 CMa \textit{(left)} and Il Cep \textit{(right)}. The data for 27 CMa and Il Cep have been binned to a polarization accuracy of $0.1\%$ and $0.5\%$, respectively, per binning element.  These Herbig Ae/Be stars show depolarization effects in H$\alpha$ emission.}
  \label{Herbig2}

\end{center}
\end{figure}

If a drop in polarization is observed across the entire line profile of H$\alpha$ emission line compared to the continuum polarization, the effect is known as depolarization \cite{vink2002probing}. This effect is predominantly seen in classical Be stars and late Herbig Be stars. Presence of this effect gives us the idea that the hydrogen emission lines are formed in an extended ionized atmosphere of the star where they suffer fewer scatterings compared to the continuum photons emitted from the stellar photosphere \cite{ababakr2016linear}. This results in a decrease in polarization across the line. In the Q-U plane, this effect manifests itself as a line excursion from the continuum knot to the line.Figure~\ref{Herbig2} shows the depolarization effect as seen in B-type stars 27 CMa and Il Cep belonging to B3 and B2 spectral types, respectively.
\par 
\textbf{27 CMa:} It is a B3 type bright Herbig star with $T_{eff}=19300\pm193$ and $v.sin(i)$  measurement of $200\pm4$ kms$^{-1}$ indicating relatively high inclination angle and hence a line effect \cite{harrington2008spectropolarimetric}. This might explain the depolarization observed across the H$\alpha$ line as observed with ProtoPol, with polarization values decreasing to $0.2\%$ from $0.8\%$ continuum polarization. \cite{harrington2008spectropolarimetric} also conducted spectro-polarimetric observations of the star.
\par 
\textbf{Il Cep:} It is a B2 type Herbig star with high continuum polarization values $4.24\%$ as reported by \cite{vink2002probing}. However, they reported a non-detection of any line effect across H$\alpha$. \cite{harrington2008spectropolarimetric} also stated similar results. However, observations from ProtoPol reveal a small depolarization across H$\alpha$, although the continuum polarization values are similar to those reported by \cite{vink2002probing}.

\subsubsection{Intrinsic line polarization} 
\begin{figure}[H]
  \begin{center}
    \includegraphics[width=\textwidth]{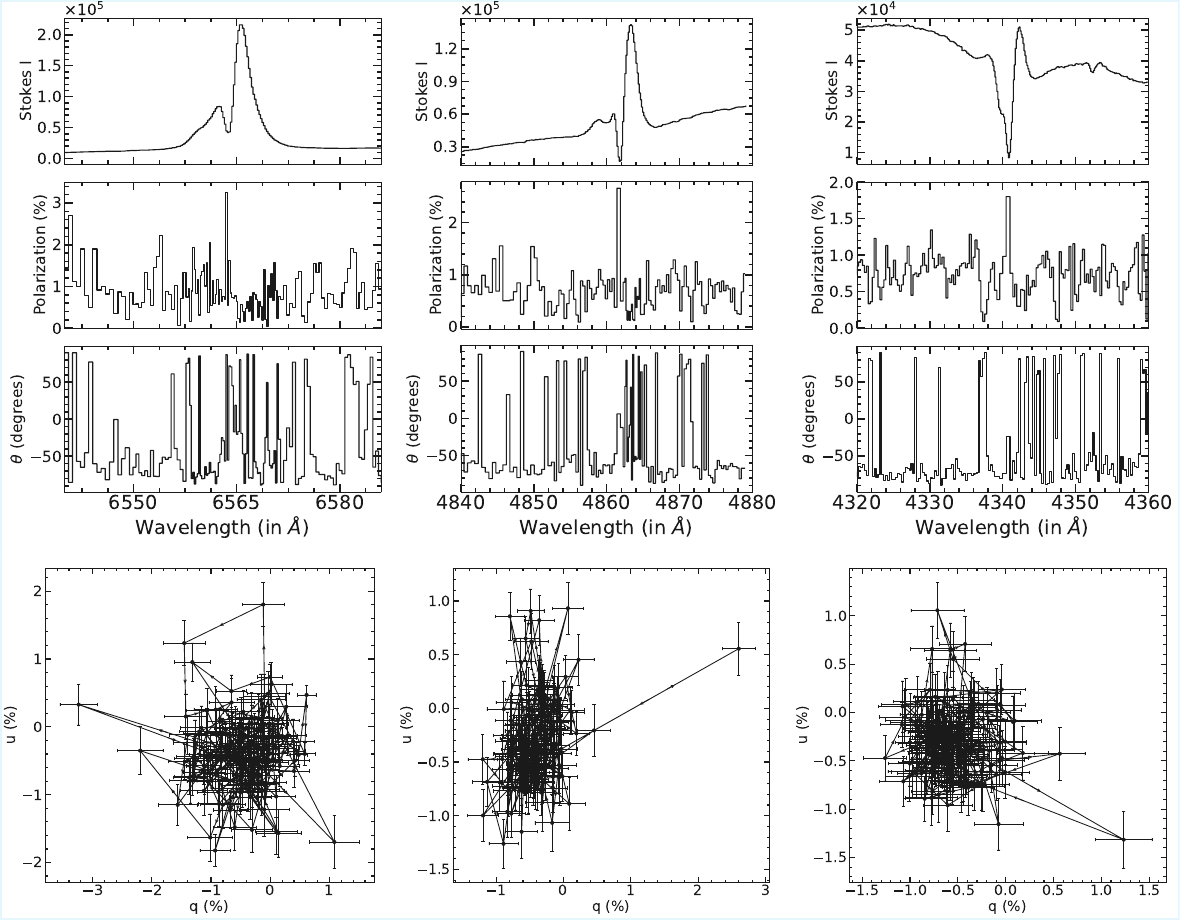}

  \caption{Same as Figure~\ref{Herbig1} but for Herbig star MWC 158 for H$\alpha$ \textit{(left)}, H$\beta$ \textit{(middle)}, and H$\gamma$ \textit{(right)} lines. The data for  H$\alpha$,  H$\beta$, and H$\gamma$ have been binned to a polarization accuracy of $0.2\%$, $0.3\%$, and $0.3\%$, respectively, per binning element. Intrinsic Line polarization is seen in MWC 158. See text for details.}
  \label{Herbig3}

\end{center}
\end{figure}

In this case, the emission line is intrinsically polarized. This effect is typically seen in the less massive, late Herbig Be stars, Herbig Ae stars, and T Tauri stars \cite{vink2002probing}. As opposed to the depolarization case, the width of the line effect in this case is typically smaller than the spectral line width. Magnetospheric accretion is thought to be the cause of line polarization where a photospheric shock is produced by in-falling disc material funneled via accretion columns \cite{ababakr2016linear}. The emission lines that are produced at this hotspot would be polarized as they suffer scattering off the disc material. If the scattering occurs in a predominantly rotating disc-like configuration, a changing position angle with wavelength is observed due to the breaking of left-right reflection symmetry in velocity space. This appears as a loop in the Q-U plane. On the other hand, if the scattering occurs in an expanding medium, there is no breaking of left-right symmetry \cite{ababakr2016linear}. This is represented by a linear track in the Q-U plane.  Figure~\ref{Herbig3} shows the intrinsic polarization in the Herbig star MWC 158 across the H$\alpha$, H$\beta$, and H$\gamma$ lines. Similar polarization profiles across all three lines indicate all three kinds of emission are probably emitted from a similar region in the circumstellar environment of the star.
\par 
\textbf{MWC 158:} MWC 158 is classified as a mid-B type star, notable for exhibiting a strong emission line with a pronounced central absorption feature. The star has been previously investigated for its spectroscopic variability through low-resolution spectro-polarimetric observations \cite{bjorkman1998first, pogodin1997circumstellar}. Signatures of both stellar winds and infalling material were identified by \cite{pogodin1997circumstellar} based on envelope lines, with the H$\alpha$ absorption spanning velocities from $-200$ kms$^{-1}$ to $+70$ kms$^{-1}$. \cite{bjorkman1998first} observed nearly constant continuum polarization across wavelengths and concluding that the polarization is caused by electron scattering rather than dust. Further H$\alpha$ spectro-polarimetric data from 1995 and 1996 were presented by \cite{oudmaijer1999halpha}. Evidence for binarity was later provided by \cite{baines2006binarity} using spectro-astrometry, which revealed shifts in both the centroid and equivalent width of the point-spread function across the H$\alpha$ line. Observations from ProtoPol reveal an increase in polarization across the absorption trough of the H$\alpha$ line. Though the polarization changes may appear to be narrow, the same change is observed across the absorption troughs of H$\beta$ and H$\gamma$ lines as well, proving the polarization change across the lines is real. \cite{vink2002probing} and \cite{harrington2008spectropolarimetric} both reported similar polarization change across the H$\alpha$ absorption trough like what is observed from ProtoPol. Furthermore, all three Balmer lines show a similar spectro-polarimetric signature, revealing that all three kinds of emission lines are probably emitted from a similar region in the circumstellar environment of the star. 

\subsubsection{McLean Effect}
\begin{figure}[H]
  \begin{center}
    \includegraphics[width=\textwidth]{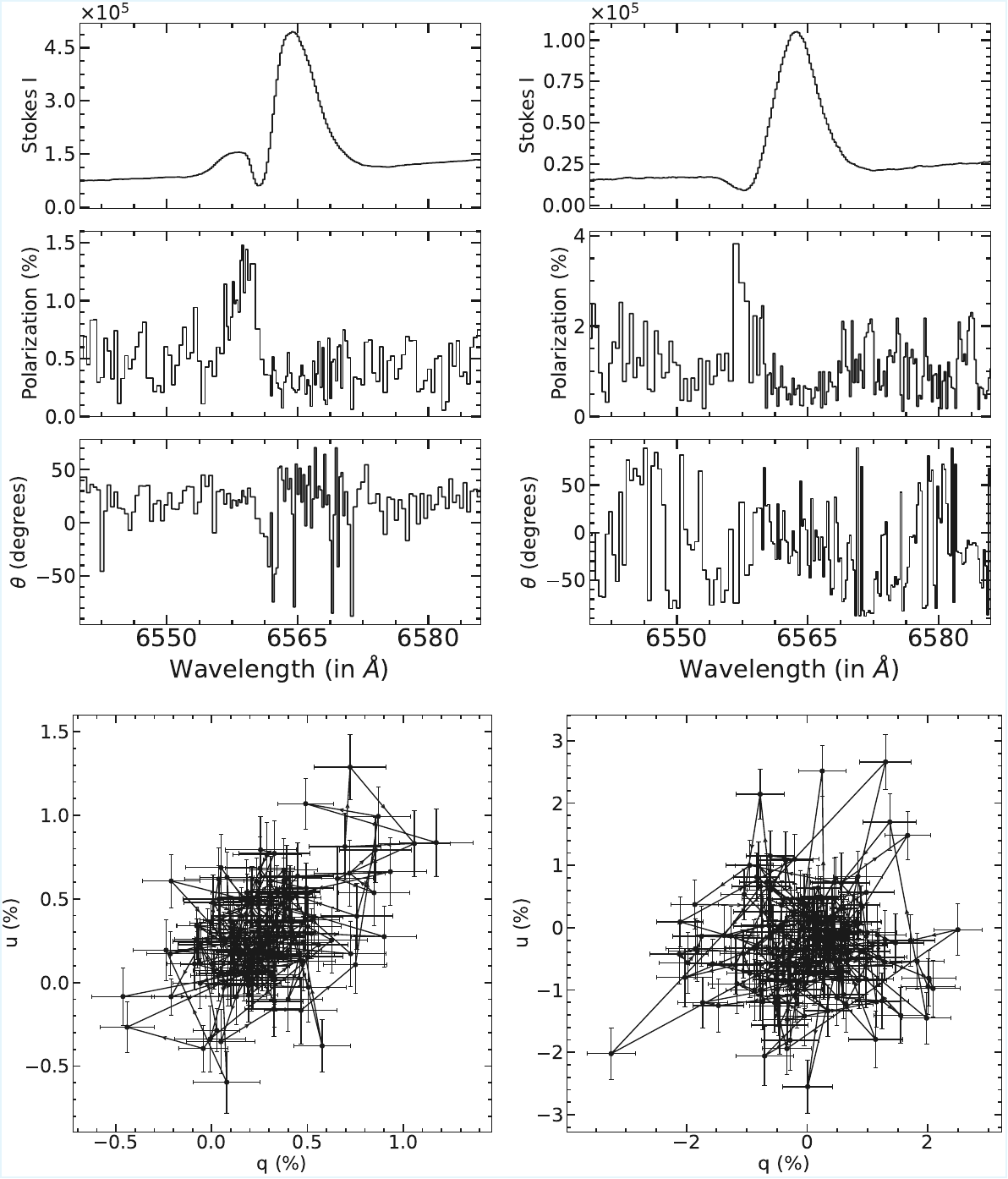}

  \caption{Same as Figure~\ref{Herbig1} but for Herbig stars MWC 480 \textit{(left)} and MWC 758 \textit{(right)}. The data for MWC 480 and MWC 758 have been binned to a polarization accuracy of $0.2\%$ and $0.5\%$, respectively, per binning element. McLean effect is noticed in the polarization spectra.}
  \label{Herbig4}

\end{center}
\end{figure}

McLean effect is observed in a number of Herbig Ae/Be stars \cite{ababakr2016linear}. In such cases, stellar wind along our line of sight blocks and removes the unscattered light from the beam. This leads to a blue-shifted absorption component superimposed on a strong emission component. However, due to the re-emission process being isotropic, the flux at the absorption component does not reach zero. This is because some photons get scattered into our line of sight, causing an increase in polarization across the absorption component \cite{ababakr2016linear}.  Figure~\ref{Herbig4} shows the McLean effect across the blue-shifted P-Cygni absorption trough of the Herbig stars MWC 480 and MWC 758. 
\par 
\textbf{MWC 480:} It is a relatively bright Herbig A3 star \cite{harrington2008spectropolarimetric}. This star exhibited a significant polarization increase of approximately $0.9\%$ in the blue-shifted absorption region, along with a $0.3\%$ decrease across the emission line, from a baseline continuum polarization of $0.4\%$, as reported by \cite{vink2002probing}. In a later study \cite{vink2005probing}, the polarization increase in the absorption region was around $0.8\%$, with the continuum level ranging from $0.18\%$ to $0.30\%$. \cite{mottram2007difference} observed a $0.4\%$ rise in polarization within the blue-shifted absorption but found no significant change across the emission line, with a continuum at $0.2\%$, indicating variability in polarization. Additionally, \cite{beskrovnaya2004active} reported R-band polarization around $0.3\%$ in most of their measurements. \cite{harrington2008spectropolarimetric} compared observations of two different epochs of the source using ESPaDOnS and HiVIS spectro-polarimeters to show clear detection of polarization variability, though with similar profiles and slightly reduced amplitude. Overall, the spectro-polarimetric data reveal strong variability in polarization, despite relatively stable line profiles. Observations from ProtoPol also reveal a similar increase in polarization across the blue-shifted absorption trough (McLean effect), with polarization increase of about $1\%$ above the continuum polarization of $0.5\%$, showing similar trends to what has been reported in the literature. 

\par 
\textbf{MWC 758:} This is a A3 type Herbig star, showing the McLean effect. \cite{harrington2008spectropolarimetric}  reported a clear spectro-polarimetric signature across the blue-shifted absorption trough of H$\alpha$ showing $\sim1\%$ change in $u$, with a smaller variation also visible in the $q$ spectrum.  In the archival data, observed with ESPaDOnS on CHFT telescope, no polarization signal was detected, and the H$\alpha$ absorption was much weaker compared to nearly all HiVIS observations \cite{harrington2008spectropolarimetric}. \cite{beskrovnaya1999spectroscopic} reported a continuum polarization of about $0.15\%$, with significant night-to-night variability reaching $0.4\%$. However, \cite{vink2002probing} did not observe a spectro-polarimetric effect in this star. The polarization variation occurs mainly on the blue-shifted transition from absorption to emission, rather than at the center of the absorption trough. The most significant decrease in $u$ occurs near the edge of the blue-shifted absorption, not at the emission peak. In $q$$u$-space, the polarization changes are nearly in phase, though not perfectly aligned, and the $q$$u$ loop shows noticeable width due to a slight increase in $q$. ProtoPol observations of the star reveal a strong increase in polarization across the blue-shifted absorption trough, $\sim2\%$ above the continuum polarization, with the polarization decreasing below the continuum level at H$\alpha$ peak. Thus, the measured polarization matches well with the reported observations of \cite{harrington2008spectropolarimetric}.


\subsection{Classical Be Stars}
\label{subsec-ClassicalBe}

Classical Be stars are B-type stars that exhibit emission lines, often spanning a range of spectral subtypes, and are associated with polarizing mechanisms \cite{oudmaijer1999halpha}. These stars are rapidly rotating and typically lie near the main sequence (MS). They are surrounded by a gaseous circumstellar disk or envelope, which contains no dust. As a result, the dominant polarizing mechanism in Classical Be stars is electron scattering \cite{harrington2008spectropolarimetric}. In contrast, B[e] stars—a distinct subtype—do exhibit strong dust signatures \cite{harrington2008spectropolarimetric}. Many B[e] stars have been spatially resolved using interferometry, revealing flattened circumstellar envelopes \cite{de2003spinning}. Observations also show intrinsic polarization that is perpendicular to the long axis of the envelope, supporting the presence of non-spherical geometries in their circumstellar environments \cite{quirrenbach1997constraints}.
\par 
Along with Herbig Ae/Be stars, several classical Be stars were observed while ProtoPol was mounted on PRL 1.2m telescope. Since, a large set of observations and theoretical work had been done on Be stars and these stars had been a target of spectro-polarimetric observations since at least the 1970s \cite{harrington2008spectropolarimetric}, they presented a good target set to test the spectro-polarimetric performance of ProtoPol across emission lines. Therefore, several bright (V $\lesssim$ 5) classical Be stars were also observed with ProtoPol on PRL 1.2m telescope. Several classical Be stars, like their Herbig Be star counterparts, show similar depolarization effects across the H$\alpha$ profile due to similar reasons as explained above. The sample of observed classical Be stars is given in Table~\ref{ClassicalBeTable}. 

\begin{table*} 
\caption{Sample of Classical Be Stars.} \label{ClassicalBeTable}
\centering
\setlength{\tabcolsep}{5pt}
\renewcommand{\arraystretch}{1.2}
	\begin{tabular}{cc cc cc cc cc cc }
           \hline
		  \hline

\textbf{Name} & \textbf{Vmag} & \textbf{Date} & \textbf{Exposure (s)} & \textbf{Line Effect?} \\
\hline
\hline
$\phi$ Per & 4.06 & 24-01-2024 & 300s x 4 x 3 sets & Yes \\
$\psi$ Per & 4.23 &  24-01-2024 & 300s x 4 x 4 sets & Yes\\
$\zeta$ Tau & 3.03 & 12-01-2024 & 120s x 4 x 5 sets & Yes \\
$\eta$ Tau & 2.87 & 25-01-2024 & 180s x 4 x 2 sets & No \\
48 Per & 4.03 & 15-01-2024 & 180s x 4 x 4 sets & No \\
$\beta$ CMi & 2.89 &  05-01-2024 & 300s x 4 x 1 set & No \\
            &      &  24-01-2024 & 180s x 4 x 2 sets & No \\
$\gamma$ Cas & 2.39 &  31-12-2023 & 300s x 4 x 4 sets & Yes \\
$\beta$ Mon & 4.60 &  16-01-2024 & 600s x 4 x 2 sets & Yes \\
            &     &  17-01-2024 & 600s x 4 x 2 sets & Yes \\
$\epsilon$ Aur & 2.99  &  05-01-2024 & 300s x 4 x 2 sets & No \\

		\hline
		\hline
	\end{tabular}
\end{table*}

\begin{figure}[H]
  \begin{center}
    \includegraphics[width=\textwidth]{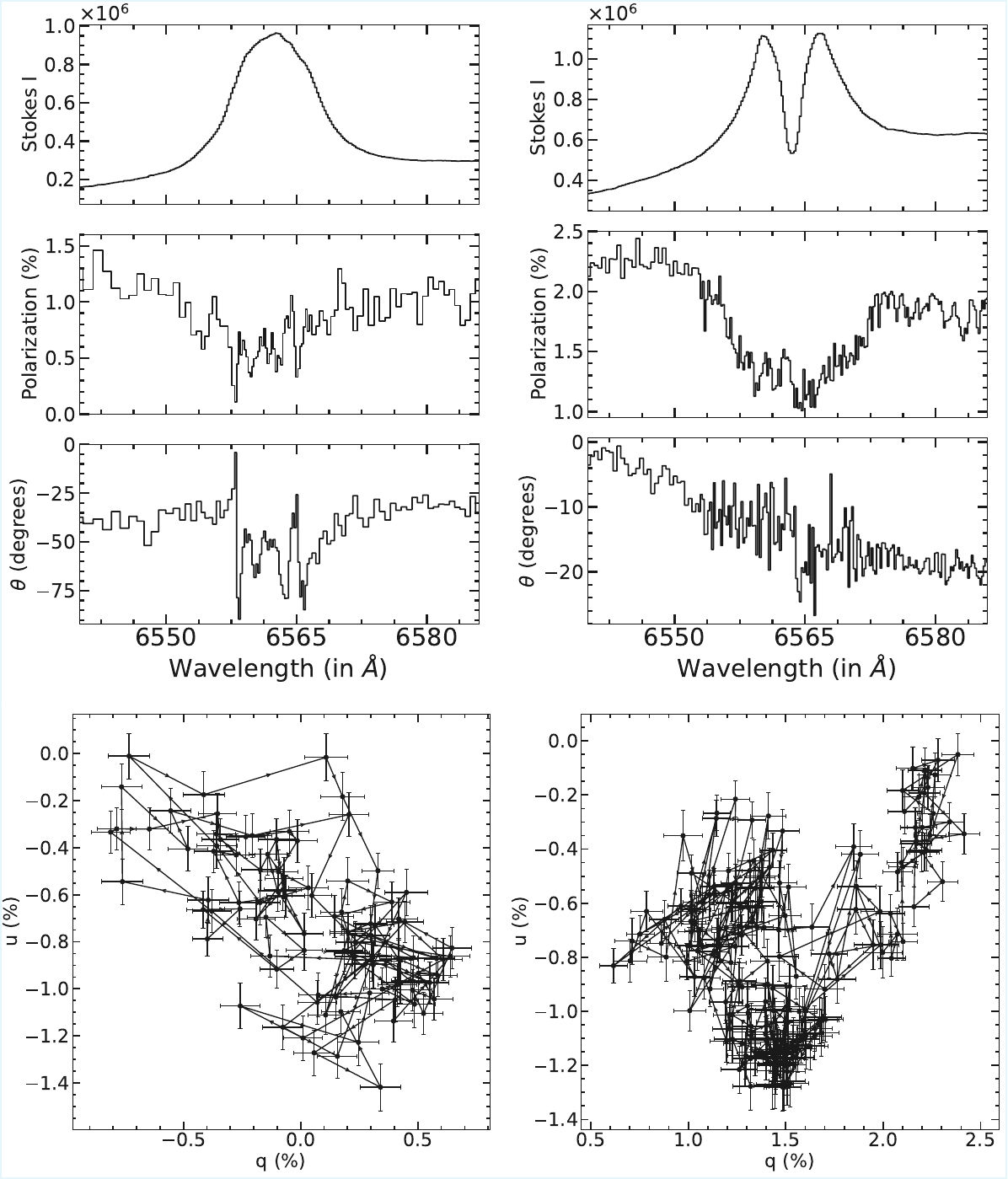}

  \caption{Polarization spectra for Classical Be stars $\phi$ Per \textit{(left)} and $\zeta$ Tau \textit{(right)}. For each star, the Stokes I spectrum is shown in the top panel of the triplot, the degree of polarization is shown in the middle panel, and the angle of polarization is shown in the bottom panel. The data for all three stars have been binned to a polarization accuracy of $0.1\%$ per binning element. The bottom panels represent the corresponding $q$-$u$ plots.}
  \label{ClassicalBe1}

\end{center}
\end{figure}

A few of the results are presented in Figure~\ref{ClassicalBe1} showing the depolarization effect across the H$\alpha$ profile of the star. The Be stars $\phi$ Per and $\zeta$ Tau show a clear depolarization effect across the H$\alpha$ profile, where the effect of depolarization is as wide as the spectral line width.

\subsection{Symbiotic Stars}

Symbiotic stars are interacting binary systems consisting of an evolved red giant and a hot radiating source, typically a white dwarf, with orbital periods of several hundred days to several decades \cite{kenyon1986symbiotic}. The temperature of the hot component is typically few tens to hundreds of thousand Kelvin \cite{kenyon1986symbiotic, munari2019symbiotic}, which ionizes the cool giant's wind, producing a dense and highly ionized nebula and resulting in an emission line spectrum of the star, wherein high excitation lines of various species are usually seen in addition to hydrogen's Balmer series lines in optical regime e.g. H$\alpha$ etc. These emission lines are rich in their polarization properties owing to an environment suitable for a variety of scattering phenomena causing polarization.
\par 
Several symbiotic stars show broad H$\alpha$ wings. This broad wing could be result of either hydrodynamical motion or inelastic Raman scattering of Lyman $\beta$ by neutral hydrogen in the surroundings of red giants. To decouple between these two scenarios, spectro-polarimetry of the H$\alpha$ profile of such stars provide a powerful diagnostic tool. If the broadening of the H$\alpha$ wing is indeed due to Raman scattering phenomena, some polarimetric signature is expected to be detected at the wings \cite{lee2000raman, yoo2002polarization, ikeda2004polarized}.  Another unique feature of the Symbiotic star spectra is the presence of the emission features at $6825\AA$ and $7082\AA$, which is not seen in any other type of stellar system \cite{schmid1994raman, harries1996raman}. It was shown in late 1980s \cite{nussbaumer1989raman, schmid1989identification} that these features are produced by the Raman scattering of O VI doublet lines ($1032\AA$ and $1038\AA$ in the far-UV regime, which originate near the hot component) by the neutral hydrogen wind of the cool star. After scattering, these features fall in the visible regime with broader widths as expected from the Raman scattering process \cite{yoo2002polarization, chang2018broad}. As these features is caused by scattering events, they show strong polarization.
\par
Several observations of Symbiotic stars had been conducted with ProtoPol on PRL 2.5m telescope. The sample selection criterion was the same as that for Herbig stars, that is, brightness of the stars (V $\lesssim$ 11) and the star must be showing H$\alpha$ emission. The observation log of the sample of symbiotics stars is given in Table~\ref{SymbioticTable}. 


\begin{table*} 
\caption{Sample of Symbiotic Stars. } \label{SymbioticTable}
\centering
\setlength{\tabcolsep}{5pt}
\renewcommand{\arraystretch}{1.2}
	\begin{tabular}{cc cc cc cc cc cc }
           \hline
		  \hline

\textbf{Name} & \textbf{Vmag} & \textbf{Date} & \textbf{Exposure (s)} \\
\hline
\hline
RW Hya & 10.0  & 31-03-2024 & 600s x 4 (3 sets) \\
       &       & 02-04-2024 & 600s x 4 (3 sets)  \\
       &       & 28-04-2024 & 600s x 4 (5 sets)  \\
Y Gem  & 9.09  & 04-03-2024 & 900s x 4 (2 sets)  \\
       &      & 05-03-2024 & 600s x 4 (2 sets)  \\
       &      & 06-03-2024 & 600s x 4 (4 sets)  \\
       &      & 10-03-2024 & 600s x 4 (2 sets)  \\
       &      & 01-03-2025 & 900s x 4 (3 sets)  \\
UV Aur & 10.41  & 06-03-2024 & 600s x 4 (3 sets)  \\
AG Peg & 8.69   & 10-05-2024 & 600s x 4 (2 sets)  \\
       &       & 29-05-2024 & 600s x 4 (1 set)  \\
       &      & 30-05-2024 & 600s x 4 (1 set)  \\
T CrB  & 10.25 & 06-01-2025 & 900s x 4 (1 set)  \\
       &       & 25-01-2025 & 1800s x 4 (1 set) \\
       &       & 01-03-2025 & 900s x 4 (3 sets)  \\
       &       & 02-03-2025 & 900s x 4 (3 sets)  \\
Z And  & 8.00 & 26-01-2025 & 900s x 4 (1 set)  \\
AG Dra & 9.74 & 08-05-2024 & 600s x 4 (5 sets) \\
       &      & 09-05-2024 & 600s x 4 (6 sets)  \\
       &       & 02-04-2025 & 900s x 4 (2 sets)  \\
        
		\hline
		\hline
	\end{tabular}
\end{table*}

\begin{figure}[H]
  \begin{center}
    \includegraphics[width=\textwidth]{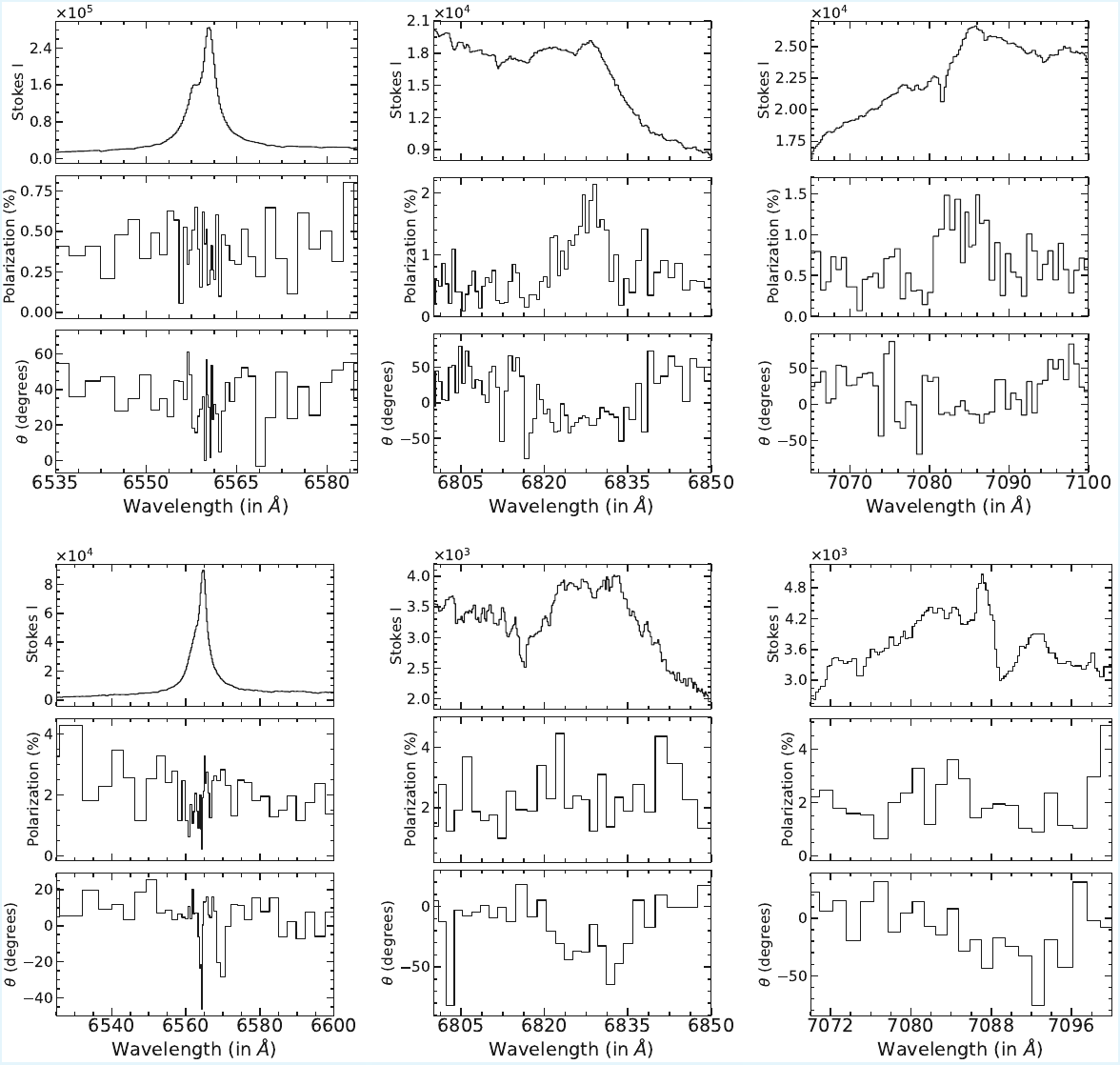}

  \caption{Polarization spectra for Symbiotic stars AG Dra \textit{(top)} and Z And \textit{(bottom)}. Spectro-polarimetry of H$\alpha$, Raman scattered $6825\AA$ and $7082\AA$ lines are shown in the left, middle, and right plots, respectively.  For each star, the Stokes I spectrum, the degree of polarization and the angle of polarization are shown in the top, middle and bottom panels respectively. The data for AG Dra is binned to a polarization accuracy of $0.15\%$ per binning element for H$\alpha$ and $0.3\%$ per binning element for the Raman $6825\AA$ and $7082\AA$ lines. The data for Z And is binned to a polarization accuracy of $0.3\%$ per binning element for H$\alpha$ and $0.5\%$ per binning element for the Raman $6825\AA$ and $7082\AA$ lines. Pseudo-ripples might be noticed in the wings of the profiles having lower SNR as pointed out in section~\ref{subsec:RelativeFluxCalib}.}
  \label{Symbiotic1}

\end{center}
\end{figure}

The polarization measurements of two symbiotic stars AG Dra and Z And are presented here. From Figure~\ref{Symbiotic1} for AG Dra, one can clearly see that the Raman scattered $6825\AA$ and $7082\AA$ lines are polarized. However, no significant polarization structure is observed in the wings of H$\alpha$. On the other hand, for Z And one can see a faint polarization signal at the Raman scattered $6825\AA$ and $7082\AA$ lines as well as the H$\alpha$ wings of the star. It is also interesting to note the change in the polarization angle at the H$\alpha$ wing of Z And. 
\par 
\textbf{AG Dra:} It is a well-studied symbiotic system for its spectro-polarimetry \cite{schmid1994raman, ikeda2004polarized}. It shows Raman scattered features at $6825\AA$ and $7082\AA$. Both the features show a polarization change as reported by \cite{schmid1994raman}. The Raman line $6825\AA$ show a significant polarization of $\sim 1.5\%$ above the continuum, whereas the $7082\AA$ line shows $\sim 0.8\%$ polarization above the continuum level, both having similar polarization profiles. On the other hand, AG Dra is also known to show a small polarization at the H$\alpha$ wings as previously reported by \cite{ikeda2004polarized}. Observations from ProtoPol reveal similar polarization in the Raman scattered features what has been reported by \cite{schmid1994raman}. However, no visible polarization signal was detected at the H$\alpha$ wings.  
\par 
\textbf{Z And:} It is also another well-studied symbiotic star and is also known to show Raman scattered features. Spectro-polarimetric observations by \cite{schmid1994raman} show a polarization increase at $6825\AA$ line and a smaller change across the $7082\AA$ line. At the H$\alpha$ wings as well, a very small polarization is reported by \cite{ikeda2004polarized}. Observations from ProtoPol reveal a very small polarization change at the H$\alpha$ line and the Raman scattered features at $6825\AA$ and $7082\AA$, although a change in angle of polarization is more prominently visible across all the lines. 

\subsection{AGB Stars}
\label{subsec-AGB}

Asymptotic Giant Branch (AGB) stars are low- to intermediate-mass stars, typically with starting masses ranging from about 0.9 - 10 $\mathrm{M}_\odot$ in a late thermally-pulsing evolutionary stage, distinguished by significant (Mira-type) pulsations and robust dust-driven winds \cite{herwig2005evolution}. In this phase, stars experience sequential helium- and hydrogen-burning shell flashes and convective dredge-up, enriching their envelopes with newly synthesized elements (e.g., carbon and s-process nuclei) while ejecting gas and dust into the interstellar medium \cite{herwig2005evolution, siess2006evolution}. AGB stars significantly contribute to galactic chemical evolution by returning nucleosynthesis products and abundant dust grains, whether carbonaceous or silicate, to the Galaxy \cite{willson2000mass}. Spectro-polarimetric studies of these stars offer distinctive diagnostics of these intricate environments \cite{boyle1986ccd, bieging2006optical}. High-resolution circular spectro-polarimetry (Stokes V) can identify Zeeman-induced polarization in AGB star spectra, uncovering weak (about gauss-level) surface magnetic fields \cite{lebre2014search}, while linear polarization of molecular lines or scattered continuum light traces the geometry of the dusty circumstellar shell \cite{bieging2006optical}. Imaging polarimetry has revealed polarized arcs and shells around AGB stars \cite{ramstedt2011imaging}. Recent spectropolarimetric studies of molecular emissions (e.g., CO, SiO) in AGB winds in the radio and millimeter domain characterize the three-dimensional structure and intensity of magnetic fields and dust alignment within the envelope \cite{vlemmings2024molecular}. Therefore, these spectro-polarimetric observations complement conventional spectroscopy and imaging, offering fresh insights into the magnetic fields, dust characteristics, and asymmetric mass-loss mechanisms that dictate AGB growth. 


\begin{table*} 
\caption{Sample of AGB/post-AGB Stars. } \label{AGBTable}
\centering
\setlength{\tabcolsep}{5pt}
\renewcommand{\arraystretch}{1.2}
	\begin{tabular}{cc cc cc cc cc }
           \hline
		  \hline

\textbf{Name} & \textbf{Vmag} & \textbf{Date} & \textbf{Exposure (s)} \\
\hline  
\hline
\multicolumn{4}{c}{\textbf{Observations from PRL 1.2m Telescope}}\\
\hline
Omi Cet & 6.53 &  25-01-2024 & 300s x 4 (4 sets) \\
R Leo   & 7.53 &  23-01-2024 & 300s x 4 (5 sets)  \\
V CVn   & 6.74 &  15-01-2024 & 300s x 4 (2 sets)  \\
R Hya   & 4.97 &  09-02-2024 & 600s x 4 (2 sets)  \\
V Hya   & 6.80 &  09-02-2024 & 600s x 4 (3 sets)  \\
W Hya   & 7.70 &  10-02-2024 & 300s x 4 (4 sets)  \\
LN Hya  & 8.69 &  10-02-2024 & 900s x 4 (2 sets)  \\
AG Ant  & 5.53 &  11-02-2024 & 900s x 4 (1 set)  \\
89 Her  & 5.36 &  11-02-2024 & 900s x 4 (1 set)  \\
\hline
\multicolumn{4}{c}{\textbf{Observations from PRL 2.5m Telescope}}\\
\hline
TV Gem  & 6.56 &  06-03-2024 & 600s x 4 (1 set)  \\
$\Omega$ Vir & 5.36 &  12-03-2024 & 300s x 4 (2 sets)  \\
RX Boo  & 8.60 &   28-03-2024 & 600s x 4 (1 set)  \\
Y CVn   & 4.87 &  29-03-2024 & 600s x 4 (1 set)  \\
U Hya   & 4.82 &  29-03-2024 & 300s x 4 (1 set)  \\
G Her   & 5.01 &  31-03-2024 & 300s x 4 (1 set)  \\
LQ Her  & 5.70 &  31-03-2024 & 300s x 4 (1 set)  \\
X Her   & 6.58 &  31-03-2024 & 300s x 4 (1 set)  \\
BQ Gem  & 5.00 &  01-04-2024 & 300s x 4 (1 set)  \\
R Gem   & 7.68 &  02-04-2024 & 600s x 4 (2 sets)  \\
U Her   & 6.70 &  02-04-2024 & 300s x 4 (2 sets)  \\
CU Dra  & 4.66 &  10-04-2024  & 120s x 4 (1 set)  \\
$\psi$ Vir  & 4.80 & 10-04-2024  & 300s x 4 (1 set)  \\
V636 Her& 5.87 &  10-04-2024  & 180s x 4 (2 sets)  \\
FS Com  & 5.61 &  26-04-2024  & 30s x 4 (1 set)  \\
RT Vir  & 7.41 &  26-04-2024  & 300s x 4 (1 set)  \\
ST UMa  & 6.28 &  26-04-2024  & 300s x 4 (1 set)  \\    
SW Vir  & 6.85 &  26-04-2024  & 180s x 4 (1 set)  \\       
TU CVn  & 5.84 &  26-04-2024  & 60s x 4 (1 set)  \\ 
BK Vir  & 7.28 &  26-04-2024  & 600s x 4 (1 set)  \\ 

		\hline
		\hline
	\end{tabular}
\end{table*}

Several AGB stars were observed to trace the continuum polarization of the targets. ProtoPol observations were conducted from PRL 1.2m and PRL 2.5m telescopes. The complete sample of target AGB/post-AGB stars are given in Table~\ref{AGBTable}. The polarization in the continuum of three 

\begin{figure}[H]
  \begin{center}
    \includegraphics[width=0.8\textwidth]{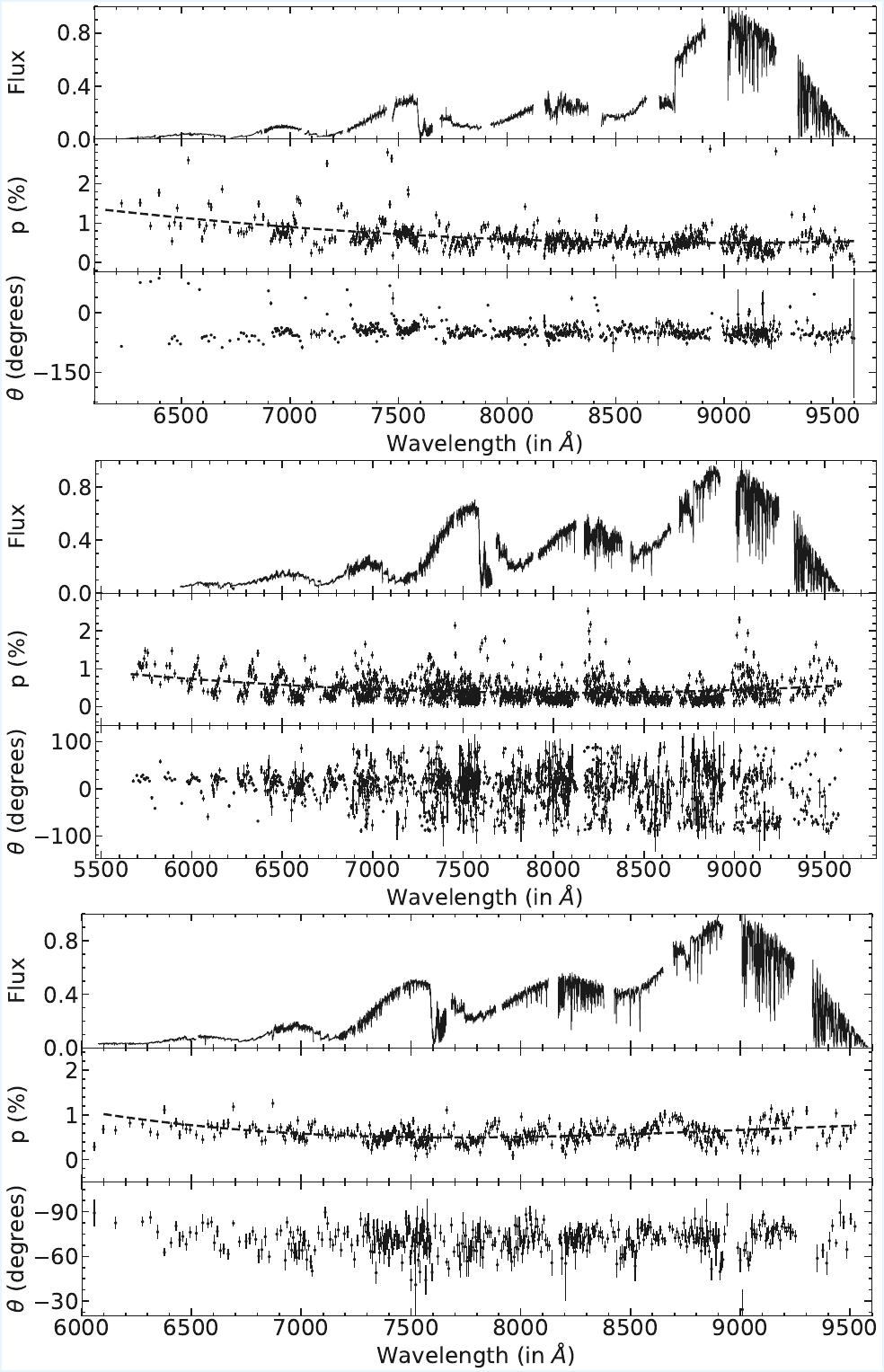}
  
  \caption{Continuum polarization of observed AGB stars BK vir \textit{(top)}, G Her \textit{(middle)} and TU CVn \textit{(bottom)}. The top panels of the triplots show relative flux calibrated spectra of the stars, the middle panel shows the degree of polarization, and the bottom panel shows the angle of polarization. The dashed line in the middle panel shows the best fit to the degree of polarization data. The data has been dynamically binned to a polarization accuracy of $0.1\%$.}
  \label{AGB_plots}

\end{center}
\end{figure}

AGB/post-AGB stars are shown in Figures~\ref{AGB_plots}. As one can see from some of the plots, the polarization is higher for the bluer wavelengths as compared to red, while in some others, the polarization remains roughly constant over the wavelength range or even increases towards the redder end.


\section{Summary}
\label{sec-summary}

This paper presents the on-sky characterization and first science results of ProtoPol, an echelle spectro-polarimeter developed for PRL 1.2m and 2.5m  telescopes. A fully automated data reduction pipeline was developed to extract science-ready data from the raw frames. Observations of standard polarized and unpolarized stars were conducted to determine the instrumental polarization and other polarization-dependent factors, while observations of Jupiter were used to calculate the instrumental throughput. Several spectro-polarimetric science observation programs were conducted to test the performance of the instrument, including Herbig Ae/Be stars, classical Be stars, Symbiotic stars, and AGB/post-AGB stars. The results obtained are also presented in this paper as first science results from ProtoPol.
\par
The fully automated spectro-polarimetric data reduction pipeline was developed to process the raw science frames obtained from ProtoPol observations and produce science-ready data at the output. The pipeline algorithm detects the echelle order peaks and subsequently traces the orders for extraction purposes. U-Ar calibration frames were used to perform wavelength calibration of the frames, while standard spectro-photometric star observations were used to perform relative flat-field corrections to the data. The extracted intensities from the four HWP position observations were used to compute the Stokes parameters and hence the polarization of the source as a function of wavelength. Any false spectral or polarization signatures due to cosmic rays, scattered background light, and misalignment of spectra were taken care of during the reduction process.
\par 
ProtoPol was thoroughly characterized on PRL 1.2m and 2.5m telescopes between December 2023 and May 2025. A verity of astrophysical targets were chosen with good sample size to evaluate its performance in varying astrophysical situations and observing conditions. ProtoPol performed exceedingly well to make the polarization measurements across the spectral line profile and in the continuum, which have been verified with objects showing known effects causing polarization. From these extensive observing campaigns, it is established that ProtoPol achieves a spectral resolution in the range of $\sim$0.40-0.75 $\AA$ across various orders, polarization accuracy $\sim 0.1-0.2\%$ with magnitude of $m_V = 8$ in 2 hours of total integration time. The instrumental polarization (including telescope optics) is determined to be $\delta p \sim 0.1\%$ per pixel. For spectroscopy, a source of $m_V\sim13.2$ magnitude was observed to achieve SNR $\sim 10$ per pixel in 1 hour of integration time.
\par 
Development of ProtoPol presents a novel way of developing spectro-polarimeters for small to medium aperture telescopes. As it is designed and developed completely with off-the-shelf components, the approach offers significant reduction in the cost and in the development time cycle. Initially, conceived only as a prototype instrument for the currently under-development purpose-specific M-FOSC-EP instrument, the subsequent design performance of ProtoPol prompted us to develop it as a full-fledged instrument for PRL telescopes. Though the expected performances of some of the components (such as off-the-shelf camera etc) could not be modeled and only guessed from their performance in other instruments, it turned out to be of desired quality. So ProtoPol also validates the use of such components in scientific instrumentation. The spectral resolution and limiting magnitude achieved with ProtoPol are also of interest to small/medium (2-3m class) aperture observatories wherein very high resolution spectrographs, (though cater to specific science cases) often limit to only bright targets, in addition to being of high cost in nature. Medium resolution instruments such as ProtoPol could thus not only be cost-effective, they will also enhance the observing parameter space and hence can be of general use. This is of particular importance for the science domains like spectro-polarimetry, which are traditionally not so common, partially due to the cost of the instrument and the requirement for large integration times, which is rather competitive to get at large aperture telescopes. We hope the successful demonstration of ProtoPol on PRL telescope would pave the way to develop similar instruments on small/medium aperture telescopes around the world and thereby bring spectro-polarimetry to the forefront of modern astrophysics.

\subsection*{Disclosures}
The authors declare that there are no financial interests, commercial affiliations, or other potential conflicts of interest that could have influenced the objectivity of this research or the writing of this paper

\subsection* {Code, Data, and Materials Availability} 
The data presented in this manuscript may be made available upon reasonable request. Corresponding authors may be contacted for that.

\subsection* {Acknowledgments}
We thank the anonymous reviewers for the review and their constructive comments and suggestion on the manuscript. Development of the ProtoPol instrument has been funded by the Department of Space, Government of India through Physical Research Laboratory (PRL), Ahmedabad. ProtoPol team is thankful to the Director, PRL, for supporting the ProtoPol and M-FOSC-EP development program. MKS expresses sincere thanks to Shyam N. Tandon (Inter-University Center for Astronomy and Astrophysics – IUCAA, Pune, India) for detailed discussions on several aspects of spectro-polarimeter design throughout the development process. AM is thankful to PRL for his Ph.D. research fellowship. RP thanks PRL for her post-doctoral fellowship. ProtoPol team expresses sincere thanks to Mt. Abu observatory staff for their sustained help and support during ProtoPol commissioning and subsequent observations.



\begin{thebibliography}{10}

\bibitem{ikeda2003development}
Y.~Ikeda, H.~Akitaya, K.~Matsuda, {\em et~al.}, ``Development of the high-resolution spectropolarimeter: Lips,'' in {\em Polarimetry in Astronomy},   {\bf 4843}, 437--447, SPIE  (2003).

\bibitem{arasaki2015very}
T.~Arasaki, Y.~Ikeda, Y.~Shinnaka, {\em et~al.}, ``The very precise echelle spectropolarimeter on the araki telescope (vespola),'' {\em Publications of the Astronomical Society of Japan} {\bf 67}(3), 35  (2015).

\bibitem{piskunov2011harpspol}
N.~Piskunov, F.~Snik, A.~Dolgopolov, {\em et~al.}, ``Harpspol—the new polarimetric mode for harps,'' {\em The Messenger} {\bf 143}(7)  (2011).

\bibitem{donati2003espadons}
J.-F. Donati, ``Espadons: An echelle spectropolarimetric device for the observation of stars at cfht,'' in {\em Solar Polarization},   {\bf 307}, 41  (2003).

\bibitem{rudy1978polarization}
R.~J. Rudy, ``Polarization from thomson scattering of the light of a spherical, limb-darkened star.,'' {\em Publications of the Astronomical Society of the Pacific} {\bf 90}(538), 688  (1978).

\bibitem{cassinelli1987polarization}
J.~P. Cassinelli, K.~Nordsieck, and M.~Murison, ``Polarization of light scattered from the winds of early-type stars,'' {\em Astrophysical Journal, Part 1 (ISSN 0004-637X), vol. 317, June 1, 1987, p. 290-302.} {\bf 317}, 290--302  (1987).

\bibitem{bastin2022mount}
C.~Bastin, G.~P. Lousberg, O.~Pirnay, {\em et~al.}, ``Mount abu 2.5 m telescope: first light and performance assessment,'' in {\em Modeling, Systems Engineering, and Project Management for Astronomy X},   {\bf 12187}, 265--275, SPIE  (2022).

\bibitem{Deshpande1995}
M.~R. {Deshpande}, ``{A brief report on the Infrared Telescope at Gurushikhar, MT Abu},'' {\em Bulletin of the Astronomical Society of India} {\bf 23}, 13  (1995).

\bibitem{kumar2022designs}
V.~Kumar, M.~K. Srivastava, V.~Dixit, {\em et~al.}, ``Designs of mt. abu faint object spectrograph and camera-echelle polarimeter (m-fosc-ep) and its prototype: spectro-polarimeters for prl 1.2 m and 2.5 m mt. abu telescopes, india,'' in {\em Ground-based and Airborne Instrumentation for Astronomy IX},   {\bf 12184}, 1696--1714, SPIE  (2022).

\bibitem{srivastava2024development}
M.~K. Srivastava, A.~Maiti, V.~Kumar, {\em et~al.}, ``Development of protopol: a medium resolution echelle spectro-polarimeter for prl 1.2 m and 2.5 m telescopes, mt abu, india,'' in {\em Ground-based and Airborne Instrumentation for Astronomy X},   {\bf 13096}, 577--592, SPIE  (2024).

\bibitem{harrington2008spectropolarimetric}
D.~Harrington and J.~R. Kuhn, ``Spectropolarimetric observations of herbig ae/be stars. i. hivis spectropolarimetric calibration and reduction techniques,'' {\em Publications of the Astronomical Society of the Pacific} {\bf 120}(863), 89  (2008).

\bibitem{baranne1996elodie}
A.~Baranne, D.~Queloz, M.~Mayor, {\em et~al.}, ``Elodie: A spectrograph for accurate radial velocity measurements,'' {\em Astronomy and Astrophysics Supplement Series} {\bf 119}(2), 373--390  (1996).

\bibitem{bernstein2015data}
R.~M. Bernstein, S.~M. Burles, and J.~X. Prochaska, ``Data reduction with the mike spectrometer,'' {\em Publications of the Astronomical Society of the Pacific} {\bf 127}(955), 911  (2015).

\bibitem{piskunov2021optimal}
N.~Piskunov, A.~Wehrhahn, and T.~Marquart, ``Optimal extraction of echelle spectra: Getting the most out of observations,'' {\em Astronomy \& Astrophysics} {\bf 646}, A32  (2021).

\bibitem{churchill1995treatment}
C.~W. Churchill and S.~Allen, ``A treatment for the background correction on the hamilton echelle spectrograph,'' {\em Publications of the Astronomical Society of the Pacific} {\bf 107}(708), 193  (1995).

\bibitem{van2001cosmic}
P.~G. Van~Dokkum, ``Cosmic-ray rejection by laplacian edge detection,'' {\em Publications of the Astronomical Society of the Pacific} {\bf 113}(789), 1420  (2001).

\bibitem{sarmiento2018comparing}
L.~Sarmiento, A.~Reiners, P.~Huke, {\em et~al.}, ``Comparing the emission spectra of u and th hollow cathode lamps and a new u line list,'' {\em Astronomy \& Astrophysics} {\bf 618}, A118  (2018).

\bibitem{lovis2007new}
C.~Lovis and F.~Pepe, ``A new list of thorium and argon spectral lines in the visible,'' {\em Astronomy \& Astrophysics} {\bf 468}(3), 1115--1121  (2007).

\bibitem{oudmaijer1999halpha}
R.~D. Oudmaijer and J.~E. Drew, ``H$\alpha$ spectropolarimetry of b [e] and herbig be stars,'' {\em Monthly Notices of the Royal Astronomical Society} {\bf 305}(1), 166--180  (1999).

\bibitem{vink2005probing}
J.~S. Vink, J.~E. Drew, T.~J. Harries, {\em et~al.}, ``Probing the circumstellar structures of t tauri stars and their relationship to those of herbig stars,'' {\em Monthly Notices of the Royal Astronomical Society} {\bf 359}(3), 1049--1064  (2005).

\bibitem{ikeda2004polarized}
Y.~Ikeda, H.~Akitaya, K.~Matsuda, {\em et~al.}, ``Polarized h$\alpha$ wings in the symbiotic stars ag draconis and z andromedae,'' {\em The Astrophysical Journal} {\bf 604}(1), 357  (2004).

\bibitem{patat2006error}
F.~Patat and M.~Romaniello, ``Error analysis for dual-beam optical linear polarimetry1,'' {\em Publications of the Astronomical Society of the Pacific} {\bf 118}(839), 146  (2006).

\bibitem{vink2002probing}
J.~S. Vink, J.~E. Drew, T.~J. Harries, {\em et~al.}, ``Probing the circumstellar structure of herbig ae/be stars,'' {\em Monthly Notices of the Royal Astronomical Society} {\bf 337}(1), 356--368  (2002).

\bibitem{srivastava2021design}
M.~K. Srivastava, V.~Kumar, V.~Dixit, {\em et~al.}, ``Design and development of mt. abu faint object spectrograph and camera--pathfinder (mfosc-p) for prl 1.2 m mt. abu telescope,'' {\em Experimental Astronomy} {\bf 51}(2), 345--382  (2021).

\bibitem{vskoda2008investigation}
P.~{\v{S}}koda, B.~{\v{S}}urlan, and S.~Tomi{\'c}, ``Investigation of residual blaze functions in slit-based echelle spectrograph,'' in {\em Ground-based and Airborne Instrumentation for Astronomy II},   {\bf 7014}, 2037--2048, SPIE  (2008).

\bibitem{suzuki2003relative}
N.~Suzuki, D.~Tytler, D.~Kirkman, {\em et~al.}, ``Relative flux calibration of keck hires echelle spectra1,'' {\em Publications of the Astronomical Society of the Pacific} {\bf 115}(811), 1050  (2003).

\bibitem{harding2016chimera}
L.~K. Harding, G.~Hallinan, J.~Milburn, {\em et~al.}, ``Chimera: a wide-field, multi-colour, high-speed photometer at the prime focus of the hale telescope,'' {\em Monthly Notices of the Royal Astronomical Society} {\bf 457}(3), 3036--3049  (2016).

\bibitem{Filippenko1982Atmospheric}
A.~V. {Filippenko}, ``{The importance of atmospheric differential refraction in spectrophotometry.},'' {\em Publications of the Astronomical Society of the Pacific} {\bf 94}, 715--721  (1982).

\bibitem{gubler1998differential}
J.~Gubler and D.~Tytler, ``Differential atmospheric refraction and limitations on the relative astrometric accuracy of large telescopes,'' {\em Publications of the Astronomical Society of the Pacific} {\bf 110}(748), 738  (1998).

\bibitem{chakraborty2003optical}
P.~Chakraborty and R.~Vasundhara, ``An optical, dual-beam, automated medium resolution spectropolarimeter for the vainu bappu telescope,'' {\em Experimental Astronomy} {\bf 16}(2), 69--84  (2003).

\bibitem{harrington2008spectropolarimetric_a}
D.~Harrington and J.~R. Kuhn, ``Spectropolarimetric observations of herbig ae/be stars. i. hivis spectropolarimetric calibration and reduction techniques,'' {\em Publications of the Astronomical Society of the Pacific} {\bf 120}(863), 89  (2008).

\bibitem{harries2002spectropolarimetry}
T.~J. Harries, I.~D. Howarth, and C.~J. Evans, ``Spectropolarimetry of o supergiants,'' {\em Monthly Notices of the Royal Astronomical Society} {\bf 337}(1), 341--355  (2002).

\bibitem{semel2003spectropolarimetry}
M.~Semel, ``Spectropolarimetry and polarization-dependent fringes,'' {\em Astronomy \& Astrophysics} {\bf 401}(1), 1--14  (2003).

\bibitem{clarke2005effects}
D.~Clarke, ``Effects in polarimetry of interference within wave plates,'' {\em Astronomy \& Astrophysics} {\bf 434}(1), 377--384  (2005).

\bibitem{harrington2017polarization}
D.~M. Harrington, F.~Snik, C.~U. Keller, {\em et~al.}, ``Polarization modeling and predictions for dkist part 2: application of the berreman calculus to spectral polarization fringes of beamsplitters and crystal retarders,'' {\em Journal of Astronomical Telescopes, Instruments, and Systems} {\bf 3}(4), 048001--048001  (2017).

\bibitem{harrington2018polarization}
D.~M. Harrington and S.~R. Sueoka, ``Polarization modeling and predictions for dkist part 3: focal ratio and thermal dependencies of spectral polarization fringes and optic retardance,'' {\em Journal of Astronomical Telescopes, Instruments, and Systems} {\bf 4}(1), 018006--018006  (2018).

\bibitem{harrington2020polarization}
D.~M. Harrington, S.~A. Jaeggli, T.~A. Schad, {\em et~al.}, ``Polarization modeling and predictions for daniel k. inouye solar telescope, part 6: fringe mitigation with polycarbonate modulators and optical contact calibration retarders,'' {\em Journal of Astronomical Telescopes, Instruments, and Systems} {\bf 6}(3), 038001--038001  (2020).

\bibitem{harrington2015correcting}
D.~M. Harrington, S.~V. Berdyugina, O.~Kuzmychov, {\em et~al.}, ``Correcting systematic polarization effects in keck lrisp spectropolarimetry to< 0.05\%,'' {\em Publications of the Astronomical Society of the Pacific} {\bf 127}(954), 757  (2015).

\bibitem{casini2012analysis}
R.~Casini, A.~G. de~Wijn, and P.~G. Judge, ``Analysis of seeing-induced polarization cross-talk and modulation scheme performance,'' {\em The Astrophysical Journal} {\bf 757}(1), 45  (2012).

\bibitem{liu2022study}
S.~Liu, J.~Su, X.~Bai, {\em et~al.}, ``A study on correcting the effect of polarization crosstalk in full-disk solar photospheric magnetic field observations,'' {\em Solar Physics} {\bf 297}(1), 6  (2022).

\bibitem{breckinridge2015polarization}
J.~B. Breckinridge, W.~S.~T. Lam, and R.~A. Chipman, ``Polarization aberrations in astronomical telescopes: the point spread function,'' {\em Publications of the Astronomical Society of the Pacific} {\bf 127}(951), 445  (2015).

\bibitem{chipman2015polarization}
R.~A. Chipman, W.~S.~T. Lam, and J.~Breckinridge, ``Polarization aberration in astronomical telescopes,'' in {\em Polarization Science and Remote Sensing VII},   {\bf 9613}, 124--134, SPIE  (2015).

\bibitem{serkowski1974many}
K.~Serkowski, ``For many astronomical objects the observed polarization is very small, making high polarimetric accuracy essential. polarimetric precision can,'' {\em Planets, Stars and Nebulae: Studied with Photopolarimetry} {\bf 23}, 135  (1974).

\bibitem{whittet1992systematic}
D.~Whittet, P.~Martin, J.~Hough, {\em et~al.}, ``Systematic variations in the wavelength dependence of interstellar linear polarization,'' {\em Astrophysical Journal, Part 1 (ISSN 0004-637X), vol. 386, Feb. 20, 1992, p. 562-577. Research supported by SERC, NSERC, and University of Toronto.} {\bf 386}, 562--577  (1992).

\bibitem{hsu1982standard}
J.-C. Hsu and M.~Breger, ``On standard polarized stars,'' {\em Astrophysical Journal, Part 1, vol. 262, Nov. 15, 1982, p. 732-738.} {\bf 262}, 732--738  (1982).

\bibitem{bouvier2007magnetospheric}
J.~Bouvier, S.~Alencar, T.~Boutelier, {\em et~al.}, ``Magnetospheric accretion-ejection processes in the classical t tauri star aa tauri,'' {\em Astronomy \& Astrophysics} {\bf 463}(3), 1017--1028  (2007).

\bibitem{alecian2013high}
E.~Alecian, G.~Wade, C.~Catala, {\em et~al.}, ``A high-resolution spectropolarimetric survey of herbig ae/be stars--i. observations and measurements,'' {\em Monthly Notices of the Royal Astronomical Society} {\bf 429}(2), 1001--1026  (2013).

\bibitem{waelkens1997comet}
C.~Waelkens, K.~Malfait, and L.~Waters, ``Comet hale-bopp, circumstellar dust, and the interstellar medium,'' {\em Earth, Moon, and Planets} {\bf 79}(1), 265--274  (1997).

\bibitem{de1994new}
D.~De~Winter, M.~Perez, {\em et~al.}, ``A new catalogue of members and candidate members of the herbig ae/be (haebe) stellar group,'' {\em Astronomy and Astrophysics Suppl., Vol. 104, p. 315-339 (1994)} {\bf 104}, 315--339  (1994).

\bibitem{ababakr2016linear}
K.~M. Ababakr, {\em Linear Spectropolarimetry Of Herbig Ae/Be Stars}.
\newblock PhD thesis, University of Leeds  (2016).

\bibitem{hillenbrand1992herbig}
L.~A. Hillenbrand, S.~E. Strom, F.~J. Vrba, {\em et~al.}, ``Herbig ae/be stars-intermediate-mass stars surrounded by massive circumstellar accretion disks,'' {\em Astrophysical Journal, Part 1 (ISSN 0004-637X), vol. 397, no. 2, p. 613-643.} {\bf 397}, 613--643  (1992).

\bibitem{bjorkman1998first}
K.~Bjorkman, A.~Miroshnichenko, J.~Bjorkman, {\em et~al.}, ``The first ultraviolet and optical spectropolarimetry of the b [e] star hd 50138,'' {\em The Astrophysical Journal} {\bf 509}(2), 904  (1998).

\bibitem{pogodin1997circumstellar}
M.~Pogodin, ``Circumstellar peculiarities in the unusual be star hd 50138.,'' {\em Astronomy and Astrophysics, v. 317, p. 185-192} {\bf 317}, 185--192  (1997).

\bibitem{baines2006binarity}
D.~Baines, R.~D. Oudmaijer, J.~M. Porter, {\em et~al.}, ``On the binarity of herbig ae/be stars,'' {\em Monthly Notices of the Royal Astronomical Society} {\bf 367}(2), 737--753  (2006).

\bibitem{mottram2007difference}
J.~C. Mottram, J.~Vink, R.~Oudmaijer, {\em et~al.}, ``On the difference between herbig ae and herbig be stars,'' {\em Monthly Notices of the Royal Astronomical Society} {\bf 377}(3), 1363--1374  (2007).

\bibitem{beskrovnaya2004active}
N.~Beskrovnaya and M.~Pogodin, ``Active phenomena in the circumstellar environment of the herbig ae star hd 31648,'' {\em Astronomy \& Astrophysics} {\bf 414}(3), 955--967  (2004).

\bibitem{beskrovnaya1999spectroscopic}
N.~Beskrovnaya, M.~Pogodin, A.~Miroshnichenko, {\em et~al.}, ``Spectroscopic, photometric, and polarimetric study of the herbig ae candidate hd 36112,'' {\em Astronomy and Astrophysics, v. 343, p. 163-174 (1999)} {\bf 343}, 163--174  (1999).

\bibitem{de2003spinning}
A.~D. de~Souza, P.~Kervella, S.~Jankov, {\em et~al.}, ``The spinning-top be star achernar from vlti-vinci,'' {\em Astronomy \& Astrophysics} {\bf 407}(3), L47--L50  (2003).

\bibitem{quirrenbach1997constraints}
A.~Quirrenbach, K.~Bjorkman, J.~Bjorkman, {\em et~al.}, ``Constraints on the geometry of circumstellar envelopes: optical interferometric and spectropolarimetric observations of seven be stars,'' {\em The Astrophysical Journal} {\bf 479}(1), 477  (1997).

\bibitem{kenyon1986symbiotic}
S.~J. Kenyon, ``Symbiotic stars,'' in {\em Interacting Binaries},  179--203, Springer  (1986).

\bibitem{munari2019symbiotic}
U.~Munari, ``The symbiotic stars,'' {\em The Impact of Binary Stars on Stellar Evolution} {\bf 54}, 77  (2019).

\bibitem{lee2000raman}
H.-W. Lee, ``Raman-scattering wings of h$\alpha$ in symbioticstars,'' {\em The Astrophysical Journal} {\bf 541}(1), L25  (2000).

\bibitem{yoo2002polarization}
J.~J. Yoo, J.-Y. Bak, and H.-W. Lee, ``Polarization of the broad h$\alpha$ wing in symbiotic stars,'' {\em Monthly Notices of the Royal Astronomical Society} {\bf 336}(2), 467--476  (2002).

\bibitem{schmid1994raman}
H.~Schmid and H.~Schild, ``Raman scattered emission lines in symbiotic stars: a spectropolarimetric survey,'' {\em Astronomy and Astrophysics (ISSN 0004-6361), vol. 281, no. 1, p. 145-160} {\bf 281}, 145--160  (1994).

\bibitem{harries1996raman}
T.~Harries and I.~Howarth, ``Raman scattering in symbiotic stars. i. spectropolarimetric observations,'' {\em Astronomy and Astrophysics Supplement Series} {\bf 119}(1), 61--90  (1996).

\bibitem{nussbaumer1989raman}
H.~Nussbaumer, H.~Schmid, and M.~Vogel, ``Raman scattering as a diagnostic possibility in astrophysics,'' {\em Astronomy and Astrophysics (ISSN 0004-6361), vol. 211, no. 2, March 1989, p. L27-L30. Research supported by SNSF.} {\bf 211}, L27--L30  (1989).

\bibitem{schmid1989identification}
H.~Schmid, ``Identification of the emission bands at 6830, 7088 a,'' {\em Astronomy and Astrophysics (ISSN 0004-6361), vol. 211, no. 2, March 1989, p. L31-L34. Research supported by SNSF.} {\bf 211}, L31--L34  (1989).

\bibitem{chang2018broad}
S.-J. Chang, H.-W. Lee, H.-G. Lee, {\em et~al.}, ``Broad wings around h$\alpha$ and h$\beta$ in the two s-type symbiotic stars z andromedae and ag draconis,'' {\em The Astrophysical Journal} {\bf 866}(2), 129  (2018).

\bibitem{herwig2005evolution}
F.~Herwig, ``Evolution of asymptotic giant branch stars,'' {\em Annu. Rev. Astron. Astrophys.} {\bf 43}(1), 435--479  (2005).

\bibitem{siess2006evolution}
L.~Siess, ``Evolution of massive agb stars-i. carbon burning phase,'' {\em Astronomy \& Astrophysics} {\bf 448}(2), 717--729  (2006).

\bibitem{willson2000mass}
L.~A. Willson, ``Mass loss from cool stars: impact on the evolution of stars and stellar populations,'' {\em Annual Review of Astronomy and Astrophysics} {\bf 38}(1), 573--611  (2000).

\bibitem{boyle1986ccd}
R.~Boyle, C.~Aspin, G.~Coyne, {\em et~al.}, ``Ccd spectropolarimetry of a sample of cool variable stars,'' {\em Astronomy and Astrophysics (ISSN 0004-6361), vol. 164, no. 2, Aug. 1986, p. 310-320.} {\bf 164}, 310--320  (1986).

\bibitem{bieging2006optical}
J.~H. Bieging, G.~D. Schmidt, P.~S. Smith, {\em et~al.}, ``Optical spectropolarimetry of asymptotic giant branch and post-asymptotic giant branch stars,'' {\em The Astrophysical Journal} {\bf 639}(2), 1053  (2006).

\bibitem{lebre2014search}
A.~L{\`e}bre, M.~Auri{\`e}re, N.~Fabas, {\em et~al.}, ``Search for surface magnetic fields in mira stars-first detection in $\chi$ cygni,'' {\em Astronomy \& Astrophysics} {\bf 561}, A85  (2014).

\bibitem{ramstedt2011imaging}
S.~Ramstedt, M.~Maercker, G.~Olofsson, {\em et~al.}, ``Imaging the circumstellar dust around agb stars with polcor,'' {\em Astronomy \& Astrophysics} {\bf 531}, A148  (2011).

\bibitem{vlemmings2024molecular}
W.~Vlemmings, B.~Lankhaar, and L.~Velilla-Prieto, ``Molecular line polarisation from the circumstellar envelopes of asymptotic giant branch stars,'' {\em Astronomy \& Astrophysics} {\bf 686}, A274  (2024).

\end{thebibliography}


\vspace{2ex}\noindent\textbf{First Author} is currently a Senior Research Fellow (SRF) - Ph.D. scholar -  at Physical Research Laboratory (PRL), Ahmedabad, India. Mr Arijit Maiti is holding a Master of Science Degree in Physics from National Institute of technology (NIT) Rourkela, India. He is responsible for the assembly-integration-test (AIT), subsequent on-sky characterization, and science verification of ProtoPol. His current research interests include optical instrumentation for ground-based telescopes, spectro-polarimetric and spectroscopic studies of symbiotic stars and novae.

\vspace{2ex}\noindent\textbf{Second Author} is currently an associate professor in the astronomy and astrophysics division at Physical Research Laboratory (PRL), Ahmedabad, India. Dr. Mudit K. Srivastava is the Principal Investigator (PI) of ProtoPol and M-FOSC-EP instruments. He is an astronomer \& an instrumentation scientist and have been involved in the development of optical instrumentation for the last two decades. 

\vspace{2ex}\noindent\textbf{Third Author} is currently a post-doctoral fellow at I. Physikalisches Institut, Universit\"at zu K\"oln. Dr. Vipin Kumar completed his doctoral thesis from the Physical Research Laboratory, Ahmedabad, India, wherein, as a part of his Ph.D. thesis, he developed the optical design of ProtoPol. He has also contributed towards the on-sky characterization of the instrument. His research focuses on developing Optical/IR instrumentation for ground-based telescopes, with a scientific emphasis on studying cool, low-mass stars.

\vspace{2ex}\noindent\textbf{Fourth Author} is a Technical Assistant (Mechanical) in the Physical Research Laboratory (PRL), Ahmedabad, India. Mr Bhaveshkumar Mistry received his Bachelor of Engineering (B.E.) degree in mechanical engineering. His expertise lies in the mechanical design and development of ground-based astronomical instrumentation. He has been responsible for the opto-mechanical design and subsequent fabrication of ProtoPol.

\vspace{2ex}\noindent\textbf{Fifth Author} is a Scientist/Engineer in the Astronomy and Astrophysics Division at the Physical Research Laboratory, Ahmedabad, India. Mrs. Ankita Patel received her Bachelor of Technology (B.Tech.) degree from Visvesvaraya Technological University, Bengaluru, India, in Electronics and Communication Engineering. She has been working on the control system and software aspects of instrumentation for ground-based astronomy. She has been responsible for the development of the control system and the instrument operation software of ProtoPol.

\vspace{2ex}\noindent\textbf{Sixth Author} is currently an engineer at Advanced Engineering Group, Azista Industries, Ahmedabad, India. Prior to this, Mr Vaibhav Dixit was a scientist/Engineer at Physical Research Laboratory, Ahmedabad, India. He has got his Bachelor's degree in Physical Sciences from Indian Institute of Space Science and Technology (IIST), Thiruvananthapuram, India and Master in Computational Sciences from University of Heidelberg, Germany. He contributed towards the optics and system design of the ProtoPol

\vspace{2ex}\noindent\textbf{Seventh Author} is currently an Associate Research Scientist at NASA Goddard Space Flight Center (Code 662) and the Department of Physics \& Astronomy, John Hopkins University, Maryland, USA.  Dr Ruchi Pandey had been a post-doctoral fellow at Physical Research Laboratory, Ahmedabad, India wherein she contributed towards the on-sky characterization and science verification of ProtoPol. Prior to that she had received her Ph.D. degree from S. N. Bose National Centre for Basic Sciences (University of Calcutta), Kolkata, India.

\vspace{2ex}\noindent\textbf{Eighth Author} is a Scientist/Engineer at Physical Research Laboratory (PRL), Ahmedabad, India. My Jay Chitroda received his dual degree of Bachelor of technology (B. Tech.) in Engineering Physics and M.Sc. in Astronomy and Astrophysics from Indian Institute of Space Science and Technology (IIST), Thiruvananthapuram, India. He has been contributing towards the operations and other operational aspects of ProtoPol.

\listoffigures
\listoftables

\end{spacing}
\end{document}